\documentclass{article}

\makeatletter
\def\thanks#1{\protected@xdef\@thanks{\@thanks
        \protect\footnotetext{#1}}}
\makeatother

\usepackage[utf8]{inputenc}

\usepackage{xcolor}
\definecolor{blueviolet}{rgb}{0.2, 0.2, 0.6}
\definecolor{webgreen}{rgb}{0,.5,0}
\definecolor{webbrown}{rgb}{.6,0,0}
\usepackage[pdftex,
  bookmarks=false,
  colorlinks=true, 
  urlcolor=webbrown,
  linkcolor=blueviolet, 
  citecolor=webgreen,
  pdfstartpage=1,
  pdfstartview={FitH},  
  bookmarksopen=false
  ]{hyperref}
\usepackage{amsmath}
\usepackage{amssymb}
\usepackage{amsfonts}
\usepackage{cleveref}

\usepackage{enumerate}
\usepackage{comment}
\usepackage{xpatch}
\usepackage{graphicx}
\usepackage{tabularx}
\usepackage{braket}
\usepackage{physics}
\usepackage{amsthm}
\usepackage{tikz}
\usepackage{commath}
\usepackage{mathtools}
\usepackage{qcircuit}
\usepackage{algorithm}
\usepackage{algorithmicx}
\usepackage{algpseudocode}
\usepackage{tabto}
\usepackage{pgf-umlsd}
\usepackage{bm}
\usepackage{multirow, multicol}
\usepackage{float}
\usepackage{pdfpages}
\usepackage{caption}
\usepackage{subcaption}
\usepackage{authblk}
\usepackage{makecell}
\usepackage{cite}
\usepackage{subcaption}

\DeclareMathOperator*{\argmax}{arg\,max}
\DeclareMathOperator*{\argmin}{arg\,min}

\newcommand{\mc}[1]{\mathcal{#1}}

\newcommand{\mrm}[1]{\mathrm{#1}}

\newcommand{\mbf}[1]{\mathbf{#1}}

\newcommand{\dtr}{\mathrm{d_{tr}}}

\newcommand{\davg}{\mathrm{d_{avg}}}
\newcommand{\supp}{\mathsf{supp}}
\newcommand{\opt}{\mathsf{opt}}
\newcommand{\pauli}{\mathcal{P}}
\newcommand{\proju}{\mathrm{Proj_U}}
\newcommand{\qacz}{\mathsf{QAC}^0}
\newcommand{\clifford}{\mrm{Cl}}
\newcommand{\infidel}{\overline{\mathrm{F}}}

\newcounter{protocol}
\makeatletter

\newtheorem{theorem}{Theorem}
\newtheorem{prop}{Proposition}
\newtheorem{lemma}{Lemma}
\newtheorem{corollary}{Corollary}
\newtheorem{claim}{Claim}
\newtheorem{definition}{Definition}

\newtheorem{example}{Example}
\newtheorem{conjecture}{Conjecture}
\theoremstyle{remark}
\newtheorem{remark}{Remark}

\numberwithin{equation}{section}
\definecolor{lauracolor}{rgb}{0.0, 0.47, 0.75} 
\definecolor{todocol}{rgb}{1,0,0.0}
\definecolor{fixmecol}{rgb}{1.0,0.3,0.3}
\definecolor{ideacol}{rgb}{1.0,0.75,0.0}
\definecolor{probcol}{rgb}{1.0,0.1,0.0}
\definecolor{green-munsell}{rgb}{0.0, 0.66, 0.47}

\usepackage[a4paper, left=1in, right=1in, top=1in, bottom=1in]{geometry}

\title{Agnostic Process Tomography} 
\author[1,*]{Chirag Wadhwa\thanks{$^*$ \href{mailto:chirag.wadhwa@ed.ac.uk}{chirag.wadhwa@ed.ac.uk}}}
\author[1]{Laura Lewis}
\author[1,2]{Elham Kashefi}
\author[1]{Mina Doosti}

\affil[1]{\small \textit{School of Informatics, University of Edinburgh, United Kingdom}}
\affil[2]{\small \textit{Laboratoire d'Informatique de Paris 6, CNRS, Sorbonne Université, Paris, France}}

\date{}

\begin{document}

\maketitle

\vspace{-2em}
\begin{abstract}
    Characterizing a quantum system by learning its state or evolution is a fundamental problem in quantum physics and learning theory with a myriad of applications. Recently, as a new approach to this problem, the task of agnostic state tomography was defined, in which one aims to approximate an arbitrary quantum state by a simpler one in a given class. Generalizing this notion to quantum processes, we initiate the study of agnostic process tomography: given query access to an unknown quantum channel $\Phi$ and a known concept class $\mathcal{C}$ of channels, output a quantum channel that approximates $\Phi$ as well as any channel in the concept class $\mathcal{C}$, up to some error. In this work, we propose several natural applications for this new task in quantum machine learning, quantum metrology, classical simulation, and error mitigation. In addition, we give efficient agnostic process tomography algorithms for a wide variety of concept classes, including Pauli strings, Pauli channels, quantum junta channels, low-degree channels, and a class of channels produced by $\qacz$ circuits. The main technical tool we use is Pauli spectrum analysis of operators and superoperators. We also prove that, using ancilla qubits, any agnostic state tomography algorithm can be extended to one solving agnostic process tomography for a compatible concept class of unitaries, immediately giving us efficient agnostic learning algorithms for Clifford circuits, Clifford circuits with few T gates, and circuits consisting of a tensor product of single-qubit gates. Together, our results provide insight into the conditions and new algorithms necessary to extend the learnability of a concept class from the standard tomographic setting to the agnostic one. 
\end{abstract}

\newpage
\tableofcontents
\newpage

\section{Introduction}
\label{sec:intro}
A fundamental problem in quantum physics and quantum learning theory is to completely characterize a quantum system by learning its evolution, a task called \emph{quantum process tomography}~\cite{mohseni2008quantum,o2004quantum,scott2008optimizing,chuang1997prescription,d2001quantum,haah2023query,poyatos1997complete}.
Quantum process tomography has numerous applications in benchmarking~\cite{mohseni2008quantum,scott2008optimizing,chuang1997prescription,o2004quantum,levy2024classical,huang2022foundations,merkel2013self,blume2017demonstration}, error mitigation \cite{endo2021hybrid,seif2023shadow, cai2023quantum}, quantum metrology \cite{giovannetti2006quantum,giovannetti2011advances}, quantum control~\cite{kim2014quantum}, and quantum cryptography \cite{arapinis2021quantum, schuster2024random}.
Unfortunately, recovering a full description of an arbitrary quantum process requires exponential resources~\cite{haah2023query}.

On the other hand, when the unknown process possesses some underlying structure, e.g., for Clifford circuits~\cite{low2009learning,lai2022learning}, Clifford circuits with few non-Clifford T gates~\cite{lai2022learning}, shallow circuits~\cite{nadimpalli2024pauli,huang2024learning}, junta unitaries/channels~\cite{chen2023testing,bao2023testing}, matrix product operators~\cite{torlai2023quantum}, circuits with a bounded number of gates~\cite{zhao2023learning}, quantum boolean functions~\cite{montanaro2008quantum,rouze2024quantum}, low-degree unitaries/channels~\cite{arunachalam2024learning}, and Pauli channels~\cite{harper2020efficient,harper2021fast,flammia2020efficient,chen2022quantum,flammia2021pauli,fawzi2023lower,chen2024tight}, or when one only wishes to predict some limited properties of this process \cite{kunjummen2023shadow,huang2023learning, levy2024classical, chen2024predicting,chung2018sample}, efficient algorithms can be developed.
However, the algorithms for the former case only apply in the \emph{realizable} setting: the unknown quantum process must \emph{exactly} satisfy this known structure. 
In practice, this assumption may not hold (e.g., due to noisy access), rendering these algorithms ineffective. 

In this work, we study how one may still be able to learn a process with a desirable underlying structure when given access to a more complex one.
In particular, we aim to answer the following central question.

\begin{center}
    \emph{Given an unknown quantum process, can we learn\\ its best approximation with a simple structure?}
\end{center}

Ideally, once we have such an approximation, the simple representation can be used to learn about the complex process.
We outline potential applications in quantum machine learning, metrology, classical simulation, and error mitigation in~\Cref{sec:appl}.
Notably, the analogous problem for quantum states~\cite{grewal2024agnostic, chen2024stabilizer,anshu2024survey,badescu2021improved,grewal2024improved} does not apply to many of these settings, motivating our generalization to quantum processes.

More precisely, the problem we aim to solve is: given access to an unknown quantum process $\Phi$ and a known concept class $\mathcal{C}$ of channels, output the nearest approximation of $\Phi$ from the class $\mathcal{C}$.
We refer to this task as \emph{agnostic process tomography}, inspired by recent developments for agnostic state tomography~\cite{grewal2024agnostic, chen2024stabilizer,anshu2024survey,badescu2021improved,grewal2024improved} and classical agnostic learning~\cite{kearns1992toward,haussler1992decision}.
We are primarily interested in cases when the concept class has a simple underlying structure and consists of processes that can be efficiently implemented. 
One may also consider \emph{improper} agnostic process tomography, in which the output of the algorithm is not in the concept class $\mathcal{C}$, but we emphasize that most applications require \emph{proper} learning.
We define the task formally in \Cref{sec:def}.

In this work, we initiate the study of agnostic learning of quantum processes and present new efficient agnostic learning algorithms for a wide variety of concept classes, including Pauli strings, Pauli channels, quantum junta channels, low-degree channels, and a class of channels produced by $\qacz$ circuits.
We also learn the class of bounded-gate circuits, and the associated analysis 
reveals that when studying agnostic process tomography, one should focus on reducing the time complexity rather than the sample complexity, analogously to the state case~\cite{grewal2024agnostic}. In addition, we prove that any agnostic state tomography algorithm can be extended to perform improper agnostic process tomography of a compatible concept class when equipped with ancilla qubits.
Applying this to the agnostic state tomography algorithms of~\cite{chen2024stabilizer}, we immediately obtain efficient improper agnostic learning algorithms for the classes of Clifford circuits, Clifford circuits with few T gates, and circuits consisting of a tensor product of single-qubit gates. 
Thus, our results answer our central question in the affirmative for these classes and also partially answer an open question of~\cite{grewal2024agnostic}, regarding whether one can suitably define a notion of agnostic learning for processes and prove guarantees for the class of Clifford circuits.

Notice that for the concept classes we consider in this work, learning algorithms have already been developed in the realizable setting~\cite{montanaro2008quantum,chen2023testing,bao2023testing,arunachalam2024learning,nadimpalli2024pauli,zhao2023learning,lai2022learning,low2009learning}. 
As a whole, our results provide new insight into when these realizable algorithms can be extended, with different technical proofs, to the agnostic setting.
In particular, for simpler concept classes such as Pauli strings, Pauli channels, bounded-gate circuits and $\qacz$ channels, our algorithms and analysis do not differ significantly from the realizable setting~\cite{montanaro2008quantum,zhao2023learning,nadimpalli2024pauli}.
On the other hand, for juntas and low-degree quantum channels, the realizable algorithms~\cite{chen2023testing,bao2023testing,arunachalam2024learning} fail to generalize straightforwardly to the agnostic case, requiring us to develop new, robust algorithms.
Specifically, for juntas, the realizable algorithms and proofs rely crucially on a quantity (namely, the influence) that only helps characterize the unknown unitary/channel if it is indeed a junta. 
Similarly, for low-degree unitaries, we show that a simple agnostic extension of a realizable algorithm~\cite{arunachalam2024learning} results in a drastically increased sample/time complexity compared to both our new algorithm and the original, realizable algorithm.
We elaborate more on these points in~\Cref{sec:proof-ideas}.

We define agnostic process tomography formally in \Cref{sec:def}.
Then, we describe general applications of agnostic process tomography in~\Cref{sec:appl} and our results in~\Cref{sec:results}.
The main technical tool we use is Pauli spectrum analysis of operators and superoperators~\cite{montanaro2008quantum,chen2023testing,nadimpalli2024pauli, arunachalam2024learning,rouze2024quantum,chen2024predicting}, and we detail the ideas behind our proofs in~\Cref{sec:proof-ideas}.

\subsection{Agnostic Process Tomography}
\label{sec:def}

\begin{figure}[htbp]
    \centering
    \includegraphics[scale=0.225]{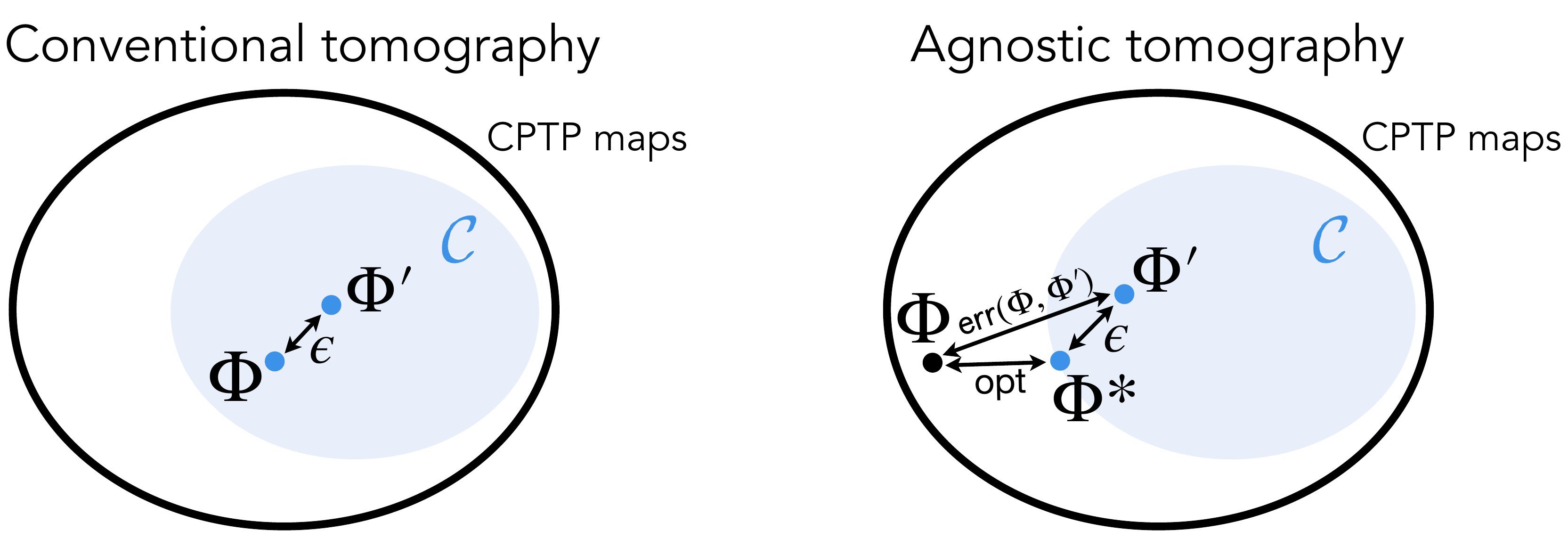}
    \caption{\textbf{(Proper) Agnostic process tomography. \textbf{(Left)} Conventional tomography.} In the realizable case, the unknown quantum process $\Phi$ is promised to fall in the class $\mathcal{C}$, and the goal is to find some $\Phi' \in \mathcal{C}$ that is close to $\Phi$. \textbf{(Right) Agnostic tomography.} Here, the unknown process $\Phi$ can be any completely positive trace-preserving (CPTP) map. The goal is to find some $\Phi' \in \mathcal{C}$ that is not far from the closest channel $\Phi^*$ to $\Phi$ in $\mathcal{C}$.}
    \label{fig:def}
\end{figure}

Formally, we define agnostic process tomography as follows, where we adopt the notation of~\cite{caro2023classical,anshu2024survey}.
We also depict the definition in \Cref{fig:def}.

\begin{definition}[Agnostic process tomography]
    Let $1 > \epsilon ,\delta > 0$ and $\alpha \geq 1$.
    Let $\mathcal{C}$ be a class of completely positive trace-preserving (CPTP) maps. A learning algorithm $\mathcal{A}$ is an \emph{$(\alpha,\epsilon,\delta)$-agnostic learner} with respect to the class $\mathcal{C}$ of CPTP maps if, given access to an arbitrary CPTP map $\Phi$, $\mathcal{A}$ outputs a hypothesis $\Phi^\prime$ such that
    \begin{equation}
        \mathsf{err}(\Phi,\Phi^\prime) \leq \alpha \cdot \mathsf{opt}(\Phi, \mathcal{C}) + \epsilon
    \end{equation}
    with probability at least $1-\delta$, where $\mathsf{err}$ denotes an error function and 
    \begin{equation}
        \mathsf{opt}(\Phi, \mathcal{C}) = \underset{\mathcal{E} \in \mathcal{C}}{\min} \text{ }\mathsf{err}(\Phi,\mathcal{E})
    \end{equation} is the optimal error of the unknown channel with respect to $\mathcal{C}$.
    \emph{Agnostic process tomography} with respect to $\mathcal{C}$ is the task of $(\alpha, \epsilon, \delta)$-agnostic learning with respect to $\mathcal{C}$.
    When the hypothesis $\Phi^\prime$ is from the concept class $\mathcal{C}$, we say that $\mathcal{A}$ is \emph{proper}. Otherwise, we say that it is \emph{improper}.
    Moreover, if $\Phi$ is a unitary channel and $\mathcal{C}$ is a class of unitaries, we refer to the task as \emph{agnostic unitary estimation}.
\end{definition}

Here, we purposefully leave some aspects of the definition flexible to capture the most general setting.
In particular, we do not specify the access model for the unknown process $\Phi$ (e.g., query access, statistical query access~\cite{nadimpalli2024pauli,angrisani2023learning,wadhwa2023learning, wadhwa2024noise}, samples from the output distribution~\cite{hinsche2023one,nietner2023average, nietner2023unifying}, etc.) nor the error function $\mathsf{err}$ (e.g., diamond distance, average-case distance, Frobenius distance, etc.).
Moreover, for brevity, we may refer to an $(\alpha,\epsilon,\delta)$-agnostic learner as an $\alpha$-agnostic learner and similarly refer to $\opt(\Phi, \mathcal{C})$ as $\opt$ when the other arguments are clear from context.

For our earlier motivation and the applications in \Cref{sec:appl}, we focus on \emph{proper} agnostic learning.
We also remark that even if one succeeds at agnostic process tomography of a given concept class, the hypothesis will not be accurate if the concept class does not model the unknown channel well.
Namely, despite finding a hypothesis $\Phi'$ that is close to the optimal process $\Phi^*$, the optimal process itself may not approximate $\Phi$ well.
One may consider, e.g., running property testing algorithms~\cite{montanaro2013survey,montanaro2008quantum,wang2011property,low2009learning,chen2023testing,bao2023testing} beforehand to check if the concept class is suitable.

Note that agnostic process tomography includes conventional process tomography~\cite{mohseni2008quantum,o2004quantum,scott2008optimizing,chuang1997prescription,d2001quantum,haah2023query,low2009learning,lai2022learning,nadimpalli2024pauli,huang2024learning,torlai2023quantum,zhao2023learning, montanaro2008quantum,rouze2024quantum,arunachalam2024learning,harper2020efficient,harper2021fast,flammia2020efficient,chen2022quantum,flammia2021pauli,fawzi2023lower,chen2024tight} as a special case, when the unknown process is guaranteed to be in $\mathcal{C}$.
However, agnostic process tomography also allows one to learn quantum processes in the presence of arbitrary noise.
This greatly contrasts realizable process tomography algorithms, which are usually extremely noise-sensitive and require the unknown process to exactly satisfy some known conditions.
While learning under noise is one motivation, agnostic process tomography is more broadly applicable, as discussed in the next section.
\subsection{Applications}
\label{sec:appl}

As discussed previously, selecting an appropriate concept class is crucial to applying agnostic learning algorithms to practical scenarios.
By carefully choosing a concept class, we highlight a wide range of problems in near-term quantum computation that can be modelled in the framework of \emph{proper} agnostic process tomography. 
We hope that this offers fresh perspectives and tools for further advancements in these fields.
For some of these applications, our proposed algorithms are already applicable, while others require new algorithms, motivating further exploration of agnostic process tomography beyond this work.
We also emphasize that these applications are unique to the agnostic process tomography framework and often cannot be addressed using agnostic state tomography~\cite{grewal2024agnostic,chen2024stabilizer} alone.

\begin{figure}[htbp]
    \centering
    \includegraphics[scale=0.27]{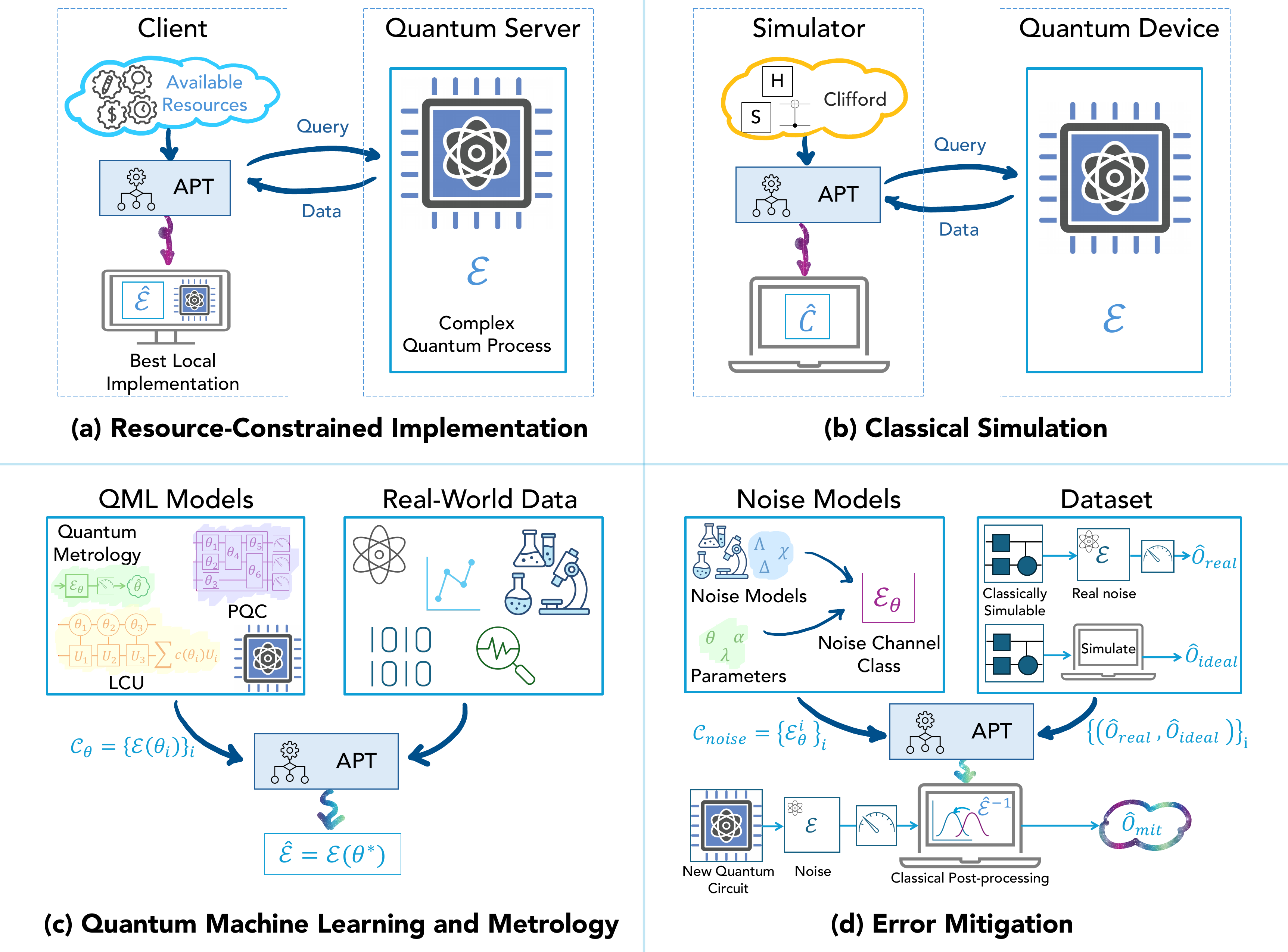}
    \caption{\textbf{Applications of agnostic process tomography.} \textbf{(a) Resource-constrained implementation}. A client with limited computational resources has access to a quantum server, which can implement the client's desired complex quantum process $\mc{E}$. The client chooses a suitable concept class based on its available resources and provides it as input to an agnostic process tomography (APT) algorithm. The APT algorithm has access to $\mc{E}$ via the quantum server and can (potentially adaptively) query $\mc{E}$ on any input state. The APT algorithm then outputs the best approximate implementation $\hat{\mc{E}}$ of $\mc{E}$, which the client can implement on their local device. \textbf{(b) Classical simulation}. Consider a user with access to a quantum device, which can implement some quantum process $\mc{E}$, and the user wishes to classically simulate $\mc{E}$. Given the class of Clifford circuits (or any classically simulable class) and after querying $\mc{E}$ from the quantum device, the APT algorithm outputs a Clifford circuit $\hat{C}$ that approximates $\mc{E}$, which can then be classically simulated. \textbf{(c) Quantum machine learning and metrology}. Consider a class $\mc{C}_\theta = \{\mc{E}(\theta_i)\}_i$ of parameterized quantum processes, which naturally occur in quantum machine learning (QML), e.g, parameterized quantum circuits (PQC) and linear combination of unitaries (LCU), and quantum metrology. Running APT with respect to $\mc{C}_\theta$ and an appropriate access model (e.g., real-world data), one can use APT to find the unknown parameter $\theta^*$ corresponding to the QML model or metrology problem. \textbf{(d) Error mitigation.} Consider the parameterized class of noise channels $\{\mc{E}_\theta\}_\theta$ that can be mitigated via classical post-processing when $\theta$ is known. In addition, suppose we obtain noisy data $\hat{o}_{\mathrm{real}}$ by running classically-simulable circuits on our noisy device and ideal data $\hat{o}_{\mathrm{ideal}}$ by classically simulating the same circuits.
    Passing the concept class and data $\{(\hat{o}_{\mathrm{real}},\hat{o}_{\mathrm{ideal}} )\}$ into an APT algorithm, we learn the parameter $\theta$ that best models the noise. Then, on a new quantum circuit run on our noisy device, we can mitigate the noise via classical post-processing from our knowledge of $\theta$.}
    \label{fig:appl}
\end{figure}

\paragraph{Resource-constrained implementation of complex quantum processes.}
A key application of agnostic process tomography is approximately implementing complex quantum processes on devices with limited resources.
Suppose a user with limited local resources has cloud access to a large fault-tolerant quantum computer and a desired complex process to run.
There are many reasons why the user may want to approximate this complex process by one that can be implemented on a weaker device.
For instance, repeated access to the remote system could be prohibitively expensive.
Instead, the user can obtain the best local approximation of the process by agnostic learning with respect to the class of circuits that can be implemented with the user's local resources.
For example, if the user's device has fewer qubits, one could consider agnostic unitary estimation of quantum juntas (\Cref{sec:aue-junta}), which learns the best few-qubit implementation of an arbitrary unitary.
Other interesting classes may include Clifford circuits, Cliffords with a few T gates and circuits consisting of tensor products of single-qubit gates (\Cref{sec:aue-clifford}).
We again emphasize that this requires carefully choosing the concept class, as the complex process may not be well-approximated by an arbitrary class.
This application is depicted in \Cref{fig:appl}(a).

\paragraph{Approximate Classical Simulation.}
We highlight a special case of the above application for approximately classically simulating complex quantum processes.
For example, given an arbitrary process, finding the closest Clifford circuit allows for efficient classical simulation~\cite{aaronson2004improved,gottesman1998theory}.
If the process is not too far from all Clifford circuits, which can be verified via property testing~\cite{montanaro2013survey,wang2011property}, this will offer accurate results.
This holds more generally for any classically simulable class admitting an efficient agnostic process tomography algorithm.
This application is depicted in \Cref{fig:appl}(b). 

\paragraph{A unifying view of quantum machine learning.}
Classically, the theoretical framework of agnostic learning captures the most general settings in machine learning (ML).
Namely, learning the best parameters for an ML model is equivalent to agnostic learning the concept class of all functions obtained by varying parameters of this ML model.
We argue that a similar equivalence holds for quantum machine learning (QML).
In QML, functions are evaluated using parameterized quantum circuits~\cite{cerezo2021variational}, or more generally, quantum channels (such as a Linear Combinations of Unitaries (LCU)~\cite{childs2012hamiltonian,heredge2024non,coyle2024training}).
By choosing an appropriate access model and error, agnostic process tomography of the class of channels expressible by an ansatz allows one to find the best parameters for the QML model.
In this sense, efficient agnostic process tomography of a model class implies efficient trainability. Thus, studying agnostic process tomography under various access models may lead to new insights into QML for both quantum and classical problems.
For example, many QML algorithms learn from expectation values of observables on output states of parameterized quantum circuits, which can be modelled as quantum statistical queries\cite{arunachalam2020quantum,arunachalam2024role,angrisani2023learning,wadhwa2023learning,nadimpalli2024pauli,nietner2023unifying}. Thus, agnostic process tomography from quantum statistical queries could be a promising direction for new insights in QML. This application is depicted in \Cref{fig:appl}(c). 

\paragraph{Agnostic Quantum Metrology.}
Quantum metrology is considered one of the most significant real-world applications of quantum technology~\cite{giovannetti2011advances}.
Here, one aims to estimate an unknown parameter $\theta$ with precision limited by the fundamental bounds of measurement~\cite{giovannetti2006quantum}, given access to a known class of parameterized quantum channels $\{\mathcal{E}_\theta\}_\theta$ called the \emph{encoding}.
By probing the device modelled by the channel 
and applying general quantum measurements to the outputs, one can construct an estimator $\hat{\theta}$ of the unknown parameter with desirable statistical properties.
However, in practice, the encoding may be corrupted by unknown environmental noise.
Due to the challenges of characterizing noise and its effect on parameter estimation, noisy quantum metrology is less well-studied, and results are restricted to specific classes of noise~\cite{gorecki2022quantum,demkowicz2012elusive,escher2011general,ji2008parameter,zhou2018achieving,demkowicz2017adaptive,zhou2024achieving}.
Many standard protocols are also highly noise-sensitive~\cite{giovannetti2011advances,ono2010effects,gilbert2008use,rubin2007loss}.
Considering the known class $\{\mathcal{E}_\theta\}_\theta$ as our concept class, one can model the problem of extracting a desirable estimator as an agnostic process tomography problem, which may provide us with new insights into metrology in noisy settings.
While noiseless metrology is close to conventional tomography, the parameter-encoding channel becomes unknown when transformed by realistic noise, which is more suitable for agnostic learning.
This perspective is also applicable in adversarial scenarios, where an unknown attack channel affects the encoding~\cite{shettell2022cryptographic,huang2019cryptographic}.
We are unsure how this approach would compare to the standard quantum limit (the Heisenberg limit is unattainable for general noise without quantum error correction~\cite{zhou2024limits}) but nevertheless believe this is interesting to explore.
This application is depicted in \Cref{fig:appl}(c).

\paragraph{Error Mitigation.}
Error mitigation protocols are often based on machine learning \cite{czarnik2021error, strikis2021learning} or require a strong characterization of the noise channel beforehand \cite{endo2021hybrid, van2023probabilistic}.
We argue that agnostic process tomography can be used to generalize a machine-learning-based error mitigation protocol called Clifford Data Regression \cite{czarnik2021error}.
In this setting, we wish to mitigate a fixed, arbitrary noise channel $\mathcal{E}$ on a quantum device.
We assume that $\mc{E}$ is well-approximated by a parameterized class of noise channels $\{\mc{E}_\theta\}_\theta$ such that, when $\theta$ is known, the errors can be mitigated via classical post-processing. This is relevant when methods that map arbitrary noise to a known class, e.g., Pauli twirling~\cite{van2023probabilistic, gupta2024probabilistic,geller2013efficient,wallman2016noise}, are unavailable.
We propose the following protocol: 1) Execute various classically-simulable circuits on the quantum device to obtain noisy data.
2) Simulate the same circuits classically to obtain ideal, noiseless data.
3) Using the noisy and ideal data, perform agnostic process tomography with respect to $\{\mc{E}_\theta\}_\theta$.
This procedure outputs an approximation $\mc{E}_{\theta'}$ of the best representative of the arbitrary noise channel $\mc{E}$ from $\{\mc{E}_\theta\}_\theta$.
Then, when implementing new circuits on this device, we can use $\mc{E}_{\theta'}$ to mitigate noise via classical post-processing.
Our new protocol can be seen as providing the pre-characterization of error often needed for error mitigation protocols such as probabilistic error cancellation \cite{van2023probabilistic}.
Note that in this setting, our access to the unknown process is through the pairs of ideal and noisy data rather than query access.

We consider the class of Pauli channels $\sum_x p_x \sigma_x (\cdot) \sigma_x$, parameterized by an underlying probability distribution $\{p_x\}_x$, as an example of our protocol for error mitigation.
Our agnostic tomography algorithm for Pauli channels in \Cref{sec:apt-pauli} only requires data obtained by performing measurements in the basis of Choi states of the Pauli operators.
The approximated noise can then be used to mitigate errors in new data using standard techniques for Pauli channel noise \cite{van2023probabilistic,ferracin2024efficiently}.
Note a limitation is that our agnostic learning algorithm for Pauli channels requires the use of ancilla qubits.
This example also indicates that for this protocol to work, agnostic process tomography cannot be used in a black box manner. Instead, the agnostic algorithm should dictate what data to gather from the noisy quantum device.
While we do not show an immediate application, we believe that this approach can result in new protocols for error mitigation for specific classes of noise.
This application is depicted in \Cref{fig:appl}(d).

\subsection{Our Results}
\label{sec:results}

We design new efficient agnostic learning algorithms for a wide variety of concept classes.
We consider the setting where we are given query access to the unknown quantum channel, i.e., we can apply the unknown channel to any (potentially adaptively-chosen) input state.
Moreover, we consider the error function to be given by either the root mean squared trace distance $\davg$ or the normalized Frobenius distance $d_F$, depending on the concept class.
The root mean squared trace distance is defined as
\begin{equation}
    \davg(\mc{E}_1, \mc{E}_2) \triangleq \sqrt{\mathop{\mathbb{E}}_{\ket{\psi} \sim \mu_S} [\dtr(\mc{E}_1(\ketbra{\psi}), \mc{E}_2(\ketbra{\psi}))^2]},
\end{equation}
where the expectation is taken over Haar-random states, and $\dtr$ denotes the trace distance $\dtr(\rho,\sigma) \triangleq \norm{\rho - \sigma}_1/2$.
For operators on an $n$-qubit space, the normalized Frobenius distance is
\begin{equation}
    d_F(A, B) \triangleq \frac{1}{\sqrt{2^n}}\norm{A- B}_F,
\end{equation}
where $\|.\|_F$ is the Frobenius norm (see \Cref{sec:pauli}).
Meanwhile, for superoperators in an $n$-qubit space, the normalized Frobenius distance is
\begin{equation}
    d_F(\Phi, \Psi) \triangleq \frac{1}{2^n\sqrt{2}}\norm{J(\Phi) - J(\Psi)}_F,
\end{equation}
where $J(\Phi)$ denotes the Choi representation of $\Phi$, i.e., the unnormalized Choi state.

For agnostic unitary estimation, we consider the classes of circuits consisting of $G$ 2-qubit gates~\cite{zhao2023learning}, Pauli strings, unitary $k$-juntas, and degree-$d$ unitaries.
Pauli strings are unitaries of the form $\{I, X, Y, Z\}^{\otimes n}$.
Unitary $k$-juntas act nontrivially only on $k$ out of $n$ qubits (\Cref{def:unitary-junta}).
Also, degree-$d$ unitaries are unitaries whose Pauli coefficients of order greater than $d$ are $0$ (\Cref{def:unitary-low-degree}).

For agnostic process tomography, we consider similar classes, namely circuits consisting of $G$ gates, Pauli strings, Pauli channels, $k$-junta channels, degree-$d$ channels, and $\qacz$ channels~\cite{nadimpalli2024pauli}.
Here, Pauli channels are defined as those of the form $\sum_x p_x \sigma_x (\cdot) \sigma_x$ for a probability distribution $\{p_x\}_x$.
$k$-junta channels and degree-$d$ channels are defined similarly to the unitary case (see \Cref{def:channel-junta,def:channel-low-degree}).
Finally, we also consider $n$-to-$1$ $\qacz$ channels~\cite{nadimpalli2024pauli}, which are defined by applying a unitary implemented by a $\qacz$ circuit~\cite{moore1999quantum} and tracing out all but one qubit (see \Cref{def:qac0-channels}).

Moreover, under compatibility conditions on the concept classes, we show that one can convert any agnostic state tomography algorithm (with respect to fidelity) into an agnostic unitary estimation/process tomography algorithm (with respect to the normalized Frobenius distance $d_F$) by learning the Choi state of the unknown unitary/channel.
More precisely, if $\mathcal{C}$ is the concept class for agnostic state tomography, this conversion holds when, for each unitary $U$ in the concept class for agnostic unitary estimation/process tomography, the Choi state of $U$ is in the class $\mathcal{C}$.
Note that for agnostic process tomography, the concept class must be unitary.
This immediately gives us $1$-agnostic unitary estimation algorithms for Clifford circuits, Clifford circuits with few T gates, and circuits consisting of tensor products of single-qubit gates from existing agnostic state tomography algorithms~\cite{chen2024stabilizer} with a quadratic increase with respect to the error $\epsilon$.
We obtain similar results for agnostic process tomography with a weaker guarantee on the error ($\mathsf{err} \leq 2^{1/4} \opt^{1/2} + \epsilon$).
We note that we can avoid the quadratic increase in $\epsilon$ and obtain 1-agnostic unitary estimation/process tomography with respect to entanglement infidelity, defined as the infidelity between two Choi states (\Cref{eq:entangle-infidel}).
This measure has been considered widely in (realizable) unitary estimation~\cite{leung2000towards,haah2023query,acin2001optimal,peres2002covariant,hayashi2006parallel,chiribella2005optimal,kahn2007fast,yang2020optimal}, but we focus on $d_F$ for consistency with our other results.

\begin{table}[h]
    \centering
    \begin{subtable}[t]{\textwidth}
    \centering
    \begin{tabular}{|c|c|c|c|}
    \hline
        Concept Class & Result & Queries (N) & Time\\
    \hline
   G-gate Circuits (\textcolor{blueviolet}{ Cor. \ref{cor:aue-gate}})& Proper 1-agnostic ($\davg$) & $\mc{\Tilde{O}}(G\log (n) /\epsilon^4)$& Inefficient \\
   AUE from AST (\textcolor{blueviolet}{Sec. \ref{sec:aue-clifford}}) & Improper 1-agnostic ($d_F$) & $N(2n, \epsilon^2)$ & $T(2n, \epsilon^2)$\\ 
    Pauli Strings (\textcolor{blueviolet}{Thm. \ref{thm:aue-pauli-davg}})&Proper 1-agnostic ($\davg$) & $\mc{O}(1/\epsilon^4)$& $\mc{O}(N)$ \\ 
    Unitary $k$-Juntas (\textcolor{blueviolet}{Thm. \ref{thm:aue-juntas-df}}) & Proper 2-agnostic ($d_F$)& $\mc{O}\left(\frac{k\log(n)}{\epsilon^4} + \frac{k16^k}{\epsilon^2}\right)$& $\mathcal{O}\left(
    \frac{n^k k \log(n)}{\epsilon^4}\right)$ \\ 
    Degree-$d$ Unitaries (\textcolor{blueviolet}{Thm. \ref{thm:aue-low-1}}) & Improper 1-agnostic $(d_F)$& $\mc{O}\left(d\cdot 9^d n^{2d} \log(n)/\epsilon^2\right)$ & $\mathcal{O}(N)$\\
    \hline
    \end{tabular}
    \caption{\textbf{Summary of agnostic unitary estimation guarantees.}}
    \label{tab:summary-aue}
    \end{subtable}
    \vspace{0.5em}

    \begin{subtable}[t]{\textwidth}
    \centering
    \begin{tabular}{|c|c|c|c|}
    \hline
        Concept Class & Result & Queries (N) & Time\\
    \hline
   \makecell{G-gate Circuits (\textcolor{blueviolet}{Cor. \ref{cor:apt-gate}})} & Proper, $\opt^{1/4} (\davg)$ & $\mc{\Tilde{O}}(G\log(n)/\epsilon^4)$& Inefficient \\
    \makecell{APT from AST  (\textcolor{blueviolet}{Sec. \ref{sec:apt-clifford}})} & Improper, $2^{1/4} \opt^{1/2}$ ($d_F$) & $N(2n, \epsilon^2)$  & $T(2n,\epsilon^2)$ \\
     \makecell{Pauli Strings (\textcolor{blueviolet}{Thm. \ref{thm:apt-pauli-df}})} &Proper 1-agnostic ($d_F$) & $\mc{O}(1/\epsilon^4)$ & $\mc{O}(N)$\\
     \makecell{Pauli Channels (\textcolor{blueviolet}{Thm. \ref{thm:apt-pauli-channel-df}})}& Proper 1-agnostic ($d_F$) & $\mc{O}(1/\epsilon^2)$ & $\mc{O}(N \log N)$ \\
     \makecell{$k$-Junta Channels (\textcolor{blueviolet}{Thm. \ref{thm:apt-juntas}})} & Improper 1-agnostic ($d_F$)& $\mc{O}\left(\frac{k 2^{12k} n^k \log n}{\epsilon^4}\right)$& $\mc{O}(N)$ \\
    \makecell{Degree-$d$ Channels (\textcolor{blueviolet}{Thm. \ref{thm:apt-low-1}})} & Improper 1-agnostic ($d_F$) &$\mc{O}\left(d \cdot 81^d n^{4d} \log(n)/\epsilon^2\right)$ & $\mc{O}(N)$\\
     \makecell{$\qacz$ Channels (\textcolor{blueviolet}{Thm. \ref{thm:qac0-apt}})} & Improper 1-agnostic ($d_F$)& $n^{\mathcal{O}(\log^d(s^4/\epsilon^4))}$ & $n^{\mathcal{O}(\log^d(s^4/\epsilon^4))}$\\
    \hline
    \end{tabular}
    \caption{\textbf{Summary of agnostic process tomography guarantees.}}
    \label{tab:summary-apt}
    \end{subtable}
    \caption{\textbf{Summary of results.} 
    We omit the $\delta$-dependence of the sample and time complexities, all of which scale with $\log(1/\delta)$. 
    Here, $n$ denotes the system size and $\epsilon$ denotes the approximation error from the optimal representative of the given concept class.
    The error is measured with respect to either $\davg$ (\Cref{def:davg}) or $d_F$ (\Cref{def:dist-frob,def:df-superop}).  AUE/APT from AST denotes our results in converting an agnostic state tomography algorithm into one for agnostic unitary estimation/process tomography, respectively. Here, $N/T$ denotes the sample/time complexity, respectively, of the underlying agnostic state tomography algorithm. The complexities are made explicit for specific cases in \Cref{sec:aue-clifford,sec:apt-clifford}. For APT of $G$-gate circuits, $\opt^{1/4}$ indicates that we get the bound $\mathsf{err} \leq \opt^{1/4} + \epsilon$ and similarly for APT from AST. For $\qacz$ channels, $d/s$ is the depth/size of the underlying $\qacz$ circuit, respectively.}
    \label{tab:summary}
\end{table}

In \Cref{tab:summary-aue,tab:summary-apt}, we summarize our results for agnostic unitary estimation/process tomography, respectively.
Aside from the $G$-gate circuit case, all of our algorithms are sample and time-efficient.

For both agnostic unitary estimation and agnostic process tomography, the algorithms for circuits consisting of $G$ gates are sample-efficient but computationally inefficient.
In fact, we prove a more general result (see Propositions \ref{prop:aue-finite-class} and \ref{prop:apt-finite-class}): for any finite-sized class $\mc{C}$, one can design an agnostic learning algorithm that utilizes $\mathcal{O}(\log |\mathcal{C}|)$ queries to the unknown quantum process but has time complexity scaling with $|\mc{C}|$.
Qualitatively, this tells us that efficient agnostic tomography is easy from a sample complexity perspective, but it can be difficult to achieve with respect to time complexity. This is similar to an analogous argument for agnostic state tomography~\cite{grewal2024agnostic} based on shadow tomography~\cite{aaronson2018shadow}.
Thus, for agnostic process tomography, one should focus on computational efficiency rather than sample efficiency.
This is in contrast to the realizable setting, where results typically focus on sample complexity.
\subsection{Proof Ideas}
\label{sec:proof-ideas}

Fourier analysis of Boolean functions~\cite{o2014analysis} is widely used in theoretical computer science and classical learning theory.
In particular, it has been used to develop efficient learning algorithms for functions with a sufficiently sparse Fourier decomposition~\cite{goldreich1989hard, kushilevitz1991learning, linial1993constant} and in classical agnostic learning~\cite{kearns1992toward,heidari2023agnostic,goldwasser2021interactive,gur2024power}.

\emph{Pauli spectrum analysis}~\cite{montanaro2008quantum} generalizes Fourier analysis for quantum operators (see \Cref{sec:pauli} for a more thorough discussion).
Analogous to the Fourier decomposition,  the key idea is that one can uniquely decompose an operator $A$ acting on an $n$-qubit system into a linear combination of Pauli strings:
\begin{equation}
    A = \sum_{x \in \{0,1,2,3\}^n} \hat{A}_x \sigma_x, \quad \hat{A}_x \triangleq \frac{1}{2^n}\tr(A \sigma_x),
\end{equation}
where $\sigma_x \in \{I, X,Y,Z\}^{\otimes n}$.
Recently, \cite{bao2023testing} further generalized this to the \emph{Fourier spectrum of superoperators} (see \Cref{sec:fourier-superop}), where a superoperator $\Phi$ in an $n$-qubit space is uniquely decomposed as
\begin{equation}
    \Phi = \sum_{x,y \in \{0,1,2,3\}^n} \hat{\Phi}(x,y) \Phi_{x,y},\quad \hat{\Phi}(x,y) = \frac{1}{4^n}\tr(J(\Phi_{x,y})^\dagger J(\Phi)),\quad \Phi_{x,y}(\rho) = \sigma_x \rho \sigma_y,
\end{equation}
where $J(\Phi)$ denotes the Choi represenation of $\Phi$, i.e., the unnormalized Choi state.
As in classical learning theory, many quantum learning algorithms utilize Pauli/Fourier spectrum analysis~\cite{montanaro2008quantum,chen2023testing,nadimpalli2024pauli, arunachalam2024learning,rouze2024quantum,chen2024predicting,bao2023testing}.

In our work, we leverage both formalisms.
Most of our algorithms involve learning the significant Pauli/Fourier coefficients.
However, we can only estimate up to an additive error, which \textit{a priori} may not satisfy the necessary properties of a valid unitary/channel, resulting in improper algorithms.

To resolve this, for agnostic unitary estimation, we ``project'' our improper hypothesis to its closest unitary using a technique from~\cite{huang2024learning}.
We show that this outputs the closest unitary with respect to the normalized Frobenius distance $d_F$, resulting in proper agnostic unitary estimation.
However, this problem is more difficult for agnostic process tomography.
We discuss a potential approach based on convex optimization~\cite{wang2024convergence} for projecting an improper hypothesis onto a valid quantum channel\footnote{Note that ideas from the quasiprobability decomposition and probabilistic error cancellation which approximate linear superoperators by linear combinations of quantum channels~\cite{jiang2021physical,regula2021operational,temme2017error,van2023probabilistic,zhao2023power,horodecki2002method,horodecki2003limits,fiuravsek2002structural,korbicz2008structural} do not help because we need to exactly output a quantum channel.} in~\Cref{sec:proper-apt}.
We are unable to fully solve the problem due to one difficult-to-verify condition for the convergence guarantees of~\cite{wang2024convergence} to hold (Conjecture~\ref{conj:unsolved-condition}).
Nevertheless, we made substantial progress towards resolving this open question, which is also recurrent in the realizable process tomography literature~\cite{nadimpalli2024pauli,chen2023testing, bao2023testing}.
Now, we summarize our results and the techniques used for each class.

\paragraph{Bounded-gate circuits.}
First, we consider any finite-sized class $\mc{C}$ of unitaries.
Similarly to~\cite{zhao2023learning}, we use Clifford classical shadows~\cite{huang2020predicting} to estimate the average infidelity between the unknown quantum channel $\Phi$ and all $U_i \in \mc{C}$.
Then, we output the unitary in $\mc{C}$ with the smallest estimated distance.
Thus, we solve proper agnostic unitary estimation for $\mc{C}$ with respect to the root mean squared trace distance $\davg$ and show this uses $\mc{O}(\log|\mc{C}|)$ queries to $\Phi$ and time $\mc{O}(|\mc{C}|)$.
For many classes of interest, this results in a sample-efficient but computationally inefficient algorithm, emphasizing the importance of computational efficiency for agnostic learning.
We show a weaker version of this result for agnostic process tomography of $\mathcal{C}$, which stems from converting from average infidelity to average trace distance for mixed inputs.
As a corollary, by utilizing known covering number bounds~\cite{zhao2023learning}, we obtain agnostic learning algorithms for the class of unitaries implementable by $G$ two-qubit gates.
Note that in the realizable setting,~\cite{zhao2023learning} proves computational lower bounds.
We prove this result in~\Cref{sec:aue-bounded,sec:apt-bounded}.

\paragraph{Agnostic process tomography from agnostic state tomography}
Recently, agnostic state tomography for several concept classes with respect to fidelity~\cite{grewal2024agnostic,chen2024stabilizer} has been studied.
We show that this translates into guarantees for agnostic unitary estimation/process tomography for a compatible concept class $\mc{C}$ of unitaries with respect to $d_F$ with a quadratic increase in the error $\epsilon$.
In particular, we require that the Choi state of each $U \in \mc{C}$ is in the concept class for agnostic state tomography.
Our algorithm is simple: apply the agnostic state tomography algorithm to the Choi state of the unknown quantum process.
Specifically, we obtain improper $1$-agnostic unitary estimation but a weaker error bound for agnostic process tomography (see \Cref{tab:summary-apt}), again due to converting between infidelity and trace distance for mixed inputs.
The key step in our proof shows a relationship between the infidelity and the normalized Frobenius distance between two Choi states.
We prove this result in~\Cref{sec:aue-clifford,sec:apt-clifford}.

\paragraph{Pauli strings.}
Next, we consider the class of Pauli strings $\{I,X,Y,Z\}^{\otimes n}$.
We use the fact that one can efficiently sample from the distribution formed by the Pauli/Fourier coefficients~\cite{montanaro2008quantum}.
After collecting samples, we simply output the most frequently sampled Pauli string.
Thus, we achieve proper 1-agnostic unitary estimation/process tomography.
We prove this in \Cref{sec:aue-pauli,sec:apt-pauli}.

\paragraph{Pauli channels.} Pauli channels are a naturally occurring and widely studied~\cite{erhard2019characterizing,carignan2023error,harper2020efficient,tuckett2018ultrahigh,van2023probabilistic,ferracin2024efficiently,kim2023evidence,harper2020efficient, flammia2020efficient,chen2022quantum,flammia2021pauli,fawzi2023lower,chen2024tight,harper2021fast} class of quantum channels.
Pauli channels are of the form $\sum_x p_x \sigma_x(\cdot)\sigma_x$ for a probability distribution $\{p_x\}_x$.
Similarly to Pauli strings, we can sample from $\{p_x\}_x$.
Then, agnostic tomography of Pauli channels reduces to empirically estimating $\{p_x\}_x$. 
Thus, we obtain proper 1-agnostic process tomography of Pauli channels.
We prove this in~\Cref{sec:apt-pauli}.

\paragraph{Junta unitaries and channels.}
A $k$-junta unitary/channel is an $n$-qubit unitary/channel that acts only on $k$ qubits (\Cref{def:unitary-junta,def:channel-junta}).
\cite{chen2023testing,bao2023testing} devised realizable algorithms for learning junta unitaries/channels.
Here, they estimate the \emph{influence} (\Cref{eq:influence}) of each qubit, then select a highly influential subset of qubits and perform full state tomography on the Choi state restricted to these qubits.

In the agnostic setting, we show that the optimal error is not tightly characterized by the most influential subset, but by the highest \emph{weight} subset (\Cref{def:pauli-weight,def:weight-superop}).
In the realizable setting, the relevant subset of qubits has the same weight and influence. However, we show that these notions are distinct in the agnostic setting (\Cref{ex:influence-weight}).
We also prove that the hypothesis's error is characterized by the chosen subset and depends on its weight, not its influence.
Thus, our algorithms aim to identify the highest weight subset of qubits.
However, identifying this subset requires more queries/time than identifying the most influential one, giving us worse complexities than the realizable algorithms.

In the unitary case, we first estimate the weights of all subsets of $[n]$ of cardinality $k$ using samples from the distribution formed by the Pauli coefficients.
Then, we consider the subset $S$ with the greatest estimated weight.
Next, we estimate all the coefficients of the unitary that are supported only on $S$.
We then construct a hypothesis using these estimates and project it to a unitary as described previously.
Despite being less efficient than tomography on qubits in $S$, our method makes it easier to obtain a proper hypothesis.
Our method results in improper 1-agnostic unitary estimation of unitary juntas.
After applying the unitary projection to the improper hypothesis, we obtain proper 2-agnostic unitary estimation.
We prove this formally in~\Cref{sec:aue-junta}.

For agnostic tomography of junta channels, we find that identifying the subset with the highest Fourier weight is much harder than in the unitary case.
Here, we can only sample from the distribution defined by the diagonal Fourier coefficients $\{\hat{\Phi}(x,x)\}_x$ while the weight also depends on the off-diagonal coefficients.
As a result, our algorithm learns \emph{all} low-degree coefficients and uses them to estimate the weights.
Explicitly, we learn all coefficients $\hat{\Phi}(x,y)$ for $|x|, |y| \leq k$, where $|x|$ denotes the size of the support of the string $x \in \{0,1,2,3\}^n$.
From here, the algorithm is similar to the unitary junta case.
We achieve improper 1-agnostic process tomography of junta channels.
We discuss a potential approach for producing a proper hypothesis in \Cref{sec:proper-apt}.
We prove this in \Cref{sec:apt-junta}.

\paragraph{Low-degree unitaries and channels.}
Low-degree unitaries/channels are defined as those with all their Pauli/Fourier weight on low-degree terms.
Explicitly, a unitary has degree $d$ if when $|x| > d$, then $\hat{U}_x = 0$ and similarly for channels (\Cref{def:unitary-low-degree,def:channel-low-degree}).
A straightforward idea is to learn all of their low-degree coefficients.
It turns out that this simple algorithm results in improper 1-agnostic learning of low-degree objects.
Our unitary projection method mentioned previously is computationally inefficient for low-degree unitaries and its output may not be low-degree, leaving us with an improper hypothesis.

\cite{arunachalam2024learning} develops realizable algorithms leveraging Bohnenblust-Hille (BH) inequalities for operators~\cite{volberg2024noncommutative}/\\superoperators~\cite{arunachalam2024learning}.
These inequalities provide upper bounds on norms of the spectra of low-degree objects.
Using these bounds, \cite{arunachalam2024learning} show that it suffices to learn significantly fewer Pauli/Fourier coefficients, resulting in algorithms with complexities independent of $n$.
Using similar arguments for agnostic unitary estimation, we show that the algorithm of~\cite{arunachalam2024learning} is inefficient and only performs better than the trivial algorithm when the optimal error of the class of low-degree unitaries is at most inverse exponential in $n$, i.e., when the problem is almost realizable.
We remark that improved bounds for the BH inequality may result in better guarantees for this algorithm and leave this open for future work.
As this algorithm does not work well in the simpler unitary case, we do not analyze its performance for agnostic process tomography.
We prove these results in~\Cref{sec:aue-low,sec:apt-low-deg}.

\paragraph{QAC0 channels.} The circuit class $\qacz$ consists of constant-depth circuits with controlled-$Z$ gates and arbitrary single-qubit gates~\cite{moore1999quantum}. In \cite{nadimpalli2024pauli}, the authors study the class of channels formed by applying a $\qacz$ circuit and tracing out all but one qubit.
\cite{nadimpalli2024pauli} shows that the Choi representations of such channels have Pauli weight concentrated on low-degree terms. 
Similarly to~\cite{nadimpalli2024pauli}, we learn the low-degree coefficients of the Choi representation of an arbitrary channel using classical shadows~\cite{huang2020predicting}, resulting in improper 1-agnostic process tomography for this class of $\qacz$ channels.
We show this in~\Cref{sec:apt-qac0}.
\subsection{Outlook}
Our work initiates the study of agnostic process tomography, which we define as learning the best approximation of an unknown process from a simple class.
We design efficient agnostic learning algorithms for various concept classes, leveraging the powerful tools of Pauli spectrum analysis.
We also propose several near-term applications for agnostic process tomography.
We believe there is still substantial work to be done in this area, and we outline some open questions below.

First, can one design an efficient algorithm to map an arbitrary superoperator to its closest quantum channel?
This is desirable because many of our improper agnostic learning algorithms can be made proper using such a subroutine, and many applications of agnostic process tomography (\Cref{sec:appl}) require proper learning algorithms.
As a significant first step towards this question, we have reduced the problem to that of proving a particular mathematical condition holds (see \Cref{sec:proper-apt}) but were unable to prove or disprove it.
Solving this question would be a significant result, as it would also lead to proper learning algorithms for realizable process tomography problems, such as those in \cite{nadimpalli2024pauli, arunachalam2024learning}.

Second, can we improve the error bounds for agnostic process tomography of bounded-gate circuits and agnostic process tomography from agnostic state tomography?
Notably, as a result of the distance inequalities we used, our learning algorithms achieve error bounds of $\opt^{1/4}$ and $2^{1/4}\opt^{1/2}$, respectively, rather than $\opt$.
We believe these bounds can be improved to $1$-agnostic process tomography and leave this as an open problem.
Also, can we avoid the quadratic increase in error when converting agnostic state tomography algorithms to those for agnostic process tomography, as in quantum process tomography~\cite{haah2023query}?

Third, is it possible to develop efficient agnostic learning algorithms for juntas that match the realizable lower bounds~\cite{chen2023testing,bao2023testing}?
Our proposed algorithms for juntas have worse complexities than the realizable setting due to needing to estimate the weight of subsets of qubits, rather than the influence.

Next, for learning low-degree quantum objects, can one obtain better complexities by using the algorithm of \cite{arunachalam2024learning} in the agnostic setting?
We show in \Cref{sec:aue-low} that straightforwardly utilizing the algorithm from~\cite{arunachalam2024learning} results in an inefficient algorithm with worse performance than ours.
However, we believe that by proving tighter upper bounds on the norms in the BH inequalities, this could be improved.

Finally, we list several broad open questions.
Could different definitions of agnostic tomography lead to new applications, e.g., comparing the performances of agnostic learners for different concept classes?
For instance, the definition of \cite{caro2023classical} may lend itself to this, where they define agnostic learning with respect to two classes rather than just one.
Can one develop agnostic learning algorithms with respect to worst-case distance metrics, e.g., diamond distance? All results in our work only hold for average-case distances.
Also, many of our proposed algorithms require preparing the Choi state of the unknown channel.
Instead, can one develop agnostic learning algorithms with fewer resources?
As agnostic process tomography is at least as hard as in the realizable setting, lower bounds in realizable process tomography can be transferred to the agnostic setting as well.
Can one prove stronger worst-case agnostic lower bounds?
Finally, are there other physically-motivated classes that admit efficient agnostic process tomography algorithms?
Some interesting classes include the evolution of well-studied Hamiltonians~\cite{clinton2021hamiltonian,schmitt2022quantum,yanay2020two,chatterjee2024comprehensive} or classes of parameterized quantum circuits~\cite{cerezo2021variational} with specific ansatz \cite{romero2017quantum,havlivcek2019supervised,abbas2021power}.

\subsection*{Acknowledgements}
The authors thank Richard R. Allen, Alexandru Cojocaru, Ryan Sweke, Bo Yang, and Sisi Zhou for insightful discussions.
The authors are supported by the Quantum
Advantage Pathfinder (QAP) with grant reference EP/X026167/1, the Hub in Quantum Computing and Simulation with grant reference EP/T001062/1, and the UK Engineering and Physical Sciences Research Council (EPSRC).
LL is also supported by a Marshall Scholarship.

\section{Preliminaries}
\label{sec:prelim}
For $k \in \mathbb{N}$, we write $[k] \triangleq \{1, \dots, k\}$. Throughout, let $n$ denote the system size. Let $\mathcal{H} = (\mathbb{C}^2)^{\otimes n}$ denote the Hilbert space on these $n$ qubits, and let $\mathcal{B}(\mathcal{H})$ denote the set of bounded linear operators on $\mathcal{H}$.
Let $\mathcal{U}_d \subseteq \mathcal{B}(\mathcal{H})$ denote the set of $d$-dimensional unitaries. A map from $\mc{B}(\mc{H})$ to itself is known as a \emph{superoperator} on $\mc{B}(\mc{H})$.
A \emph{quantum channel} on $\mc{B}(\mc{H})$ is a superoperator that is completely positive and trace-preserving (CPTP). For any unitary, we sometimes use a calligraphic font to denote the associated quantum channel.
\subsection{Choi state}
\label{sec:choi}

A useful tool we will need is the Choi-Jamiolkowski isomorphism~\cite{choi1975completely,jamiolkowski1972linear}, which maps between unitaries acting on the Hilbert space $\mathcal{H}$ and pure states in $\mathcal{H}\otimes \mathcal{H}$.

\begin{definition}[Choi state of a unitary]
\label{def:choi-state}
    Let $U$ be a unitary acting on an $n$-qubit Hilbert space $\mathcal{H}$.
    Then, the \emph{Choi state} is defined by applying $U$ on one half of a maximally entangled state over $2n$ qubits.
    In particular, define the maximally entangled state
    \begin{equation}
        |\Phi^+\rangle = \frac{1}{\sqrt{2^n}} \sum_{i = 0}^{2^n - 1} |i,i\rangle.
    \end{equation}
    Then, the Choi state is defined as
    \begin{equation}
        |v(U)\rangle \triangleq (U \otimes I) |\Phi^+\rangle = \frac{1}{\sqrt{2^n}}\sum_{i,j=0}^{2^n - 1}U_{i,j} \ket{i,j}.
    \end{equation}
\end{definition}
The Choi state can be further generalized for quantum channels, mapping quantum channels to mixed states.
\begin{definition}[Choi state of a quantum channel]
   Let $\mc{E}$ be a quantum channel acting on $\mc{B}(\mc{H})$. Then, the associated \emph{Choi state} is defined by applying $\mc{E}$ on one half of a maximally entangled state over $2n$ qubits. In particular, the Choi state is defined as
   \begin{equation}
       v(\mc{E}) = (\mc{E} \otimes \mc{I}) (|\Phi^+\rangle\langle\Phi^+|) = \frac{1}{2^n} \sum_{i,j = 0}^{2^n-1} \mc{E} (|i\rangle\langle j |) \otimes |i\rangle\langle j |.
   \end{equation}
   For an arbitrary superoperator $\Psi$ on $\mc{B}(\mc{H})$, we will also consider the \emph{Choi representation}, obtained by applying $\Psi$ on one half of the unnormalized, maximally entangled state, i.e.,
   \begin{equation}
       J(\Psi) = \sum_{i,j = 0}^{2^n-1} \Psi (|i\rangle\langle j |) \otimes |i\rangle\langle j |.
   \end{equation}
\end{definition}
The Choi representation of a quantum channel satisfies the following necessary and sufficient properties.
\begin{lemma}[Properties of Choi representations of quantum channels~\cite{watrous2018theory}]
    \label{lem:channel-choi-equiv}
    Let $\Psi$ be a superoperator from $\mc{B}(\mc{H}_1)$ to $\mc{B}(\mc{H}_2)$. Then $\Psi$ is a quantum channel if and only if
    \begin{equation}
        J(\Psi) \succcurlyeq 0 \quad and \quad \tr_1(J(\Psi)) = I_2. 
    \end{equation}
    where $\tr_{1}$ denotes the partial trace over $\mc{H}_1$ and $I_2$ denotes the identity on $\mc{H}_2$.
\end{lemma}

\subsection{Pauli spectrum of operators}
\label{sec:pauli}

A key technique we use throughout our proofs is the analysis of the Pauli spectrum of quantum operators, first introduced in~\cite{montanaro2008quantum} to generalize Boolean Fourier analysis~\cite{o2014analysis} to the quantum setting.
Pauli analysis was also studied in~\cite{rouze2024quantum,chen2023testing, nadimpalli2024pauli}.
Throughout this work, we mostly follow the notation of~\cite{montanaro2008quantum,o2014analysis}.

The set of bounded linear operators $\mathcal{B}(\mathcal{H})$ can be viewed as a Hilbert space when endowed with the Hilbert-Schmidt inner product
\begin{equation}
    \langle A, B \rangle \triangleq \tr(A^\dag B) = \sum_{i,j \in [2^n]} A_{i,j}^*B_{i,j},
\end{equation}
where we use $A_{i,j}$ to denote the $ij$-th entry of $A$ as a matrix.
We use $\norm{\cdot}_F$ to denote the norm induced by this inner product, namely the Hilbert-Schmidt or Frobenius norm.
In particular, we can write
\begin{equation}
    \norm{A}_F^2 = \langle A, A \rangle = \sum_{i,j\in [2^n]} |A_{i,j}|^2.
\end{equation}
Let $\{\sigma_0, \sigma_1, \sigma_2, \sigma_3\}$ denote the Pauli matrices, i.e.,
\begin{equation}
    \sigma_0 = I = \begin{pmatrix}
        1 & 0\\0&1
    \end{pmatrix},\quad \sigma_1 = X = \begin{pmatrix}
        0 & 1\\1&0
    \end{pmatrix},\quad \sigma_2 = Y = \begin{pmatrix}
        0 & -i\\i&0
    \end{pmatrix},\quad \sigma_3 = Z = \begin{pmatrix}
        1 & 0\\0 & -1
    \end{pmatrix}.
\end{equation}
For a vector $x \in \{0,1, 2, 3\}^n$, we denote $\sigma_x \triangleq \sigma_{x_1} \otimes \cdots \otimes \sigma_{x_n}$ and write $\mathsf{supp}(x) \triangleq \{i \in [n] : x_i \neq 0\}$. We will denote the \emph{degree} of $x$ by $|x| = |\supp(x)|$. Because the Paulis $\{\sigma_x\}_{x \in \{0,1,2,3\}^n}$ form an orthogonal basis for $\mathcal{B}(\mathcal{H})$ with respect to the Hilbert-Schmidt inner product, then we can uniquely decompose any operator with respect to this basis.

\begin{definition}[Pauli spectrum~\cite{montanaro2008quantum}]
\label{def:pauli-spectrum}
    For any $A \in \mathcal{B}(\mathcal{H})$, there exists a unique decomposition as
    \begin{equation}
        A = \sum_{x \in \{0,1,2,3\}^n} \hat{A}_x \sigma_x,\quad \hat{A}_x \triangleq \frac{1}{2^n}\langle \sigma_x, A \rangle.
    \end{equation}
    Here, $\hat{A}_x$ is called the \emph{Pauli coefficient} of $A$, and $\{\hat{A}_x\}_{x\in \{0,1,2,3\}^n}$ is called the \emph{Pauli spectrum} of $A$.
\end{definition}

As in classical Fourier analysis, there are quantum analogues of Parseval and Plancherel's identities.

\begin{lemma}[Proposition 10 in~\cite{montanaro2008quantum}]
\label{lem:parseval-plancherel}
    Let $A, B \in \mathcal{B}(\mathcal{H})$.
    Then, Parseval's identity is
    \begin{equation}
        \frac{1}{2^n} \|A\|_F^2 = \sum_{x\in \{0,1,2,3\}^n} |\hat{A}_x|^2.
    \end{equation}
    Moreover, Plancherel's identity is
    \begin{equation}
        \frac{1}{2^n} \langle A,B \rangle = \sum_{x \in \{0,1,2,3\}^n} \hat{A}^*_x \hat{B}_x.
    \end{equation}
\end{lemma}
Following the notation in~\cite{montanaro2008quantum,nadimpalli2024pauli}, we define the \emph{Pauli weight of an operator at level $k$}.
This is similar to the classical definition of Fourier weight~\cite{o2014analysis}.
\begin{definition}[Pauli weight at level $k$~\cite{montanaro2008quantum,nadimpalli2024pauli}]
    \label{def:pauli-weight-level}
     Let $A \in \mathcal{B}(\mathcal{H})$ and $0 \leq k \leq n$. Then, the \emph{Pauli weight of $A$ at level $k$} is defined by
     \begin{equation}
         w^{=k}(A) \triangleq \sum_{x \in \{0,1,2,3\}^n, |x| = k} |\hat{A}_x|^2.
     \end{equation}
     We similarly define $w^{\leq k}(A), w^{> k}(A)$, etc.
\end{definition}
For a subset $S \subseteq [n]$, we also denote
\begin{equation}
    \pauli_S \triangleq \{x \in \{0,1,2,3\}^n :  \supp(x) \subseteq S\}.
\end{equation}
Notice that
\begin{equation}
    |\pauli_S| = 4^{|S|}.
\end{equation}
We also use the shorthand $\mathcal{P}_n \triangleq \mathcal{P}_{[n]} = \{0,1,2,3\}^n$.
We also define the Pauli weight of $A \in \mathcal{B}(\mathcal{H})$ on a given subset $S \subseteq [n]$, similar to the Pauli weight at level $k$.

\begin{definition}[Pauli weight on a subset]
\label{def:pauli-weight}
    Let $A \in \mathcal{B}(\mathcal{H})$ and $S \subseteq [n]$.
    Then, the \emph{Pauli weight of $A$ on the subset $S$} is defined by
    \begin{equation}
        w_S(A) \triangleq \sum_{x \in \pauli_S} |\hat{A}_x|^2.
    \end{equation}
\end{definition}
This definition is reminiscent of the influence on a set of qubits.
Recall from~\cite{chen2023testing} that the influence of a unitary $U$ on a subset $S \subseteq [n]$ is defined as
\begin{equation}
    \label{eq:influence}
    \mathbf{Inf}_S[U] = \sum_{\substack{x \in \{0,1,2,3\}^n\\ \mathsf{supp}(x) \cap S \neq \emptyset} }|\hat{U}_x|^2.
\end{equation}
Here, we see that the Pauli weight on $S$ differs from the influence due to the set of $x \in \{0,1,2,3\}^n$ that the summation is over.

The Choi state of a unitary has a nice relationship with its Pauli decomposition.
Let $U = \sum_{x \in \{0,1,2,3\}^n} \hat{U}_x \sigma_x$ be the Pauli decomposition of some unitary $U$.
Then, the Choi state of $U$ decomposes as
\begin{equation}
    |v(U)\rangle = \sum_{x \in \{0,1,2,3\}^n} \hat{U}_x |v(\sigma_x)\rangle,
\end{equation}
where $\ket{v(\sigma_x)}$ denotes the Choi state of the Pauli $\sigma_x$.
\subsection{Fourier spectrum of superoperators}
\label{sec:fourier-superop}

One can also extend the Pauli analysis from the previous section to superoperators, which was initiated in~\cite{bao2023testing}.
This is also explored in~\cite{nadimpalli2024pauli}, but we mainly follow the presentation of~\cite{bao2023testing}.
The space of superoperators on $\mc{B}(\mc{H})$ can be viewed as a Hilbert space when endowed with the inner product
\begin{equation}
    \langle\Phi, \Psi\rangle \triangleq \langle J(\Phi), J(\Psi) \rangle = \tr(J(\Phi)^\dagger J(\Psi)).
\end{equation}
This inner product induces the norm
\begin{equation}
    \| \Phi\|_F \triangleq \sqrt{\langle J(\Phi), J(\Phi)  \rangle } = \| J(\Phi)\|_F.
\end{equation}
Consider the superoperators $\{\Phi_{x,y}\}_{x,y\in \mathcal{P}_n}$
\begin{equation}
        \Phi_{x,y} (\rho) = \sigma_x \rho \sigma_y,\quad\forall x,y \in \pauli_n.
\end{equation}
These superoperators form an orthogonal basis for the space of superoperators on $\mc{B}(\mc{H})$ with norm given by
\begin{equation}
     \|\Phi_{x,y}\|_F = 2^n, \quad \forall x,y \in \pauli_n.
\end{equation}
Because $\{\Phi_{x,y}\}_{x,y}$ forms an orthogonal basis, we can uniquely decompose any superoperator with respect to this basis.
\begin{definition}[Fourier expansion of superoperators; Definition 7 in~\cite{bao2023testing}]
    For any superoperator $\Phi$ on $\mc{B}(\mc{H})$, there exists a unique decomposition as
    \begin{equation}
        \Phi = \sum_{x,y \in \pauli_n} \hat{\Phi}(x,y) \Phi_{x,y} , \quad \hat{\Phi}(x,y)  = \frac{1}{4^n} \langle \Phi_{x,y}, \Phi \rangle.
    \end{equation}
    Here, $\hat{\Phi}_{x,y}$ is a Fourier coefficient of $\Phi$, and $\{\hat{\Phi}_{x,y}\}_{x,y \in \pauli_n}$ is called the \emph{Fourier spectrum} of $\Phi$.  Denote by $\hat{\Phi}$ the $4^n \times 4^n$-matrix with entries $\hat{\Phi}(x,y)$.
\end{definition}
The matrix of Fourier coefficients $\hat{\Phi}$ satisfies the following properties.
\begin{lemma}[Properties of Fourier coefficients; Lemma 8 in~\cite{bao2023testing}]
    \label{lem:superop-fourier-properties}
    For any superoperator $\Phi$ on $\mc{B}(\mc{H})$, the matrix of Fourier coefficients $\hat{\Phi}$ satisfies the following properties.
    \begin{enumerate}
        \item There exists a unitary $U$ such that
        \begin{equation}
            \hat{\Phi} = \frac{1}{2^n} U^\dag J(\Phi) U.
        \end{equation}
        \item Then, $\hat{\Phi} \succcurlyeq 0$ if and only if $J(\Phi) \succcurlyeq 0$ .
        \item If $\Phi$ is a quantum channel, then
        \begin{equation}
            0 \leq \hat{\Phi}(x,x) \leq 1, \;\forall x \in \pauli_n \quad \text{ and } \quad \sum_{x \in \pauli_n} \hat{\Phi}(x,x) = \tr(\hat{\Phi}) = 1.
        \end{equation}
    \end{enumerate}
\end{lemma}
We remark that for a unitary with Pauli spectrum $\{\hat{U}_x\}_x$, the Fourier coefficients of the associated quantum channel $\Phi$ are given by
\begin{equation}
    \hat{\Phi}(x,y) = \hat{U}_x \hat{U}_y^*.
\end{equation}
Similar to the case of the Pauli spectrum, we once again have analogues of Parseval's and Plancherel's identities.
\begin{lemma}[Parseval's and Plancherel's for superoperators]
    \label{lem:superop-parseval-plancherel} 
    Let $\Phi, \Psi$ be quantum channels on $\mc{B}(\mc{H})$. Then, Parseval's identity is
    \begin{equation}
        \frac{1}{4^n}\| \Phi\|_F^2 = \sum_{x,y \in \pauli_n} |\hat{\Phi}(x,y)|^2. 
    \end{equation}
    Moreover, Plancherel's identity is
    \begin{equation}
        \frac{1}{4^n} \langle \Phi, \Psi \rangle = \sum_{x,y \in \pauli_n} \hat{\Phi}(x,y)^* \hat{\Psi}(x,y).
    \end{equation}
\end{lemma}
Also similarly to the Pauli spectrum, we will define the weight at level $k$ of a superoperator. 
\begin{definition}[Fourier weight at level $k$]
    \label{def:weight-superop}
     Let $\Phi$ be a superoperator on $\mathcal{B}(\mathcal{H})$ and $0 \leq k \leq n$. Then, the \emph{Fourier weight of $\Phi$ at level $k$} is defined by
     \begin{equation}
         w^{=k}(A) \triangleq \sum_{x,y \in \pauli_n, |x| = |y| = k} |\hat{\Phi}(x,y)|^2.
     \end{equation}
\end{definition}
The weight of a subset, the weights of order at least $k$, and so on can be defined similarly to \Cref{def:pauli-weight,def:pauli-weight-level}.

The Choi state of a quantum channel $\Phi$ also has a nice relationship with the Fourier coefficients. In particular, the Choi representation is given by
\begin{equation}
    J(\Phi) = \sum_{x,y \in \pauli_n} \hat{\Phi}(x,y) J(\Phi_{x,y}),
\end{equation}
and the Choi state is given by
\begin{equation}
    v(\Phi) = \frac{1}{2^n}J(\Phi) = \frac{1}{2^n}\sum_{x,y \in \pauli_n} \hat{\Phi}(x,y) J(\Phi_{x,y}) = \sum_{x,y \in \pauli_n} \hat{\Phi}(x,y) | v(\sigma_x)\rangle \langle v(\sigma_y) |,
\end{equation}
which can be seen by normalizing $J(\Phi)$ by $2^n$, as $\Phi$ is trace-preserving.
\subsection{Distance metrics and covering nets}
\label{sec:dist}

In this section, we review definitions of various distance metrics and important properties needed for our proofs.
In \Cref{sec:pauli}, we introduced the Frobenius norm $\norm{A}_F^2 = \tr(A^\dagger A)$.
We now define normalized variants of the Frobenius distance.
\begin{definition}[Normalized Frobenius distance between operators]
\label{def:dist-frob}
    Let $A, B \in \mathcal{B}(\mathcal{H})$.
    We define the \emph{normalized Frobenius distance} as
    \begin{equation}
        d_F(A,B) \triangleq \frac{1}{\sqrt{2^{n+1}}}\|A-B\|_F.
    \end{equation}
\end{definition}
We have the following relation between the Pauli spectra of two operators and their normalized Frobenius distance.
\begin{lemma}[Frobenius distance and Pauli coefficients]
    \label{lem:df-pauli}
    For any $A,B \in \mathcal{B}(\mathcal{H})$,
    \begin{equation}
        d_F(A,B)^2 = \frac{1}{2} \sum_{x \in \pauli_n} \left|\hat{A}_x - \hat{B}_x\right|^2.
    \end{equation}
\end{lemma}
\begin{proof}
    This follows by direct computation.
    \begin{align}
            d_F(A,B)^2 &= \frac{1}{2^{n+1}}\|A-B\|_F^2
            \\&= \frac{1}{2^{n+1}}  \left(\sum_{i,j \in [2^n]} |A_{i,j}-B_{i,j}|^2\right)
            \\&= \frac{1}{2^{n+1}}  \left(\sum_{i,j \in [2^n]} |A_{i,j}|^2+|B_{i,j}|^2- A_{i,j}^*B_{i,j}-A_{i,j}B_{i,j}^*\right)
            \\&= \frac{1}{2^{n+1}}  \left(\|A\|_F^2 + \|B\|_F^2 - \langle A,B\rangle- \langle B, A\rangle\right)
            \\&= \frac{1}{2} \left(\sum_{x \in \pauli_n} |\hat{A}_x|^2 + |\hat{B}_x|^2 - \hat{A}^*_x\hat{B}_x - \hat{A}_x\hat{B}^*_x\right)
            \\&= \frac{1}{2} \sum_{x \in \pauli_n} \left|\hat{A}_x - \hat{B}_x\right|^2.
    \end{align}
    Here, the second and fourth lines follow by definition of the Frobenius norm and the Hilbert-Schmidt inner product.
    The second to last line follows from~\Cref{lem:parseval-plancherel}.
\end{proof}
Based on the notion of a Frobenius norm for superoperators introduced in~\Cref{sec:fourier-superop}, we now define a similar distance between superoperators.
\begin{definition}[Normalized Frobenius distance between superoperators~\cite{bao2023testing}]
\label{def:df-superop}
    Let $\Phi, \Psi$ be two superoperators on $\mc{B}(\mc{H})$. Then, we define the normalized Frobenius distance as
    \begin{equation}
        d_F(\Phi,\Psi) \triangleq \frac{1}{2^n\sqrt{2}} \| \Phi - \Psi\|_F \triangleq \frac{1}{2^n\sqrt{2}} \|J(\Phi) - J(\Psi) \|_F.
    \end{equation}
\end{definition}
Note that the normalization ensures that the distance is in $[0,1]$ for two quantum channels.
We now show a relation between this distance and the Fourier spectrum of superoperators.
\begin{lemma}[Normalized Frobenius distance of superoperators and Fourier coefficients; Corollary 28 in~\cite{bao2023testing}]
    \label{lem:df-superop-fourier}
    Let $\Phi, \Psi$ be two superoperators on $\mc{B}(\mc{H})$. Then,
    \begin{equation}
        d_F(\Phi, \Psi)^2 = \frac{1}{2}\sum_{x,y, \in \pauli_n} |\hat{\Phi}(x,y) - \hat{\Psi}(x,y)|^2.
    \end{equation}
\end{lemma}
\begin{proof}
    We give an explicit proof via computation.
    \begin{align}
        d_F(\Phi,\Psi)^2 &= \frac{1}{2^{2n+1}}\|\Phi-\Psi\|_F^2
        \\&= \frac{1}{2^{2n+1}} \langle \Phi - \Psi, \Phi-\Psi\rangle
        \\&= \frac{1}{2^{2n+1}} \left(\|\Phi\|_F^2 + \|\Psi\|_F^2 -  \langle \Phi, \Psi\rangle -  \langle \Psi, \Phi\rangle  \right)
        \\&= \frac{1}{2} \sum_{x,y \in \pauli_n}\left( |\hat{\Phi}(x,y)|^2 + |\hat{\Psi}(x,y)|^2 - \hat{\Phi}(x,y)^*\hat{\Psi}(x,y) - \hat{\Phi}(x,y)\hat{\Psi}(x,y)^*\right)
        \\&= \frac{1}{2} \sum_{x,y \in \pauli_n} |\hat{\Phi}(x,y) - \hat{\Psi}(x,y)|^2,
    \end{align}
    where the first three lines follow from the definitions of the normalized Frobenius distance and Hilbert-Schmidt norm for superoperators, and the fourth line follows from~\Cref{lem:superop-parseval-plancherel}.
\end{proof}
We now define distances between two quantum states, which we will shortly use to define average-case distances between quantum channels.
\begin{definition}[Trace Distance and Fidelity]
\label{def:dtr-infidelity}
    The \emph{trace distance} between two quantum states is given by
    \begin{equation}
        \dtr(\rho, \sigma) \triangleq \frac{1}{2}\norm{\rho -\sigma}_1.
    \end{equation}
    The \emph{fidelity} between two quantum states is given by
    \begin{equation}
        F(\rho,\sigma) \triangleq \tr \left(\sqrt{\sqrt{\rho} \sigma \sqrt{\rho} }\right)^2.
    \end{equation}
    We call $1-F$ the \emph{infidelity} of two states.
\end{definition}
The trace distance and fidelity are related in the following ways~\cite{nielsen2010quantum}.
For any two quantum states $\rho, \sigma$,
\begin{equation}
    1 - \sqrt{F(\rho, \sigma)} \leq \dtr(\rho, \sigma) \leq \sqrt{1-F(\rho,\sigma)}.
\end{equation}
When one of the states is pure, e.g., $\sigma = \ketbra{\psi}$, then
\begin{equation}
    1 - F(|\psi\rangle\langle\psi|, \rho) \leq \dtr(|\psi\rangle\langle\psi|, \rho).
\end{equation}
When both states are pure, then
\begin{equation}
    1 - F(|\psi\rangle, |\phi\rangle) = \dtr(|\psi\rangle, |\phi\rangle)^2.
\end{equation}

In addition to the normalized Frobenius distance, we can also consider average-case distances.
We follow~\cite{zhao2023learning} to define an average distance between two quantum channels with respect to a measure. 

\begin{definition}[Average distance~\cite{zhao2023learning}]
\label{def:davg}
    For any measure $\mu$ over $d$-dimensional states, we define the \emph{average distance} between two channels $d_\mu$ as root mean squared trace distance of output states over inputs from $\mu$
    \begin{equation}
         d_{\mu}(\mathcal{E}_1,\mathcal{E}_2) \triangleq \sqrt{ \mbf{E}_{\rho \sim \mu} [\dtr(\mathcal{E}_1(\rho),\mathcal{E}_2(\rho))^2]}.
    \end{equation}
    In particular, when the average is taken over the Haar measure $\mu_S$, we denote the distance by $\davg$
        \begin{equation}
        \davg(\mathcal{E}_1,\mathcal{E}_2) \triangleq \sqrt{ \mbf{E}_{|\psi\rangle \sim \mu_S} [\dtr(\mathcal{E}_1(|\psi\rangle\langle\psi|),\mathcal{E}_2(|\psi\rangle\langle\psi|))^2]}.
    \end{equation}
\end{definition}

However, because Haar-random states are difficult to generate, one may turn to other notions of random states with respect to which to measure an average-case distance.
One such physically motivated class of states is locally scrambled ensembles~\cite{kuo2020markovian,hu2023classical,zhao2023learning}.

We will make use of the fact that for any measure $\mu$ over states, $d_\mu$ satisfies the triangle inequality. We will follow the proof of~\cite{zhao2023learning}, where this was shown for unitaries.
\begin{lemma}[Triangle inequality of $\davg$; Lemma 6 in~\cite{zhao2023learning}]
    \label{lem:davg-triangle-ineq}
        \item For any quantum channels $\mc{E}_1, \mc{E}_2,\mc{E}_3$ acting on $\mc{B}(\mc{H})$, and any measure $\mu$ over states in $\mc{B}(\mc{H})$, we have
        \begin{equation}
            d_\mu(\mc{E}_1, \mc{E}_3) \leq d_\mu(\mc{E}_1, \mc{E}_2) + d_\mu(\mc{E}_2, \mc{E}_3).
        \end{equation}
\end{lemma}
\begin{proof}
    \begin{align}
        d_\mu(\mc{E}_1, \mc{E}_3)^2 &= \mbf{E}_{\rho \sim \mu}\left[\dtr(\mathcal{E}_1(\rho), \mathcal{E}_3(\rho))^2 \right]
        \\&\leq \mbf{E}_{\rho \sim \mu}\left[(\dtr(\mathcal{E}_1(\rho), \mathcal{E}_2(\rho)) + (\dtr(\mathcal{E}_2(\rho), \mathcal{E}_3(\rho)) )^2 \right]
        \\&= d_\mu(\mc{E}_1, \mc{E}_2)^2 + d_\mu(\mc{E}_2, \mc{E}_3)^2 + 2\mbf{E}_{\rho \sim \mu}\left[\dtr(\mathcal{E}_1(\rho), \mathcal{E}_2(\rho))\dtr(\mathcal{E}_2(\rho), \mathcal{E}_3(\rho))\right]
        \\&\leq d_\mu(\mc{E}_1, \mc{E}_2)^2 + d_\mu(\mc{E}_2, \mc{E}_3)^2 + 2\sqrt{\mbf{E}_{\rho \sim \mu}\left[\dtr(\mathcal{E}_1(\rho), \mathcal{E}_2(\rho))^2\right]}\sqrt{\mbf{E}_{\rho \sim \mu}\left[\dtr(\mathcal{E}_2(\rho), \mathcal{E}_3(\rho))^2\right]}
        \\&= (d_\mu(\mc{E}_1, \mc{E}_2) + d_\mu(\mc{E}_2, \mc{E}_3))^2,
    \end{align}
    where in the first inequality we use the triangle inequality of trace distance, and in the last inequality we use the Cauchy-Schwarz inequality.
\end{proof}
We can also relate the average-case distance $\davg$ to the Pauli coefficients via the following lemma.

\begin{lemma}[Average distance and Pauli coefficients]
    \label{lem:davg-pauli}
    For any $n$-qubit unitaries $U,V$, we have
    \begin{equation}
        \davg(U,V)^2 = \frac{2^n}{2^n+1}\left(1-\bigg|\sum_{x \in \pauli_n} \hat{U}^*_x \hat{V}_x\bigg|^2\right),
    \end{equation}
\end{lemma}
\begin{proof}
    By properties of the Haar integral (e.g., Example 48 in~\cite{mele2024introduction}), we have
    \begin{equation}
    \label{eq:davg-trace}
        \davg(U,V)^2 = \frac{4^n-\left|\tr(U^\dag V)\right|^2}{2^n(2^n+1)}.
    \end{equation}
    Then, by Plancherel's identity (\Cref{lem:parseval-plancherel}), we have
    \begin{equation}
        \tr(U^\dagger V) = \langle U, V \rangle = 2^n\sum_{x \in \pauli_n} \hat{U}_x^* \hat{V}_x.
    \end{equation}
    The result clearly follows from this.
    \end{proof}

A related topic is that of covering nets from high-dimensional probability theory.
We can define a covering net with respect to any of the distance metrics introduced above.
    
\begin{definition}[Covering net/number and metric entropy]
    \label{def:covering}
    Let $(X,d)$ be a metric space. Let $K \subseteq X$ be a subset and $\epsilon > 0$. Then, define the following.
    \begin{itemize}
        \item $N \subseteq K$ is an \emph{$\epsilon$-covering net} of $K$ if for any $x \in K$, there exists a $y \in N$ such that $d(x,y) \leq \epsilon$.
        \item The \emph{covering number} $\mathcal{N}(K, d, \epsilon)$ of $K$ is the smallest possible cardinality of an $\epsilon$-covering net of $K$.
        \item The \emph{metric entropy} is $\log \mathcal{N}(K,d,\epsilon)$.
    \end{itemize}
\end{definition}

A useful result about covering nets that we will need in our proofs is bounds on the metric entropy of covering nets for unitaries of bounded gate complexity~\cite{zhao2023learning}.

\begin{lemma}[Covering nets for unitaries of bounded gates with respect to $d_F$; Theorem 8 and Corollary 1 in~\cite{zhao2023learning}]
    \label{lem:cover-unitary-gate}
    Let $U^G \subseteq U(2^n)$ be the set of $n$-qubit unitaries that can be implemented by $G$ two-qubit gates. 
    Then for any $\epsilon \in (0, 1]$, there exist universal constants $c_1, c_2, C>0$ such that for $1\leq G/C\leq 4^{n+1}$, the metric entropy of $U^G$ with respect to the normalized Frobenius distance $d_F$ can be bounded as
    \begin{equation}
        \frac{G}{4C}\log(\frac{c_1}{\epsilon})\leq \log\mathcal{N}(U^G, d_F, \epsilon) \leq 16G\log(\frac{c_2G}{\epsilon}) + 2G\log n.
    \end{equation}
    Moreover, the metric entropy of $U^G$ with respect to the average-case distance $d_{\mathrm{avg}}$ can be bounded as
    \begin{equation}
        \frac{G}{4C}\log(\frac{c_1}{8\epsilon}) - \log \left(\frac{c_2}{2\epsilon}\right)\leq \log\mathcal{N}(U^G, d_{\mathrm{avg}}, \epsilon) \leq 16G\log(\frac{c_2G}{\epsilon}) + 2G\log n.
    \end{equation}
\end{lemma}

\section{Useful Lemmas}
\label{sec:lemmas}
In this section, we compile several lemmas that we use repeatedly throughout our proofs.
Our algorithms for both agnostic unitary estimation and agnostic process tomography rely on sampling from a probability distribution to estimate it empirically.
In our agnostic unitary estimation algorithms, we will often sample from the distribution $\{|\hat{U}_x|^2\}_x$ formed by the Pauli coefficients to estimate it empirically. On the other hand, for APT, we will sample from the distribution formed by the diagonal Fourier coefficients $\{\hat{\Phi}(x,x)\}_x$.
For guarantees on the estimation, we make use of the following well-known result from distribution learning theory.
We refer to Theorem 9 in~\cite{canonne2020short} for a proof and follow the presentation of Lemma 16 in~\cite{arunachalam2024learning}.

\begin{lemma}[Empirical estimation of a distribution; Theorem 9 in~\cite{canonne2020short}]
    \label{lem:dist-learning}
    Let $p = \{p(x)\}_x$ be a probability distribution over some set $\mc{X}$. Let $p^\prime$ be the empirical probability distribution obtained after sampling $N$ times from $p$. Then, $N = \mc{O}(\log(1/\delta)/\epsilon^2)$ samples suffice to have, with probability at least $1-\delta$,
    \begin{enumerate}
        \item  $|p(x)-p^\prime(x)| \leq \epsilon, \forall x \in \mc{X}$,
        \item $ \sqrt{\sum_{x \in \mc{X}} |p(x)-p^\prime(x)|^2} \leq \epsilon$.
    \end{enumerate}
\end{lemma}
Many of our agnostic unitary estimation algorithms require learning a subset of the Pauli coefficients of a unitary.
In the following lemma, we show how any Pauli coefficient can be estimated.
This algorithm extends the technique introduced in Lemma 24 of~\cite{montanaro2008quantum} for learning Hermitian unitary operators, where the Pauli coefficients are all guaranteed to be real.
However, in general, the Pauli coefficients may be complex.
We resolve this by using the approach from~\cite{montanaro2008quantum} to separately estimate the real and imaginary parts of each Pauli coefficient.
This extension is also proposed in Lemma 17 of~\cite{arunachalam2024learning} without proof.
For completeness, we provide a proof here.

\begin{lemma}[Estimating any Pauli coefficient of a unitary; Lemma 17 in~\cite{arunachalam2024learning}]
\label{lem:estimate-pauli-coeff-unitary}
Given query access to an $n$-qubit unitary $U$ and a string $x \in \pauli_n$ as input, there exists an algorithm that produces an estimate $\alpha_x$ of the Pauli coefficient $\hat{U}_x$ using $N = \mathcal{O}\left(\frac{\log(1/\delta)}{\epsilon^2}\right)$ queries to $U$ such that
\begin{equation}
    |\alpha_x-\hat{U}_x| \leq \epsilon,
\end{equation}
with probability at least $1-\delta$. Moreover, the algorithm runs in computational time $\mathcal{O}\left(\frac{\log(1/\delta)}{\epsilon^2}\right)$.
\end{lemma}
\begin{proof}
    As discussed above, the algorithm estimates the real and imaginary parts of each Pauli coefficient separately by the same approach as in Lemma 24 of~\cite{montanaro2008quantum}.
    To do so, we need access to controlled-$U$.
    This can be implemented using query access to $U$ alone by preparing $n$ ancilla qubits, applying a controlled-\textsf{SWAP} between the ancilla and main register, applying $U$ to the ancillas, another controlled-\textsf{SWAP}, and finally discarding the ancillas. 
    
    First, we estimate the real part.
    We proceed in the same way as Lemma 24 of~\cite{montanaro2008quantum}.
    We first prepare $\ket{0}\ket{\Phi^+}$, where $\ket{\Phi^+}$ denotes a Bell pair.
    Then, we apply a Hadamard gate to get
    \begin{equation}
        \frac{1}{\sqrt{2}}\ket{0}\ket{\Phi^+} + \frac{1}{\sqrt{2}}\ket{1}\ket{\Phi^+}.
    \end{equation}
    Applying controlled-$\sigma_x$ controlled on $\ket{0}$, we have
    \begin{equation}
        \frac{1}{\sqrt{2}}\ket{0}\ket{v(\sigma_x)} + \frac{1}{\sqrt{2}}\ket{1}\ket{\Phi^+}.
    \end{equation}
    Next, we apply controlled-$U$ as described above to obtain the state
    \begin{equation}
        \label{eq:2}
        \frac{1}{\sqrt{2}}|0\rangle|v(\sigma_x)\rangle+\frac{1}{\sqrt{2}}|1\rangle|v(U)\rangle.
    \end{equation}
    Then, we apply another Hadamard gate to the first qubit to obtain
    \begin{equation}
        \frac{1}{2}|0\rangle(|v(\sigma_x)\rangle+|v(U)\rangle)+\frac{1}{2}|1\rangle(|v(\sigma_x)\rangle-|v(U)\rangle).
    \end{equation}
    Finally, we measure the first qubit in the computational basis, giving us outcome probabilities
    \begin{equation}
        \mathbf{Pr}[0] = \frac{1}{4}(\langle v(\sigma_x)|+ \langle v(U)|)(|v(\sigma_x)\rangle+|v(U)\rangle) = \frac{1}{2}+\frac{(\hat{U}_x+\hat{U}_x^*)}{4} = \frac{1}{2}+\frac{\mathrm{Re}(\hat{U}_x)}{2},
    \end{equation}
    and similarly
    \begin{equation}
        \mathbf{Pr}[1] = \frac{1}{2}-\frac{\mathrm{Re}(\hat{U}_x)}{2}.
    \end{equation}
    Using Hoeffding's inequality, we see that by repeating this procedure $\mathcal{O}(\log (1/\delta)/\epsilon^2)$ many times, we can estimate $\mathrm{Re}(\hat{U}_x)$ up to accuracy $\epsilon/\sqrt{2}$ with probability at least $1-\delta/2$.

    To estimate the imaginary part, we replace \Cref{eq:2} with the following
    \begin{equation*}
         \frac{1}{\sqrt{2}}|0\rangle|v(\sigma_x)\rangle+\frac{i}{\sqrt{2}}|1\rangle|v(U)\rangle.
    \end{equation*}
    We again perform a Hadamard operation followed by a measurement of the control qubit.
    Using a similar analysis to the real case, we obtain measurement probabilities
    \begin{equation}
        \mathbf{Pr}[0] = \frac{1}{2}-\frac{1}{2}\mathrm{Im}(\hat{U}_x) \text{ and } \mathbf{Pr}[1] = \frac{1}{2}+\frac{1}{2}\mathrm{Im}(\hat{U}_x).
    \end{equation}
    Then, by repeating this procedure $\mathcal{O}(\log (1/\delta)/\epsilon^2)$ many times, we can now estimate $\mathrm{Im}(\hat{U}_x)$ up to accuracy $\epsilon/\sqrt{2}$ with probability at least $1-\delta/2$. Using a union bound, we see that with probability at least $1-\delta$, we have an $\epsilon$-accurate estimate of $\hat{U}_x$. 

    As we spend $\mathcal{O}(1)$ time processing each sample in both procedures, the computational time is of the same order as the sample complexity.
\end{proof}

Similarly, we can also learn any Fourier coefficient of a quantum channel.
The same result is in Lemma 15 in~\cite{arunachalam2024learning}, but we give a different proof that allows for an easier analysis of the computational time.
This is because our proof makes the state preparation and measurement steps more explicit than that of~\cite{arunachalam2024learning}.

\begin{lemma}[Estimating any Fourier coefficient of a quantum channel]
\label{lem:estimate-fourier-coeff-channel}
Given query access to a quantum channel $\Phi$ on $\mc{B}(\mc{H})$, and strings $x,y \in \pauli_n$ as input, there exists a method to produce an estimate $\alpha_{x,y}$ of the Fourier coefficient $\hat{\Phi}(x,y)$ using $N = \mathcal{O}\left(\frac{\log(1/\delta)}{\epsilon^2}\right)$ queries to $\Phi$ such that
\begin{equation}
    |\alpha_{x,y}-\hat{\Phi}(x,y)| \leq \epsilon
\end{equation}
with probability at least $1-\delta$. Moreover, the algorithm runs in computational time $\mathcal{O}\left(\frac{\log(1/\delta)}{\epsilon^2}\right)$.
\end{lemma}

\begin{proof}
    Recall that the Choi state of a quantum channel can be expanded in terms of its Fourier coefficients and the Choi states of the Pauli operators as
    \begin{equation}
    \label{eq:fourier-superop-coeff-1}
        v(\Phi) = \sum_{x, y \in \pauli_n} \hat{\Phi}(x,y) |v(\sigma_x)\rangle \langle v(\sigma_y)|.
    \end{equation}
    We first estimate the diagonal coefficients $\hat{\Phi}(x,x)$ and $\hat{\Phi}(y,y)$. From~\Cref{eq:fourier-superop-coeff-1}, we have
    \begin{equation}
        \langle v(\sigma_x) | v(\Phi) | v(\sigma_x) \rangle = \hat{\Phi}(x,x).
    \end{equation}
    Then, similarly to~\Cref{lem:estimate-pauli-coeff-unitary}, by conducting $\mc{O}(\log(1/\delta)/\epsilon^2)$ SWAP tests~\cite{buhrman2001quantum,barenco1997stabilization,kobayashi2003quantum} between $v(\Phi)$ and $|v(\sigma_x)\rangle\langle v(\sigma_x)|$, we can estimate $\hat{\Phi}(x,x)$ up-to accuracy $\epsilon/(2\sqrt{2})$ with probability at least $1-\delta/4$, and similarly for $\hat{\Phi}(y,y)$. Then, consider the state
\begin{equation}
    |\psi(x,y)^+\rangle \triangleq \frac{|v(\sigma_x)\rangle + |v(\sigma_y)\rangle}{\sqrt{2}}.
\end{equation}
Using~\Cref{eq:fourier-superop-coeff-1}, we have
\begin{align}
    \langle\psi(x,y)^+| v(\Phi) |\psi(x,y)^+\rangle &= \frac{(\langle v(\sigma_x) | + \langle v(\sigma_y) |) v(\Phi) (| v(\sigma_x) \rangle + | v(\sigma_y) \rangle)}{2}
    \\&= \frac{\hat{\Phi}(x,x)+\hat{\Phi}(y,y)+\hat{\Phi}(x,y)+\hat{\Phi}(y,x)}{2}
    \\&= \frac{\hat{\Phi}(x,x)+\hat{\Phi}(y,y)+2\mathrm{Re}(\hat{\Phi}(x,y))}{2},
\end{align}
    where the last line uses the fact that $\hat{\Phi}(y,x) = \hat{\Phi}(x,y)^*$, as $\hat{\Phi}$ is Hermitian. Now, with probability at least $1-\delta/4$, we estimate the overlap between $v(\Phi)$ and $|\psi(x,y)^+\rangle\langle\psi(x,y)^+|$ up to accuracy $\epsilon/(2\sqrt{2})$, using $\mc{O}(\log(1/\delta)/\epsilon^2)$ SWAP tests. This allows us to estimate $\mathrm{Re}(\hat{\Phi}(x,y))$ up to accuracy $\epsilon/\sqrt{2}$ since we already estimated $\hat{\Phi}(x,x)$ and $\hat{\Phi}(y,y)$ from the previous step.

    Similarly, to estimate $\mathrm{Im}(\hat{\Phi}(x,y))$, instead of $|\psi(x,y)^+\rangle$, we use the state
    \begin{equation}
        |\psi(x,y)^-\rangle \triangleq \frac{|v(\sigma_x)\rangle -i |v(\sigma_y)\rangle}{\sqrt{2}},
    \end{equation}
    which satisfies
    \begin{equation}
         \langle\psi(x,y)^-| v(\Phi) |\psi(x,y)^-\rangle = \frac{\hat{\Phi}(x,x)+\hat{\Phi}(y,y)+2\mathrm{Im}(\hat{\Phi}(x,y))}{2}.
    \end{equation}
    As in the case of the real component, we use $\mc{O}(\log(1/\delta)/\epsilon^2)$ SWAP tests to estimate this overlap, allowing us to estimate the imaginary component up to error $\epsilon/\sqrt{2}$. Thus, we obtain an estimate of $\hat{\Phi}(x,y)$ up to error $\epsilon$. From a union bound, we have a total failure probability of at most $\delta$.
\end{proof}

Another tool we utilize often is the classical shadow formalism~\cite{huang2020predicting}, which allows for the efficient prediction of expectation values of many observables from copies of a particular state.
We use classical shadows to estimate Pauli coefficients as a subroutine in some of our learning algorithms.
\begin{lemma}[Classical Shadows; Theorem 1 in~\cite{huang2020predicting}]
    \label{lem:classical-shadows}
    Given copies of a quantum state $\rho \in \mc{B}(\mc{H})$ and a list of Hermitian observables $\{O_1,\dots,O_M\} \subset \mc{B}(\mc{H})$, satisfying $\|O_i\|_\infty \leq 1$ for all $i \in [M]$, there exists an algorithm that produce estimates $\{\hat{o}_i\}$ such that
    \begin{equation}
    \label{eq:cs-guarantee}
        |\hat{o}_i - \tr(O_i\rho)| \leq \epsilon, \forall i \in [M]
    \end{equation}
     with probability at least $1-\delta$.
    The method employs randomized measurements on copies of $\rho$, followed by classical post-processing.
    Depending on the measurement primitive, this aforementioned algorithm has the following sample and time complexity.
    \begin{enumerate}
        \item To achieve \Cref{eq:cs-guarantee}, it suffices to perform \emph{random Clifford measurements} on
        \begin{equation}
            N = \mc{O}\left(\frac{\log(M/\delta)}{\epsilon^2} \max_{i \in [M]} \tr(O_i^2)\right)
        \end{equation}
        copies of the unknown quantum state, but the classical post-processing time may be inefficient in general.
        When the observables have efficient representations, the algorithm is computationally efficient.
        For example, when the observable is a stabilizer state, the computational time is
        \begin{equation}
            \mc{O}\left(n^2 MN\right).
        \end{equation}
        \item When all observables $O_i$ are $k$-local, to achieve~\Cref{eq:cs-guarantee}, it suffices to perform \emph{random Pauli measurements} on
        \begin{equation}
            N = \mc{O}\left(\frac{4^k \log(M/\delta)}{\epsilon^2} \right)
        \end{equation}
        copies of the unknown quantum state, and the classical post-processing can be performed in time
        \begin{equation}
            \mc{O}\left(kNM\right).
        \end{equation}
        \item When all observables $O_i$ are $k$-local Paulis, to achieve~\Cref{eq:cs-guarantee}, it suffices to perform \emph{random Pauli measurements} on
        \begin{equation}
        N = \mc{O}\left(\frac{3^k\log(M/\delta)}{\epsilon^2}\right)
        \end{equation} copies of the state, and the classical post-processing can be performed in time
        \begin{equation}
            \mc{O}\left(kNM\right).
        \end{equation}
    \end{enumerate}
\end{lemma}

Finally, the last tool we need is used to ensure that the output of our agnostic learning algorithms is unitary or CPTP.
This is because many of our learning algorithms rely on estimating Pauli coefficients; hence these estimates may not result in an actual unitary or quantum channel.
Thus, in order to achieve \emph{proper} learning, in which the output of our learning algorithm lies in the concept class $\mathcal{C}$ that we want to learn with respect to, we need to approximate our learned operator by the nearest unitary or quantum channel.
We refer to this as ``projection.''
The following lemma shows how to achieve this for unitary projection via the approach of Lemma 8 in~\cite{huang2024learning}.
Note that Lemma 8 in~\cite{huang2024learning} uses the same algorithm but shows that it is a unitary projection with respect to the operator norm.
We show that this also gives a unitary projection with respect to the (normalized) Frobenius distance.

\begin{lemma}[Unitary projection]
\label{lem:unitary-projection}
There exists a map $\proju$ that takes as input any operator $A \in \mathcal{B}(\mathcal{H})$ and outputs a unitary such that
\begin{equation}
    d_F(A,\proju(A)) \leq d_F(A,B)
\end{equation}
for any $2^n$-dimensional unitary $B$ and runs in time $\mathcal{O}(2^{3n})$. In other words, $\proju(A)$ is the closest unitary projection of $A$ within normalized Frobenius distance:
\begin{equation}
    \proju(A) = \argmin_{B \in \mathcal{U}_{2^n}} d_F(A, B).
\end{equation}
\end{lemma}
\begin{proof}
    The projection $\proju$ is defined as follows.
    Let $A = U\Sigma V^\dag$ be the singular value decomposition of $A$, where $\Sigma$ is a diagonal matrix consisting of non-negative real entries and $U,V \in \mathcal{U}_{2^n}$.
    Then, $\proju$ is defined as
    \begin{equation}
        \proju(A) \triangleq UV^\dag.
    \end{equation}
    Intuitively, the projection computes the singular value decomposition of $A$ and replaces all the singular values with $1$.
    This technique was originally proposed in~\cite{huang2024learning}, where the authors showed that this is the closest unitary projection for any operator with respect to the operator norm.
    We instead show that this also gives a unitary projection for the (normalized) Frobenius distance. It suffices to show that for all $B \in \mathcal{U}_d$
    \begin{equation}
        \|A-B\|_F^2 \geq \|A-UV^\dag\|_F^2,
    \end{equation}
    or equivalently, by the unitary invariance of the Frobenius norm,
    \begin{equation}
        \|\Sigma-U^\dagger BV\|_F^2 \geq \|\Sigma-I\|_F^2.
    \end{equation}
    Denote $W \triangleq U^\dagger B V$. Now,
    \begin{align}
            \|\Sigma-W\|_F^2 &= \tr\left((\Sigma-W^\dag)(\Sigma-W)\right)
            \\&= \tr(\Sigma^2 - W^\dag\Sigma-\Sigma W + I)
    \end{align}
    where the first equality uses the fact that $\Sigma$ is diagonal and non-negative, while the second line uses the fact that $W$ is unitary.
    
    Let $\{\Sigma_{ii}\}_{i \in [2^n]}$ denote the diagonal elements of $\Sigma$.
    Then, we have
    \begin{align}
        \tr(\Sigma W + W^\dag \Sigma) &= \sum_{i \in [2^n]} \langle i | (\Sigma W + W^\dag \Sigma) | i \rangle
        \\&= \sum_{i \in [2^n]} \Sigma_{ii}  \langle i |(W + W^\dag)| i \rangle
        \\&= \sum_{i \in [2^n]} 2\Sigma_{ii} \cdot \mathrm{Re}(\langle i | W | i \rangle)
        \\& \leq \sum_{i \in [2^n]} 2\Sigma_{ii}
        \\&= 2 \cdot \tr(\Sigma).
    \end{align}
    Here, in the second line, we use that $\Sigma$ is diagonal and so commutes with $W, W^\dag$.
    In the inequality, we use that $W$ is unitary.
    Finally, we get
    \begin{equation}
         \|\Sigma-W\|_F^2 \geq \tr(\Sigma^2-2\Sigma+I) = \|\Sigma-I\|_F^2,
    \end{equation}
    concluding the proof.
    The runtime can be seen as the runtime of any algorithm computing the SVD and multiplying $U, V^\dag$.
    We refer to Lecture 31 of \cite{trefethen2022numerical} for the analysis of algorithms based on the Golub-Kahan bidiagonalization procedure \cite{golub1965calculating}.
    Namely, computing the SVD takes time $\mathcal{O}(m^3)$ for an $m \times m$ matrix.
    While such algorithms are only approximate, they demonstrate superlinear convergence, allowing one to efficiently obtain an output within machine precision and thus ignore this error in practice.
\end{proof}

Showing such a result for CPTP projection is significantly more difficult, and we discuss it in detail in \Cref{sec:proper-apt}.

\section{Agnostic Unitary Estimation (AUE)}
In this section, we consider the task of agnostic unitary estimation (AUE).
Recall that this is a special case of agnostic process tomography (APT) in which the unknown channel is assumed to be unitary and the concept class $\mathcal{C}$ consists entirely of unitaries.
This section serves as a warm-up to~\Cref{sec:apt}, as we will generalize many of the techniques presented here to APT later.
\label{sec:aue}
\subsection{Bounded-gate circuits}
\label{sec:aue-bounded}
In this section, we consider AUE when the concept class $\mathcal{C}$ consists of unitaries with a bounded gate complexity.
In particular, we consider $n$-qubit unitaries generated by $G$ two-qubit gates.
This class of unitaries was studied in the realizable case in~\cite{zhao2023learning}.

First, we prove a more general result about AUE for any class $\mathcal{C}$ of unitaries with finite size $|\mathcal{C}| < +\infty$. In the following proposition, we show that AUE with respect to the average-case distance metric $\davg$ and the class $\mathcal{C}$ can be accomplished using a sample complexity scaling as $\log |\mathcal{C}|$.
Thus, if $\mathcal{C}$ is any class of unitaries generated by $\mathrm{poly}(n)$ gates, $\log |\mathcal{C}| \leq \mathrm{poly}(n)$ so that the sample complexity is efficient.
Qualitatively, this says that efficient AUE is easy from a sample complexity perspective, but it can be difficult to achieve with respect to time complexity.
This is similar in spirit to an analogous argument by~\cite{grewal2024agnostic} for agnostic state tomography.
Their argument relies on shadow tomography~\cite{aaronson2018shadow}, while ours is based on classical shadows and takes inspiration from a proof of~\cite{zhao2023learning}.

\begin{prop}[Agnostic unitary estimation of finite-sized classes]
    \label{prop:aue-finite-class}
    There exists a learning algorithm for proper $(1,\epsilon,\delta)$-agnostic unitary estimation with respect to error in average-case distance $\davg$ for any finite-sized class $\mathcal{C}$ of unitaries using 
    \begin{equation}
        N = \mathcal{O}\left(\frac{\log (|\mathcal{C}|/\delta)}{\epsilon^4}\right)
    \end{equation}
    queries to the unknown unitary $U$ and running in time
    \begin{equation}
    \mathcal{O}\left(T \cdot \frac{|\mathcal{C}|\log(|\mathcal{C}|/\delta)}{\epsilon^4}
        \right),
    \end{equation}
    where $T$ is the time complexity of computing the inner product of two states given their exact description.
\end{prop}

We note that the time complexity in the proposition depends on the complexity of computing the inner products of pairs of states.
This is due to the use of Clifford classical shadows~\cite{huang2020predicting}, which is not computationally efficient for arbitrary observables (see \Cref{lem:classical-shadows}).
The resulting complexity depends on the class of unitaries one aims to learn and how efficiently their output states can be represented. However, since the time complexity already scales linearly with the size of the concept class, this algorithm would be computationally inefficient for most classes of interest, so we do not analyze the time complexity in detail.

\begin{proof}
    The algorithm is the same as that of Section C.2(a) of~\cite{zhao2023learning}.
    Namely, we use Clifford classical shadows~\cite{huang2020predicting} to estimate the distance $\davg(U_i, U)$ between the unknown unitary $U$ and all unitaries $U_i \in \mathcal{C}$.
    Then, our algorithm outputs the one with the minimal distance.

    The analysis is the same as Section C.2(a) of~\cite{zhao2023learning}, so we do not reproduce the full argument here. We start by preparing random input states from the ensemble
    \begin{equation}
        P = \mathrm{Uniform}(\{|\psi\rangle = U |0\rangle, U \in \clifford(n) \})
    \end{equation}
    of random stabilizer states, where $\clifford(n)$ denotes the $n$-qubit Clifford group. Then, we use classical shadows~\cite{huang2020predicting} to estimate distances with respect to this ensemble.
    
    While~\cite{zhao2023learning} used random stabilizer product states as input, we use random stabilizer states instead. We note that changing the input distribution only changes the ensemble over which the distance is estimated, not the sample complexity for successful estimation. Now, by Equation (C.8) of~\cite{zhao2023learning}, using
    \begin{equation}
        N = 204 \frac{\log (2|\mathcal{C}|/\delta)}{\epsilon^{\prime2}} = \mathcal{O}\left(\frac{\log(|\mathcal{C}|/\delta)}{\epsilon'^2}\right)
    \end{equation}
    queries to the unknown unitary, we see that we can construct unbiased estimators $\hat{o}_i$, such that $1-\hat{o}_i$ estimates the square of the average distance with respect to $P$ (\Cref{def:davg}).
    \begin{equation}
        \label{eq:shadow-est}
        |d_{P}(U_i,U)^2 - (1-\hat{o}_i)| \leq \epsilon^\prime, \quad \forall i \in [|\mathcal{C}|]
    \end{equation}
    with probability at least $1-\delta$ for some $\delta, \epsilon' > 0$.     
    Then, we can select $i^\star = \arg\min_i (1 - \hat{o}_i)$ and output $U_{i^\star}$, i.e., the $U_i \in \mathcal{C}$ with the smallest estimated distance from $U$ with respect to $d_P$.
    To relate this back to $\davg$, recall that $P$ forms a $3$-design~\cite{kueng2015qubit,webb2015clifford,zhu2017multiqubit}, implying that the average infidelity over $P$ is the same as that over the Haar measure. Thus for any $2^n$-dimensional unitaries $U_1, U_2$,
    \begin{align}
         \davg(U_1, U_2)^2 &= \mathbf{E}_{|\psi\rangle \sim \mu_S} [\dtr(U_1|\psi\rangle, U_2|\psi\rangle)^2] \\&= 
         \mathbf{E}_{|\psi\rangle \sim \mu_S}[1-F(U_1|\psi\rangle, U_2|\psi\rangle)]
         \\&= \mathbf{E}_{|\psi\rangle \sim P}[1-F(U_1|\psi\rangle, U_2|\psi\rangle)]
         \\&= \mathbf{E}_{|\psi\rangle \sim P} [\dtr(U_1|\psi\rangle, U_2|\psi\rangle)^2]
         \\&= \label{eq:davg-haar-random-clifford}  d_P(U_1, U_2)^2,
    \end{align} where in the second and fourth lines we use~\Cref{def:dtr-infidelity}, and in the third line we use that $P$ is a $3$-design.
    We can now bound the error of our hypothesis with respect to $\davg$.
    \begin{align}
        \davg(U_{i^\star},U)^2 &= d_P(U_{i^\star},U)^2
        \\& \leq (1-\hat{o}_{i^\star}) + \epsilon^\prime
        \\& = \min_i (1 - \hat{o}_i) + \epsilon'
        \\& \leq  \min_i(d_P(U_i,U)^2+\epsilon^\prime) + \epsilon^\prime
        \\& \leq \min_i (d_P(U_i,U)^2) + 2\epsilon^\prime
        \\&= \min_i (\davg(U_i,U)^2)+2\epsilon^\prime
        \\&= \mathsf{opt}(U, \mathcal{C})^2+2\epsilon^\prime.
    \end{align}
    In the first line, we use~\Cref{eq:davg-haar-random-clifford}.
    In the second line, we use \Cref{eq:shadow-est}.
    In the third line, we use the definition of $i^\star$.
    In the fourth line, we use \Cref{eq:shadow-est} again.
    In the second to last line, we use~\Cref{eq:davg-haar-random-clifford} again. Finally, in the last line, we use the definition of $\mathsf{opt}$.
    Thus, by setting $\epsilon^\prime = \epsilon^2/2$, we obtain
    \begin{equation}
         \davg(U_{i^\star},U) \leq \mathsf{opt}(U, \mathcal{C})+\epsilon.
    \end{equation}
    This gives us a 1-agnostic learner with the claimed sample complexity. The time complexity is due to the computation of the $\hat{o}_{i}$ for all elements of $\mathcal{C}$, each of which requires computing inner products between $\mathcal{O}(N)$ pairs of states.
\end{proof}

While Proposition~\ref{prop:aue-finite-class} only holds for concept classes of finite size, we now show that this can be easily extended to any class as long as it has bounded metric entropy (see \Cref{def:covering}).
\begin{corollary}
    \label{prop:aue-bounded-metric}
    Let $\epsilon, \epsilon^\prime, \delta > 0$.
    There exists a learning algorithm for proper $(1,\epsilon+\epsilon^\prime,\delta)$-agnostic unitary estimation with respect to error in average-case distance $\davg$ for any class $\mathcal{C}$ with covering number $\mathcal{N}(\mathcal{C},\davg,\epsilon^\prime)$, using  
    \begin{equation}
        N = \mathcal{O}\left(\frac{\log (\mathcal{N}(\mathcal{C},\davg,\epsilon^\prime)/\delta)}{\epsilon^4}\right)
    \end{equation}
    queries to the unknown unitary $U$ and running in time
    \begin{equation}
    \mathcal{O}\left(T \cdot \mathcal{N}(\mathcal{C},\davg,\epsilon^\prime) \cdot N\right),
    \end{equation}
     where $T$ is the time complexity of computing the inner product of two states given their exact description.
\end{corollary}

\begin{proof}
    Because $\mathcal{C}$ has covering number $\mathcal{N}(\mathcal{C},\davg,\epsilon^\prime)$, there exists some $\epsilon'$-covering net $M$ with cardinality $\mathcal{N}(\mathcal{C},\davg,\epsilon^\prime)$.
    Consider applying Proposition~\ref{prop:aue-finite-class} to the $\epsilon^\prime$-cover $M$.
    This gives us a hypothesis $V \in M$ satisfying
    \begin{equation}
        \davg(U, V) \leq \min_{W \in M} \davg(U, W) + \epsilon.
    \end{equation}
    Moreover, to find this hypothesis, we used
    \begin{equation}
        N = \mathcal{O}\left(\frac{\log(|M|/\delta)}{\epsilon^4}\right) = \mathcal{O}\left(\frac{\log(\mathcal{N}(\mathcal{C},\davg,\epsilon^\prime)/\delta)}{\epsilon^4}\right)
    \end{equation}
    queries to the unknown unitary. The time complexity is also $\mathcal{O}(T \cdot \mathcal{N}(\mathcal{C},\davg,\epsilon^\prime) \cdot N)$ by Proposition~\ref{prop:apt-finite-class}.
    
    Let $W^*$ be the optimal hypothesis from $\mathcal{C}$, i.e.,
    \begin{equation}
        W^* \triangleq \argmin_{W \in \mathcal{C}} \davg(U,W)\quad \text{ and }\quad \opt(U, \mathcal{C}) = \davg(U, W^*).
    \end{equation}
    Then,
    \begin{align}
        \davg(U, V) &\leq \min_{W \in M} \davg(U, W) + \epsilon
        \\&\leq\min_{W \in M} \left( \davg(U, W^*) +  \davg(W^*, W)\right) + \epsilon
        \\& \leq \opt(U, \mathcal{C}) + \min_{W \in M} \davg(W^*, W) + \epsilon
        \\&\leq \opt(U, \mathcal{C}) + \epsilon^\prime + \epsilon,
    \end{align}
    where the first inequality follows from the guarantees of Proposition~\ref{prop:aue-finite-class}, the second inequality follows from the triangle inequality for $\davg$, the third inequality follows by the definition of $W^*$, and the fourth inequality uses that $M$ is an $\epsilon^\prime$-covering net for $\mathcal{C}$.
\end{proof}

Finally, we can obtain a sample complexity upper bound for AUE of unitaries with bounded gate complexity.
This follows immediately from \Cref{lem:cover-unitary-gate}, which gives a bound on the metric entropy of unitaries with bounded gate complexity.
However, by \Cref{prop:aue-bounded-metric}, the time complexity scales with the cardinality of the covering net, which is exponentially large (\Cref{lem:cover-unitary-gate}).
This is not surprising, as even in the realizable case, there are lower bounds on the computational complexity of learning this class of unitaries~\cite{zhao2023learning} based on the hardness of \textsf{RingLWE}~\cite{lyubashevsky2010ideal}.

\begin{corollary}[Agnostic unitary estimation of bounded-gate unitaries]
\label{cor:aue-gate}
    There exists a (computationally inefficient) learning algorithm for proper $(1,\epsilon,\delta)$-agnostic unitary estimation of the class of $n$-qubit unitaries of $G$ 2-qubit gates with respect to error in $\davg$ using
    \begin{equation}
        N = \Tilde{\mathcal{O}}\left(
            \frac{G \log( n/\delta)}{\epsilon^4}
        \right)
    \end{equation}
    queries.
\end{corollary}
\subsection{AUE from agnostic state tomography}
\label{sec:aue-clifford}

In this section, we prove a general result which shows that we can obtain an algorithm for improper AUE from existing agnostic state tomography algorithms with only a quadratic increase in sample/time complexity with respect to the error $\epsilon$.
This is a fairly general result that holds whenever the concept classes for our desired AUE algorithm and the agnostic state tomography are compatible, in the sense of Proposition~\ref{prop:ast-to-aue}.
To properly state this proposition, we first give the definition of agnostic state tomography.

\begin{definition}[Agnostic state tomography~\cite{grewal2024agnostic}]
    \label{def:ast}
    Let $0 < \epsilon, \delta < 1$. Let $\mathcal{D}$ be a class of quantum states. A learning algorithm $\mathcal{A}$ is an \emph{$(1, \epsilon, \delta)$-agnostic learner} with respect to fidelity and the concept class $\mathcal{D}$ if, given access to copies of an arbitrary quantum state $\rho$, $\mathcal{A}$ outputs a hypothesis state $\rho'$ such that
    \begin{equation}
        F(\rho', \rho) \geq \max_{\sigma \in \mathcal{D}} F(\sigma, \rho) - \epsilon
    \end{equation}
    with probability at least $1-\delta$, where $F$ denotes the fidelity. When the hypothesis $\rho'$ is from the concept class $\mathcal{D}$, we say that $\mathcal{A}$ is \emph{proper}. Otherwise, we say that it is \emph{improper}.
\end{definition}

Algorithms for agnostic state tomography with respect to fidelity are known for a variety of classes, e.g., stabilizer product states~\cite{grewal2024agnostic}, stabilizer states, states with high stabilizer dimension, and discrete product states~\cite{chen2024stabilizer}.

In the following proposition, we show how to convert any agnostic state tomography algorithm to an improper agnostic unitary estimation algorithm with effectively the same runtime, as long as the concept classes are compatible.
This is satisfied for the corresponding unitary classes of all previously mentioned agnostic state tomography algorithms.
Moreover, we note that since our algorithm is improper, the underlying agnostic state tomography algorithm does not have to be proper.
However, we expect this to be helpful for obtaining proper learners, which we leave as an open question.

\begin{prop}[Improper AUE from agnostic state tomography]
    \label{prop:ast-to-aue}
    Suppose there exists a learning algorithm for (proper) $(1,\epsilon^2, \delta)$-agnostic state tomography with respect to fidelity $F$ for some class $\mathcal{D}$ of pure states.
    Let $\mathcal{C}$ be a concept class of unitaries.
    If $\ket{v(U)} \in \mathcal{D}$ for all $U \in \mathcal{C}$, then there exists a learning algorithm for improper $(1,\epsilon,\delta)$-agnostic unitary estimation with respect to error in normalized Frobenius distance $d_F$ for the class $\mathcal{C}$.
    Moreover, this requires the same sample and time complexity as the $(1,\epsilon^2, \delta)$-agnostic state tomography algorithm.
\end{prop}

We note that our result holds for agnostically learning the unitary channel associated to the unknown unitary.
Namely, if $U$ is our unknown unitary, then our algorithm agnostically learns the channel $U(\cdot ) U^\dagger$ with respect to $d_F$ (the normalized Frobenius distance for superoperators from \Cref{def:df-superop}).
We note that this does not take into account any phase differences and so is slightly weaker than learning the unitary $U$ itself with respect to $d_F$ (the normalized Frobenius distance for operators from \Cref{def:dist-frob}).
This is because our algorithm simply applies agnostic state tomography to the Choi state of the unknown process, inspired by~\cite{leung2000towards}, which does not contain this phase information.

Moreover, our analysis can be seen as extending that of~\cite{leung2000towards} to the agnostic setting with respect to the normalized Frobenius distance $d_F$, rather than the (trivial) metric of entanglement infidelity defined as the infidelity between two Choi states:
\begin{equation}
    \label{eq:entangle-infidel}
    \overline{F}(\Phi, \Psi) \triangleq 1 - F(v(\Phi), v(\Psi)),
\end{equation}
where we adopt the (non-standard) notation of~\cite{haah2023query} and $v(\Phi)$ denotes the Choi state of $\Phi$.
We remark that we can also trivially obtain $(1,\epsilon,\delta)$-agnostic unitary estimation with respect to the entanglement infidelity from $(1,\epsilon,\delta)$-agnostic state tomography with respect to fidelity (notably, without the quadratic blowup with respect to $\epsilon$) under the same compatibility condition.
Entanglement infidelity has been considered widely in (realizable) unitary estimation~\cite{leung2000towards,haah2023query,acin2001optimal,peres2002covariant,hayashi2006parallel,chiribella2005optimal,kahn2007fast,yang2020optimal}.
However, we focus on the normalized Frobenius distance $d_F$ for consistency with our other results.

\begin{proof}
    Let $U$ denote the unknown unitary.
    The algorithm is simple: first, prepare the Choi state $\ket{v(U)}$ of $U$ and then apply the $(1,\epsilon^2,\delta)$-agnostic state tomography algorithm to $\ket{v(U)}$.
    The output of our algorithm is the density matrix of the pure state output by the state tomography algorithm.

    The agnostic state tomography algorithm outputs some state $\ket{\psi}$ such that
    \begin{equation}
        F(v(U), \ketbra{\psi}) \geq \max_{\ket{\phi} \in \mathcal{D}} F(v(U), \ketbra{\phi}) - \epsilon^2,
    \end{equation}
    where we use $v(U) \triangleq \ketbra{v(U)}$ to denote the density matrix corresponding to the pure Choi state.
    Note that this is consistent with the Choi state defined by the channel $\mathcal{E}(\rho) = U\rho U^\dagger$.
    We can equivalently write the above guarantee in terms of the infidelity
    \begin{equation}
        1 - \left| \braket{v(U)}{\psi} \right|^2 \leq \min_{\ket{\phi} \in \mathcal{D}} \left(1 - \left|\braket{v(U)}{\phi}\right|^2\right) + \epsilon^2.
    \end{equation}
    We can convert this into a bound with respect to the normalized Frobenius distance $d_F$.
    Note that
    \begin{equation}
        \norm{\ketbra{\psi} - \ketbra{\phi}}_F^2 = 2\left(1 - \left|\braket{\psi}{\phi}\right|^2\right)
    \end{equation}
    for any pure states $\ket{\psi},\ket{\phi}$.
    Moreover,
    \begin{equation}
        \frac{1}{2}\norm{\ketbra{\psi} - \ketbra{\phi}}_F^2 = 4^n d_F(\ketbra{\psi}, \ketbra{\phi})^2
    \end{equation}
    by definition of $d_F$ (\Cref{def:dist-frob}), where $\ket{\phi}, \ket{\psi}$ are each on $2n$ qubits.
    Then, the above guarantee can be written as
    \begin{equation}
        4^n d_F(v(U), \ketbra{\psi})^2 \leq 4^n \min_{\ket{\phi} \in \mathcal{D}} d_F(v(U), \ketbra{\phi})^2 + \epsilon^2.
    \end{equation}
    Taking the square root of both sides, we obtain
    \begin{equation}
        2^nd_F(v(U), \ketbra{\psi}) \leq 2^n\min_{\ket{\phi} \in \mathcal{D}} d_F(v(U), \ketbra{\phi}) + \epsilon.
    \end{equation}
    Moreover, because $\ket{v(V)} \in \mathcal{D}$ for all $V \in \mathcal{C}$, we can upper bound this minimum by the minimum over all $V$ in $\mathcal{C}$ instead.
    Thus, we have
    \begin{equation}
        \label{eq:ast-to-aue}
        2^n d_F(v(U), \ketbra{\psi}) \leq 2^n\min_{V \in \mathcal{C}} d_F(v(U), v(V)) + \epsilon.
    \end{equation}
    
    Now, we want to argue that agnostically learning the Choi state $v(U)$ of a unitary $U$ with respect to $2^n d_F$ is equivalent to agnostically learning the unitary channel $U(\cdot) U^\dagger$ with respect to $d_F$.
    Note that the first $d_F$ is the normalized Frobenius norm for operators (\Cref{def:dist-frob}) while the second $d_F$ is the normalized Frobenius norm for superoperators (\Cref{def:df-superop}).
    Then, notice
    \begin{align}
        d_F(\mathcal{U}_1, \mathcal{U}_2) &= \frac{1}{2^n\sqrt{2}}\norm{\mathcal{U}_1 - \mathcal{U}_2}_F\\
        &= \frac{1}{2^n\sqrt{2}}\norm{J(U_1) - J(U_2)}_F\\
        &= \frac{1}{\sqrt{2}}\norm{v(U_1) - v(U_2)}_F\\
        &= 2^n d_F(v(U_1), v(U_2)).
    \end{align}
    Here, we use $\mathcal{U}$ to denote the unitary channel $U (\cdot) U^\dagger$.
    In the first and second lines, we use the definition of $d_F$ for superoperators (\Cref{def:df-superop}).
    In the third line, we use that the Choi representation is simply the unnormalized Choi state.
    Here, recall that $v(U) = \ketbra{v(U)}$, so the normalization factor is $2^n$ for this density matrix.
    Finally, in the last line, we use the definition of the normalized Frobenius distance for operators (\Cref{def:dist-frob}).
    Thus, this tells us that, indeed, agnostically learning the Choi state of a unitary with respect to $2^n d_F$ is the same as learning the corresponding unitary channel with respect to $d_F$.
    Hence, \Cref{eq:ast-to-aue} is exactly the guarantee needed for improper $(1,\epsilon,\delta)$-AUE with respect to the class $\mathcal{C}$.
\end{proof}

This gives a fairly general result for converting an agnostic state tomography algorithm with respect to $\mathcal{D}$ into an improper agnostic unitary estimation algorithm with respect to $\mathcal{C}$ when $\ket{v(U)} \in \mathcal{D}$ for all $U \in \mathcal{C}$.
As remarked earlier, this holds for many classes of unitaries for which there already exist agnostic state tomography algorithms with respect to a suitable class.
In particular, there exist agnostic state tomography algorithms for stabilizer states, states with high stabilizer dimension, and discrete product states~\cite{grewal2024agnostic,chen2024stabilizer}.
Note that the corresponding classes of unitaries, i.e., Clifford circuits~\cite{gottesman1997stabilizer}, Clifford + T circuits, and unitaries consisting of a tensor product of single-qubit gates, respectively, all satisfy the required property of Proposition~\ref{prop:ast-to-aue}.
Thus, we can directly apply Proposition~\ref{prop:ast-to-aue} with the state-of-the-art sample/time complexity bounds for agnostic state tomography to obtain improper AUE algorithms for the respective unitary classes.
These are each direct corollaries of Corollary 6.3, Theorem 7.1, Corollary 8.3, and Corollary 9.3 in~\cite{chen2024stabilizer}.
The results in~\cite{chen2024stabilizer} assume that $\opt \geq \tau$ for some $\tau \geq \epsilon$, and their sample complexities depend on this parameter $\tau$.
When applying their results, we use that $\tau \geq \epsilon$ to eliminate the dependence on $\tau$.

\begin{corollary}[Improper AUE of Clifford Circuits; via Corollary 6.3 in~\cite{chen2024stabilizer}]
    There exists a learning algorithm for improper $(1,\epsilon,\delta)$-agnostic unitary estimation of the class of $n$-qubit Clifford unitaries with respect to error in $d_F$ using
    \begin{equation}
        N = 2n\log(1/\delta)\left(\frac{1}{\epsilon}\right)^{\mathcal{O}(\log 1/\epsilon)} + \mathcal{O}\left(\frac{\log(1/\delta) + \log^2(1/\epsilon)}{\epsilon^4}\right)
    \end{equation}
    queries to the unknown unitary and running in time
    \begin{equation}
        \mathcal{O}\left(n^2\log(1/\delta)\left(n + \frac{\log(1/\delta)}{\epsilon^4}\right)\right) \cdot \left(\frac{1}{\epsilon}\right)^{\mathcal{O}(\log 1/\epsilon)}.
    \end{equation}
\end{corollary}

\begin{corollary}[Improper AUE of Clifford + T Circuits; via Theorem 7.1 in~\cite{chen2024stabilizer}]
    There exists a learning algorithm for improper $(1,\epsilon,\delta)$-agnostic unitary estimation of the class of $n$-qubit unitaries consisting of Clifford gates and $t$ T gates with respect to error in $d_F$ using
    \begin{equation}
        N = 2n\log(1/\delta)\left(\frac{2^t}{\epsilon^2}\right)^{\mathcal{O}(\log 1/\epsilon)}
    \end{equation}
    queries to the unknown unitary and running in time
    \begin{equation}
        \mathcal{O}(n^2 \log(1/\delta) N).
    \end{equation}
\end{corollary}

\begin{corollary}[Improper AUE of Product Circuits; via Corollary 8.3 in~\cite{chen2024stabilizer}]
    Let $\mathcal{K}$ be a set of single-qubit gates for which $|\braket{v(U_1)}{v(U_2)}|^2 \leq 1 - \mu$ for any distinct $U_1, U_2 \in \mathcal{K}$.
    There exists a learning algorithm for improper $(1,\epsilon,\delta)$-agnostic unitary estimation of the class $\mathcal{C} = \mathcal{K}^{\otimes n}$ with respect to error in $d_F$ using
    \begin{equation}
        N = \frac{\log^2(1/\delta)(2n|\mathcal{K}|)^{\mathcal{O}\left(\log(1/\epsilon)/\mu\right)}}{\epsilon^4}
    \end{equation}
    queries to the unknown unitary and running in time $\mathcal{O}(N)$.
\end{corollary}

\begin{corollary}[Improper AUE of Clifford Product Circuits; via Corollary 9.3 in~\cite{chen2024stabilizer}]
    There exists a learning algorithm for improper $(1,\epsilon,\delta)$-agnostic unitary estimation of the class of tensor products of single-qubit Clifford gates with respect to error in $d_F$ using
    \begin{equation}
        N = \log (2n) \log(1/\delta) \left(\frac{1}{\epsilon}\right)^{\mathcal{O}(\log 1/\epsilon)} + \mathcal{O}\left(\frac{\log^2(1/\epsilon) + \log(1/\delta)}{\epsilon^4}\right)
    \end{equation}
    and running in time
    \begin{equation}
        \frac{4n^2 \log^2(1/\delta) (1/\epsilon)^{\mathcal{O}(\log 1/\epsilon)}}{\epsilon^4}.
    \end{equation}
\end{corollary}

Note that all of these algorithms are improper, while for applications, one would prefer proper learning algorithms.
We leave it to future work to make these algorithms proper.
For example, for proper AUE of Clifford circuits, one approach would be to efficiently find the closest stabilizer Choi state to a given stabilizer state.
\subsection{Pauli strings}
\label{sec:aue-pauli}
In this section, we consider agnostic learning with respect to the class of $n$-qubit Pauli strings $\mathcal{C}_{\pauli_n} \triangleq \{I, X, Y, Z\}^{\otimes n}$ and the distance metric $\davg$.
Here, we achieve a proper $1$-agnostic unitary estimation algorithm, detailed in \Cref{alg:agnostic-unitary-paulis}, with an efficient sample complexity and runtime.
In fact, the sample complexity and runtime for this class are both independent of the system size $n$.
Moreover, our algorithm only requires access to copies of the Choi state of the unknown unitary $U$.

\begin{algorithm}
   \caption{1-Agnostic Learning Pauli Strings} 
   \label{alg:agnostic-unitary-paulis}
   \begin{algorithmic}[1]
   \State Prepare $N$ copies of $|v(U)\rangle$ and measure them in the $\{|v(\sigma_x)\rangle\}_x$ basis to get strings $\{x_i\}_{i = 1}^{N}$
    \State \Return the most frequent string
   \end{algorithmic}
\end{algorithm}

\begin{theorem}[Agnostic Unitary Estimation of Pauli strings in $\davg$]\label{thm:aue-pauli-davg} 
There exists a learning algorithm for proper $(1,\epsilon,\delta)$-agnostic unitary estimation of the class of $n$-qubit Pauli strings $\mathcal{C}_{\pauli_n}$ with respect to error in $\davg$ using 
\begin{equation}
    N = 
    \mathcal{O}\left(
        \frac{ \log(1/\delta)}{\epsilon^4}
    \right)
\end{equation}
queries to the unknown unitary and running in time
\begin{equation}
    \mathcal{O}\left(
        N
    \right).
\end{equation}
\end{theorem}
\begin{proof}
    The algorithm is detailed in \Cref{alg:agnostic-unitary-paulis}.
    First, we prepare $N$ copies of the Choi state $|v(U)\rangle$ (\Cref{def:choi-state}).
    Then, we measure each one in the $\{|v(\sigma_x)\rangle\}_{x\in \pauli_n}$ basis and output $\sigma_z$ associated with the most frequent outcome $z$.
    Using the fact that 
    \begin{equation}
        |v(U)\rangle = \sum_{x \in \pauli_n} \hat{U}_x |v(\sigma_x)\rangle,
    \end{equation}
    each measurement returns a string $x$ with probability 
    \begin{equation}
        p(x) \triangleq |\hat{U}_x|^2.
    \end{equation}
    Denote the true most likely Pauli string by
    \begin{equation}
        x^\star \triangleq \arg \max_{x \in \pauli_n} p(x).
    \end{equation}
    We compute the optimal error for the class of Pauli strings. Recall from \Cref{lem:davg-pauli} that
    \begin{equation}
        \davg(U,V)^2 = \frac{2^n}{2^n+1}\left(1 - \left|\sum_{x \in \pauli_n} \hat{U}^*_x \hat{V}_x\right|^2\right).
    \end{equation}
    Then, we have
    \begin{align}
            \opt(U, \mathcal{C}_{\pauli_n})^2
            &= \min_{x \in \pauli_n} \davg(U, \sigma_x) ^2
            \\&= \min_{x \in \pauli_n} \frac{2^n}{2^n+1}(1 - |\hat{U}_x|^2) 
            \\&= \frac{2^n}{2^n+1} \left(1 - \max_{x \in \pauli_n} |\hat{U}_x|^2\right).
            \label{eq:opt-pauli-davg}
    \end{align}
    Hence, we see that the hypothesis $\sigma_{x^\star}$ achieves the optimal error for the class $\mathcal{C}_{\mathcal{P}_n}$.
    Our algorithm aims to estimate the true most likely Pauli string $x^\star$ by the one most frequently occurring in our measurements.
    Let $\{\hat{p}(x)\}_x$ be the empirical distribution obtained from our measurement outcomes, which estimates $p$.
    Then, denote the most frequent outcome out of our measurements by
    \begin{equation}
        z  \triangleq \argmax_{x \in \{x_i\}_{i \in [N]}} \hat{p}(x).
    \end{equation}
     To bound the error, consider the empirical distribution $\{\hat{p}(x)\}_x$. Using~\Cref{lem:dist-learning}, we can construct estimates $\hat{p}(x)$ of $|\hat{U}_x|^2$ such that with probability at least $1-\delta$,
    \begin{equation}
        \left|\hat{p}(x) - |\hat{U}_x|^2\right| \leq \epsilon_1, \quad \forall x \in \pauli_n
    \end{equation}
    for some $\epsilon_1 > 0$ using $\mathcal{O}\left(\frac{ \log(1/\delta)}{\epsilon_1^2}\right)$ copies of the Choi state $|v(U)\rangle$. Now, since $z$ is the most frequent string observed,
    \begin{equation}
        \hat{p}(z) \geq \hat{p}(x^\star).
    \end{equation}
    Conditioned on obtaining estimates $\hat{p}(x)$ that are $\epsilon_1$-close to $|\hat{U}_x|^2$ for all $x \in \pauli_n$, we have
    \begin{equation}
        |\hat{U}_z|^2 + \epsilon_1 \geq \hat{p}(z)\quad \text{ and }\quad \hat{p}(x^\star) \geq |\hat{U}_{x^\star}|^2 - \epsilon_1.
    \end{equation}
    Thus, we have
    \begin{equation}
        \label{eq:1}
        |\hat{U}_{x^\star}|^2 - |\hat{U}_z|^2 \leq 2\epsilon_1.
    \end{equation}
    Now, we can bound the error of the hypothesis as follows
    \begin{align}
        \davg(U, \sigma_z)^2 &= \frac{2^n}{2^n+1}(1 - |\hat{U}_z|^2)
        \\&= \frac{2^n}{2^n+1}(1 - |\hat{U}_{x^\star}|^2 + |\hat{U}_{x^\star}|^2 - |\hat{U}_z|^2) 
        \\& \leq \opt(U, \mathcal{C}_{\pauli_n})^2 + \frac{2^n}{2^n+1}\cdot 2\epsilon_1,
    \end{align}
    where the first line is true by \Cref{lem:davg-pauli}, and the last line follows by \Cref{eq:opt-pauli-davg} and \Cref{eq:1}.
    Thus, by setting $\epsilon_1 = \epsilon^2/2$, we obtain
    \begin{equation}
        \davg(U, \sigma_z)^2 \leq \opt(U, \mathcal{C}_{\pauli_n})^2 + \epsilon^2 \leq (\opt(U, \mathcal{C}_{\pauli_n}) + \epsilon)^2,
    \end{equation}
    and the desired bound can be obtained by taking the square root.
    Now, by our choice of $\epsilon_1 = \epsilon^2/2$, we use
    \begin{equation}
        N = \mathcal{O}\left(\frac{\log (1/\delta)}{\epsilon_1^2}\right) = \mathcal{O}\left(\frac{ \log(1/\delta)}{\epsilon^4}\right)
    \end{equation}
    samples of the Choi state or queries to the unitary $U$. The time complexity is that of identifying the most frequent string out of $N$ strings, which can be done in time $\mathcal{O}(N)$.
\end{proof}
\subsection{Unitary $k$-juntas}
\label{sec:aue-junta}
Now, we consider agnostic learning with respect to the class of $n$-qubit unitary $k$-juntas $\mathcal{J}_{k,n}$~\cite{montanaro2008quantum}.
An $n$-qubit $k$-junta is a $2^n$-dimensional unitary that acts nontrivially on $k$ qubits.
More formally, we have the following definition.

\begin{definition}[$k$-junta]
    \label{def:unitary-junta}
    A unitary $U$ on $n$ qubits is a \emph{quantum $k$-junta} if there exists a set $S\subseteq [n]$ with $|S| = k$ such that $U = V_S \otimes I_{S^c}$, for some $k$-qubit unitary $V_S$.
\end{definition}

We consider learning with respect to $d_F$, and obtain an efficient proper 2-agnostic unitary estimation algorithm, detailed in \Cref{alg:agnostic-unitary-juntas}.

\begin{algorithm}
   \caption{2-Agnostic Proper Learning Unitary k-Juntas} 
   \label{alg:agnostic-unitary-juntas}
   \begin{algorithmic}[1]
   \State Prepare $N_1 = \mathcal{O}\left(\frac{k \log(n/\delta)}{\epsilon^4}\right)$ copies of $|v(U)\rangle$ and use them to compute estimates of weights of all subsets of $[n]$ with cardinality $k$.
    \State Select $S$ to be the subset with the greatest estimated weight.
    \State Using $N_2 = \mathcal{O}\left(\frac{k 16^k\log(1/\delta)}{\epsilon^2}\right)$ queries to $U$, construct estimates $\alpha_x$ of the Pauli coefficients for $x \in \pauli_S$.
    \State Compute $V_1 = \sum_{x \in \pauli_S} \alpha_x \sigma_x$.
    \State Project $V_1$ onto a unitary, $V_2 = \proju(V_1)$.
    \State \Return $V_2$.
   \end{algorithmic}
\end{algorithm}

The algorithm is reminiscent of other junta learning algorithms~\cite{chen2023testing,atici2007quantum} in structure.
Namely, we first identify the qubits $S$ that the $k$-junta acts nontrivially on and then learn the unitary on the set $S$.
In contrast to~\cite{chen2023testing}, we use the Pauli weight (\Cref{def:pauli-weight}) rather than the influence to measure how nontrivially a unitary acts on a given qubit. In the realizable setting, both the weight and influence of a subset are equivalent. However, in the agnostic setting, these notions are quite distinct, as demonstrated by the following example.
\begin{example}
    \label{ex:influence-weight}
    Consider the unitary $U = \frac{1}{\sqrt{3}}X \otimes Z \otimes I + \sqrt{\frac{2}{3}} Z \otimes Z \otimes Z$. One can show that the subset of size 2 with the most weight is formed by the first 2 qubits, $\{1,2\}$, and has weight $w_{\{1,2\}}(U) = 1/3$. However, the most influential subset of size 2 is formed by the last two qubits, \{2,3\} with influence 1. Note that the corresponding weight $w_{\{2,3\}}(U) = 0$. 
\end{example}
As we will show in our proof of~\Cref{thm:aue-juntas-df}, the optimal error depends on the subset with the highest weight, not the most influential subset. So, our algorithm starts by identifying this subset instead of the most influential one.

Moreover, rather than using a tomography algorithm in the second part, which would make proper learning more difficult, we directly estimate the Pauli coefficients corresponding to $\pauli_S$, where recall we use $\pauli_S$ to denote
\begin{equation}
    \pauli_S = \{x \in \{0,1,2,3\}^n : \mathsf{supp}(x) \subseteq S \}
\end{equation}
for a subset $S \subseteq [n]$.
The resulting learned operator $V_1 = \sum_{x \in \mathcal{P}_S}  \alpha_x \sigma_x$ may not be unitary, so in order to achieve \emph{proper} agnostic unitary estimation, we need an additional projection which approximates $V_1$ by a unitary.
For this projection, we use \Cref{lem:unitary-projection}, which takes influence from~\cite{huang2024learning}.
This algorithm achieves the following guarantee.

\begin{theorem}[Agnostic Unitary Estimation of k-juntas]\label{thm:aue-juntas-df} \Cref{alg:agnostic-unitary-juntas} performs proper $(2,\epsilon,\delta)$-agnostic unitary estimation of the class of $n$-qubit unitary $k$-juntas $\mathcal{J}_{k,n}$ with respect to error in $d_F$ using 
\begin{equation}
    N = \mathcal{O}\left(
    \frac{k \log (n/\delta)}{\epsilon^4} + \frac{k16^k \log(1/\delta)}{\epsilon^2}
    \right)
\end{equation}
queries to the unknown unitary $U$ and running in time
\begin{equation}
    \mathcal{O}\left(
    \frac{n^k k \log(n/\delta)}{\epsilon^4}\right).
\end{equation}
\end{theorem}

As part of the proof of Theorem~\ref{thm:aue-juntas-df},
we first show how to estimate the Pauli weights of all subsets of cardinality $k$, i.e., $S \subseteq [n]$ with $|S| = k$.

\begin{lemma}[Estimating weights of a unitary on many subsets]
\label{lem:estimate-weight-unitary}
Given query access to a unitary $U$ and $M$ subsets $\{S_j\}_{j \in [M]}, S_j \subseteq [n]$, there exists a method to produce estimates $\hat{w}_{S_j}$ of the Pauli weights $w_{S_j}(U)$ using $N = \mathcal{O}(\log(M/\delta)/\epsilon^2)$ queries to $U$ such that
\begin{equation}
    |\hat{w}_{S_j}-w_{S_j}(U)| \leq \epsilon \quad \forall j \in [M]
\end{equation}
with probability at least $1-\delta$. Moreover, the algorithm runs in computational time $\mathcal{O}(NM)$.
\end{lemma}

\begin{proof}
    Using $N$ queries to the unitary $U$, the algorithm first prepares $N$ copies of the Choi state $|v(U)\rangle$ and measures each copy in the basis $\{|v(\sigma_x)\rangle\}_x$.
    Suppose the outcomes are $\{x_i\}_{i \in [N]}$.
    Then, for each subset $S$, one can construct unbiased estimators
    \begin{equation}
        \hat{w}_S \triangleq \frac{1}{N} \sum_{i \in [N]} b_{S,i}, \quad b_{S,i} \triangleq \mathbf{1} [x_i \in \pauli_S].
    \end{equation}
    Here, $b_{S,i}$ indicates whether $x_i$ is supported entirely in $S$, i.e., $\mathsf{supp}(x_i) \subseteq S$.
    We claim that $\hat{w}_S$ has the desired properties given our choice of $N$.
    Clearly, since the measurements are independent, then $b_{S_i}$ are i.i.d. random variables so that
    \begin{align}
            \mathbf{E}[\hat{w}_S] &= \mathbf{E}[b_{S,i}]
            \\&= \mathbf{Pr}(x_i \in \pauli_S)
            \\&= \sum_{x \in \pauli_S} |\hat{U}_x|^2
            \\&= w_S(U).
    \end{align}
    Here, the third line follows because measuring $\ket{v(U)}$ in the basis $\{\ket{v(\sigma_x)}\}_{x \in \pauli_n}$ gives outcome $x$ with probability $|\hat{U}_x|^2$.
    Thus, using 
     \begin{equation}
        N = \mathcal{O}\left(\frac{\log (M/\delta)}{\epsilon^2}\right)
    \end{equation}
    copies of the Choi state, we can see via Hoeffding's inequality and a union bound that
    \begin{equation}
        |\hat{w}_{S_j} - w_{S_j}(U)| \leq \epsilon \quad \forall j \in [M] 
    \end{equation}
    with probability at least $1-\delta$.
    For each subset, we use $\mathcal{O}(N)$ computational time to compute $\hat{w}_{S_j}$, giving an overall computational time of $\mathcal{O}(NM)$.
\end{proof}
With this lemma, we can now prove~\Cref{thm:aue-juntas-df}.

\begin{proof}[Proof of Theorem~\ref{thm:aue-juntas-df}]
    First, we start by lower-bounding $\opt(U, \mathcal{J}_{n,k})$.
    \begin{align}
            \opt(U, \mathcal{J}_{n,k})^2 &= \min_{V \in \mathcal{J}_{n,k}} d_F(U,V)^2
            \\&= \min_{V \in \mathcal{J}_{n,k}} \frac{1}{2}\sum_{x \in \pauli_n} |\hat{U}_x- \hat{V}_x|^2
            \\&= 1 - \max_{V \in \mathcal{J}_{n,k}} \sum_{x \in \pauli_n} \mathrm{Re}(\hat{U}_x^*\hat{V}_x)
            \\&\geq 1 - \max_{V \in \mathcal{J}_{n,k}} \sum_{x \in \pauli_n} |\hat{U}_x||\hat{V}_x|
            \\&\geq 1 -\max_{S \subseteq [n], |S| = k} \max_{\substack{V \in \mathcal{J}_{n,k}\\ w_S(V) = 1}} \sum_{x \in \pauli_S} |\hat{U}_x||\hat{V}_x|
            \\& \geq 1 - \max_{S \subseteq [n], |S| = k} \max_{\substack{V \in \mathcal{J}_{n,k}\\ w_S(V) = 1}} \sqrt{w_S(U)} \sqrt{w_S(V)}
            \\& = 1 - \max_{S \subseteq [n], |S| = k} \sqrt{w_S(U)},
    \end{align}
    where the second line follows from Lemma~\ref{lem:df-pauli} and the second to last line follows from Cauchy-Schwarz. 
    
    Denote the heaviest cardinality-$k$ set by
    \begin{equation}
        S^\star \triangleq \argmax_{S \subseteq [n], |S| = k} w_S(U).
    \end{equation}
    Thus, we have
    \begin{equation}
        \label{eq:opt-junta}
        \opt(U, \mathcal{J}_{n,k})^2 \geq 1 - \sqrt{w_{S^\star}(U)}.
    \end{equation}
    Hence, in order to learn the optimal $k$-junta, we first identify the subset of cardinality $k$ with the highest Pauli weight.
    To do this, we estimate the weight of all $\binom{n}{k} = \mathcal{O}(n^k)$ subsets of $[n]$ of cardinality $k$ up to accuracy $\epsilon_1$ with failure probability at most $\delta/2$.
    In particular, we find estimates $\{\hat{w}_S\}_S$ such that
    \begin{equation}
        \label{eq:estimate-weight}
        |\hat{w}_S - w_S(U)| \leq \epsilon_1,
    \end{equation}
    with probability at least $1-\delta/2$, for all subsets $S \subseteq [n]$ of cardinality $|S| = k$.
    By \Cref{lem:estimate-weight-unitary}, this requires
    \begin{equation}
        N_1 = \mathcal{O}\left(\frac{k \log(n/\delta)}{\epsilon_1^2}\right)
    \end{equation}
    queries to $U$ and computational time $\mathcal{O}\left(\frac{n^kk \log(n/\delta)}{\epsilon_1^2}\right)$.
    We now focus on the subset $S^\prime$ with the heaviest estimated weight, i.e., the largest value of $\hat{w}_{S}$.
    \begin{equation}
        S' \triangleq \argmax_{S\subseteq [n], |S| = k} \hat{w}_S.
    \end{equation}
    In the next step of \Cref{alg:agnostic-unitary-juntas}, we estimate all $4^k$ Pauli coefficients for all Pauli strings supported on $S^\prime$.
    We use Lemma~\ref{lem:estimate-pauli-coeff-unitary} $4^k$ times, to obtain estimates $\alpha_x$ such that with probability at least $1-\delta/2$, 
    \begin{equation}
        \label{eq:estimate-pauli-coeff}
        |\alpha_x-\hat{U}_x| \leq \epsilon_2 ,\quad \forall x \in \pauli_{S^\prime}
    \end{equation}
    Using a union bound over failure probabilities of repeated applications of Lemma~\ref{lem:estimate-pauli-coeff-unitary} for each Pauli coefficient, we see that this requires
    \begin{equation}
        N_2 = \mathcal{O}\left(\frac{4^k \log(4^k/\delta)}{\epsilon_2^2} \right)
    \end{equation}
    queries to $U$ and runs in time $\mathcal{O}\left(\frac{4^k \log(4^k/\delta)}{\epsilon_2^2} \right)$.
    Now, we construct the hypothesis
    \begin{equation}
        V_1 \triangleq \sum_{x \in \pauli_{S^\prime}} \alpha_x \sigma_x.
    \end{equation}
    This simply aims to approximate $U$ by approximating the Pauli coefficients of $U$ on the subset $S'$ with the heaviest estimated weight.
    As this hypothesis is not unitary in general, we apply Lemma~\ref{lem:unitary-projection} to project $V_1$ onto a unitary and obtain the final hypothesis
    \begin{equation}
        V_2 \triangleq \proju(V_1).
    \end{equation}
    As $V_1$ acts trivially on all qubits outside of $S^\prime$, the projection only needs to be applied to an operator of dimension $2^k \times 2^k$, needing time $\mathcal{O}(8^k)$.
    It remains to bound the error of this hypothesis, i.e., $d_F(U,V_2)$.
    Denote
    \begin{equation}
        \Tilde{U}_{S^\prime} \triangleq \sum_{x \in \pauli_{S^\prime}} \hat{U}_x \sigma_x,
    \end{equation}
    which is the operator obtained by truncating the Pauli coefficients of $U$ not supported on $S$. Now, we have
    \begin{align}
        d_F(U, V_2) & \leq d_F(U,V_1)+d_F(V_1,V_2)
        \\& \leq 2 d_F(U,V_1)
        \\& \leq 2d_F(U,\Tilde{U}_{S^\prime} ) + 2d_F(\Tilde{U}_{S^\prime}, V_1),
    \end{align}
    where the first and last lines follow from the triangle inequality for $d_F$, and the second line follows from Lemma~\ref{lem:unitary-projection}.
    
    For our final bound, we use the following claims on these distances that will be proven later.
    The claims are conditioned on the estimations for the weights and the Pauli coefficients succeeding. A union bound tells us that the total failure probability is at most $\delta/2+\delta/2 = \delta$.
    \begin{claim}
        \label{clm:dist-unitary-subset}
        \begin{equation}
            d_F(U, \Tilde{U}_{S^\prime}) \leq \opt(U, \mathcal{J}_{n,k})+\sqrt{\epsilon_1}.
        \end{equation}
    \end{claim}
    \begin{claim}
        \label{clm:dist-unitary-coefficients}
        \begin{equation}
            d_F(\Tilde{U}_{S^\prime}, V_1) \leq \frac{2^k\epsilon_2}{\sqrt{2}}.
        \end{equation}
    \end{claim}
    Using these claims in the above inequality, we have
    \begin{equation}
        d_F(U,V_2) \leq 2\cdot \opt(U, \mathcal{J}_{n,k})+2\sqrt{\epsilon_1}+2^k\sqrt{2}\epsilon_2
    \end{equation}
    Then, by setting $\epsilon_1 = \epsilon^2/16 , \epsilon_2 = \epsilon/(2\sqrt{2} \cdot2^k)$, we obtain
    \begin{equation}
         d_F(U,V_2) \leq 2\cdot \opt(U, \mathcal{J}_{n,k})+\epsilon,
    \end{equation}
    which is the desired result.
    The stated sample complexity can be obtained by plugging in the values of $\epsilon_1,\epsilon_2$ to the previous expressions for estimating weights and Pauli coefficients:
    \begin{equation}
        N = N_1 + N_2 = \mathcal{O}\left(\frac{k\log(n/\delta)}{\epsilon_1^2} + \frac{4^k\log(4^k/\delta)}{\epsilon_2^2}\right) = \mathcal{O}\left(\frac{k\log(n/\delta)}{\epsilon^4} + \frac{k16^k\log(1/\delta)}{\epsilon^2}\right).
    \end{equation}
    The runtime is dominated by the procedures for finding the heaviest subset and for estimating the $4^k$ Pauli coefficients:
    \begin{equation}
        \mathcal{O}\left(\frac{n^k k\log(n/\delta)}{\epsilon_1^2} + \frac{4^k\log(4^k/\delta)}{\epsilon_2^2}\right) = \mathcal{O}\left(\frac{n^k k\log(n/\delta)}{\epsilon^4} + \frac{k16^k\log(1/\delta)}{\epsilon^2}\right).
    \end{equation}
    As the first time dominates the time complexity, we obtain the stated complexity. It remains to prove the two claims above.
    \end{proof}
    \begin{proof}[Proof of Claim~\ref{clm:dist-unitary-subset}]
        Because the algorithm selects $S^\prime$, it is the subset with the heaviest estimated weight, i.e.,
        \begin{equation}
            \hat{w}_{S^\prime} \geq \hat{w}_{S^\star}.
        \end{equation}
        Further, by \Cref{eq:estimate-weight},
        \begin{equation}
            w_{S^\prime}(U)+\epsilon_1 \geq \hat{w}_{S^\prime}\quad \text{ and } \quad \hat{w}_{S^\star} \geq w_{S^\star}(U)-\epsilon_1.
        \end{equation}
        Thus, 
        \begin{equation}
            \label{eq:3}
            w_{S^\star}(U)-w_{S^\prime}(U) \leq 2\epsilon_1
        \end{equation}
        and $w_{S^\star} \geq w_{S^\prime}$.
        For ease of notation, denote $W \triangleq \Tilde{U}_{S^\prime}$.
        Then, we have
        \begin{align}
                d_F(U, \Tilde{U}_{S^\prime})^2 &= \frac{1}{2}\sum_{x \in \pauli_n}|\hat{U}_x - \hat{W}_x|^2 \\&= \frac{1}{2} \left(\sum_{x \in \pauli_n \backslash \pauli_{S^\prime}}|\hat{U}_x - 0|^2 + \sum_{x \in \pauli_{S^\prime}}|\hat{U}_x - \hat{U}_x| \right) \\&= \frac{1-w_{S^\prime}(U)}{2}
                \\&= \frac{1-w_{S^*}(U)}{2} + \frac{w_{S^*}(U)-w_{S^\prime}(U)}{2}
                \\&\leq \frac{(1-\sqrt{w_{S^*}(U)})(1+\sqrt{w_{S^*}(U)})}{2} + \epsilon_1
                \\& \leq 1-\sqrt{w_{S^*}(U)} + \epsilon_1
                \\&\leq \opt(U, \mathcal{J}_{n,k})^2 + \epsilon_1
                \\&\leq \left(\opt(U, \mathcal{J}_{n,k}) + \sqrt{\epsilon_1}\right)^2.
        \end{align}
        In the first line, we use \Cref{lem:df-pauli}.
        In the second line, we use that $W$ is only supported on $S'$, for which its Pauli coefficients are $\hat{U}_x$.
        In the third line, we use the definition of the Pauli weight (\Cref{def:pauli-weight}).
        In the fifth line, we use \Cref{eq:3}.
        In the seventh line, we use \Cref{eq:opt-junta}.
    \end{proof}
    \begin{proof}[Proof of Claim~\ref{clm:dist-unitary-coefficients}]
    As in the proof of Claim~\ref{clm:dist-unitary-subset}, for ease of notation, we denote $W \triangleq \tilde{U}_{S'}$.
        \begin{align}
            d_F(\Tilde{U}_{S^\prime}, V_1)^2 &= \frac{1}{2}\sum_{x \in \pauli_n}|\hat{W}_x - \hat{V}_{1,x}|^2
            \\&= \frac{1}{2}\sum_{x \in \pauli_{S'}}|\hat{U}_x - \hat{\alpha}_x|^2 + \frac{1}{2}\sum_{x \in \pauli_n \backslash \pauli_{S'}}|0-0|^2 
            \\&\leq \frac{4^k \epsilon_2^2}{2}.
        \end{align}
        In the second line, we use that both $W$ and $V_1$ are only supported on $S'$, and the Pauli coefficients of $W$ on $S'$ are $\hat{U}_x$.
        In the last line, we use \Cref{eq:estimate-pauli-coeff} and that $|\pauli_{S'}| = 4^k$ since $|S'| = k$.
    \end{proof}

\begin{remark}[Improper 1-agnostic learner]
    In the setting of improper agnostic learning, we do not need our final hypothesis to be unitary and can simply output $V_1$. This would give an improper $(1,\epsilon,\delta)$-agnostic learning algorithm with the same sample and time complexity.
\end{remark}
\subsection{Low-degree unitaries}
\label{sec:aue-low}
Now, we will present two algorithms for agnostic unitary estimation of the class of low-degree unitaries. A low-degree unitary is one whose high-order Pauli coefficients are 0. More formally, we have the following definition. 
\begin{definition}[Degree-$d$ unitary]
    \label{def:unitary-low-degree}
    A unitary $U$ has \emph{degree} $d$ if 
    \begin{equation}
       w^{> d}(U) = 0,
    \end{equation}
    where $w^{>d}$ is defined in \Cref{def:pauli-weight-level}.
    In other words, the degree of a unitary $U$ is the minimum integer $d$ such that if $|x| > d$, then $\hat{U}_x = 0$.
\end{definition}
We use $\mathcal{L}_{n,d}$ to denote the class of $n$-qubit unitaries of degree $d$. We note that there are ${n \choose d} 3^d \leq n^d3^d$ Pauli strings of degree exactly $d$, and thus $\mc{O}(n^d3^d)$ strings of degree at most $d$.

Our first algorithm given in \Cref{alg:agnostic-unitary-low-1} is inspired by the algorithm of~\cite{nadimpalli2024pauli}.
It involves estimating every low-degree Pauli coefficient, resulting in a query complexity of $\Tilde{\mathcal{O}}(n^{2d})$.
Our second algorithm given in \Cref{alg:agnostic-unitary-low-2} is the low-degree learning algorithm from~\cite{arunachalam2024learning}, which uses the non-commutative Bohnenblust-Hille (BH) inequality of~\cite{volberg2024noncommutative} to obtain a sample complexity independent of $n$. However, in the agnostic setting, \Cref{alg:agnostic-unitary-low-2} does not achieve a sample complexity independent of $n$. From the guarantees we show in \Cref{thm:aue-low-bh}, \Cref{alg:agnostic-unitary-low-2} is only efficient when the optimal error is at most inverse-exponential in the system size. In particular, when the unknown unitary is a member of the concept class, we recover the complexity of the realizable setting. In most other regimes for $\opt$, it is better to use \Cref{alg:agnostic-unitary-low-1}.
We discuss \Cref{alg:agnostic-unitary-low-2} to illustrate that straightforwardly modifying algorithms from the realizable setting does not always give good agnostic learning algorithms.

Note that unlike in the case of unitary juntas, our method for unitary projection is computationally inefficient when acting on arbitrary low-degree operators, as such operators still have an exponential dimension.
As a result, both our algorithms are \emph{improper} learners.
We provide our algorithms and their corresponding guarantees below.
\begin{algorithm}
   \caption{1-agnostic improper learning low-degree unitaries by learning all coefficients} 
   \label{alg:agnostic-unitary-low-1}
   \begin{algorithmic}[1]
   \State Using $N$ queries to $U$, construct estimates $\alpha_x$ of the Pauli coefficients for all $x \in \pauli_n$ s.t. $|x| \leq d$.
    \State \Return $V_1 = \sum_{x \in \pauli_n} \alpha_x \sigma_x$.
   \end{algorithmic}
\end{algorithm}
\begin{theorem}[AUE of low-degree unitaries by learning all low-degree coefficients]
\label{thm:aue-low-1}
    \Cref{alg:agnostic-unitary-low-1} performs improper $(1,\epsilon,\delta)$-agnostic unitary estimation of the class of $n$-qubit unitaries of degree $d$ $\mathcal{L}_{n,d}$ with respect to error in $d_F$ using 
\begin{equation}
    N = \mc{O}\left(\frac{d\cdot 9^d n^{2d} \log(n/\delta)}{\epsilon^2}\right)
\end{equation}
queries to the unknown unitary $U$ and running in time
\begin{equation}
    \mathcal{O}(N).
\end{equation}
\end{theorem}
\begin{algorithm}
   \caption{1-Agnostic Improper Learning Low-Degree Unitaries Using BH Inequality} 
   \label{alg:agnostic-unitary-low-2}
   \begin{algorithmic}[1]
   \State Set $c$ as in~\Cref{eq:low-deg-unitary-c}.
   \State Prepare $N_1 = \mathcal{O}(\log(1/\delta)/c^4)$ copies of $|v(U)\rangle$ and use them to construct estimates $|\hat{U}_x'|^2$ of $|\hat{U}_x|^2$.
   \State Construct $\chi_c = \{x \in \pauli _n : |\hat{U}_x'| \geq c, |x| \leq d \}$.
   \State Using $N_2 = \mathcal{O}\left(\frac{\log(1/c^2\delta)}{c^4\epsilon^2}\right)$ queries to $U$, construct estimates $\alpha_x$ of the Pauli coefficients for all $x \in \chi_c$.
    \State \Return $V_1 = \sum_{x \in \chi_c} \alpha_x \sigma_x$.
   \end{algorithmic}
\end{algorithm}
\begin{theorem}[AUE of low-degree unitaries using BH inequality]
    \label{thm:aue-low-bh}
    \Cref{alg:agnostic-unitary-low-2} performs improper $(1,\epsilon,\delta)$-agnostic unitary estimation of the class of $n$-qubit unitaries of degree $d$ $\mathcal{L}_{n,d}$ with respect to error in $d_F$ using 
\begin{equation}
    N = \exp \left(\Tilde{\mc{O}} \left( d^2 + d\log(1/\epsilon) + \min\left[d^2 \log n, d \log(1 + 2^{(n+1)/2}\cdot \opt) \right] \right) \right) \cdot \log(1/\delta)
\end{equation}
queries to the unknown unitary $U$ and running in time
\begin{equation}
   \mc{O}(N).
\end{equation}
\end{theorem}
In order to prove Theorems~\ref{thm:aue-low-1} and~\ref{thm:aue-low-bh}, we will use the following lemma on the optimal error of the class of low-degree unitaries.
\begin{lemma}[Optimal error for low-degree unitaries]
    \label{lem:aue-low-opt}
    The optimal error in $d_F$ of the class of low-degree unitaries $\mathcal{L}_{n,d}$ is lower-bounded by
    \begin{equation}
        \opt (U, \mc{L}_{n,d}) \geq \sqrt{1 - \sqrt{w^{\leq d}(U)}}.
    \end{equation}
\end{lemma}
\begin{proof}
From the definition of $\opt$, we have
    \begin{align}
        \opt(U, \mc{L}_{n,d})^2 &= \min_{V \in \mc{L}_{n,d}} d_F^2(U,V)  
        \\&= \frac{1}{2} \min_{V \in \mc{L}_{n,d}} \sum_{x \in \pauli_n} |\hat{U}_x - \hat{V}_x|^2
        \\&= 1 - \max_{V \in \mc{L}_{n,d}} \sum_{x \in \pauli_n} \mathrm{Re}(\hat{U}_x\hat{V}_x^*)
        \\&= 1 - \max_{V \in \mc{L}_{n,d}} \sum_{|x| \leq d} \mathrm{Re}(\hat{U}_x\hat{V}_x^*)
        \\& \geq 1 - \max_{V \in \mc{L}_{n,d}} \sum_{|x| \leq d} |\hat{U}_x\hat{V}_x^*|
        \\& \geq 1 - \max_{V \in \mc{L}_{n,d}} \sqrt{ \sum_{|x| \leq d} |\hat{U}_x|^2} \sqrt{ \sum_{|x| \leq d} |\hat{V}_x|^2}
        \\&= 1 - \max_{V \in \mc{L}_{n,d}} \sqrt{w^{\leq d}(U)}\sqrt{w^{\leq d}(V)}
        \\&= 1 - \sqrt{w^{\leq d}(U)},
    \end{align}
    where the second line follows from Lemma~\ref{lem:df-pauli}, the fourth line uses that $V$ has degree at most $d$, the first inequality uses the fact that the real component of a complex number is smaller than its absolute value, the second inequality uses Cauchy-Schwarz, and the final equality follows from the fact that $V$ is a degree-$d$ unitary.
    This concludes the proof.
\end{proof}

With this, we can prove Theorem~\ref{thm:aue-low-1}.
\begin{proof}[Proof of Theorem~\ref{thm:aue-low-1}]
     We start by estimating all $\mc{O}(n^d 3^d)$ Pauli coefficients of degree $d$. We use~\Cref{lem:estimate-pauli-coeff-unitary} $\mc{O}(n^d 3^d)$ times, to  obtain estimates $\alpha_x$ such that with probability at least $1-\delta$,
     \begin{equation}
         |\alpha_x-\hat{U}_x| \leq \epsilon_1, \quad \forall x \in \pauli_n \text{ s.t. } |x| \leq d.
     \end{equation}
     Using a union bound over failure probabilities of repeated applications of Lemma~\ref{lem:estimate-pauli-coeff-unitary} for each Pauli coefficient, we see that this requires
     \begin{equation}
         N = \mc{O}\left(\frac{d\cdot 3^d n^d \log(n/\delta)}{\epsilon_1^2}\right).
     \end{equation}
     queries to $U$ and runs in time $\mc{O}(N)$. Now, we construct our hypothesis
     \begin{equation}
         V \triangleq \sum_{|x| \leq d} \alpha_x \sigma_x.
     \end{equation}
     This simply aims to approximate $U$ by estimating all low-degree coefficients. It remains to bound the error of this hypothesis, i.e., $d_F(U, V)$.
     Denote 
     \begin{equation}
         \Tilde{U}_d = \sum_{|x| \leq d} \hat{U}_x \sigma_x
     \end{equation} which is the low-degree Pauli truncation of $U$.
     Now, using the triangle inequality for $d_F$, we have
     \begin{equation}
         d_F(U, V) \leq d_F(U,  \Tilde{U}_d) + d_F(\Tilde{U}_d, V)
     \end{equation}
      For ease of notation, let $\hat{W}_x$ denote the Pauli coefficients of $\Tilde{U}_d$. We upper bound $d_F(U,  \Tilde{U}_d)$ as follows.
     \begin{align}
          d_F(U,  \Tilde{U}_d)^2 &= \frac{1}{2}\sum_{x \in \pauli_n} |\hat{U}_x - \hat{W}_x|^2
          \\&= \frac{1}{2}\left(\sum_{|x| \leq d} |\hat{U}_x - \hat{U}_x|^2 + \sum_{|x| > d} |\hat{U}_x - 0|^2\right)
          \\&= \frac{w^{>d}(U)}{2}
          \\&= \frac{1-w^{\leq d}(U)}{2}
          \\&= \frac{(1 - \sqrt{w^{\leq d}(U)})(1 + \sqrt{w^{\leq d}(U)})}{2}
          \\& \leq 1 - \sqrt{w^{\leq d}(U)}
          \\&\leq \opt(U, \mc{L}_{n,d})^2,
     \end{align}
     where in the last line, we use \Cref{lem:aue-low-opt}.
     Finally, conditioned on the successful estimation of $\alpha_x$, we can also bound $d_F(\Tilde{U}_d, V)$.
     \begin{align}
         d_F(\Tilde{U}_d, V)^2 &= \frac{1}{2}\sum_{x \in \pauli_n}|\hat{W}_x - \hat{V}_x|^2
         \\&= \frac{1}{2}\sum_{|x| \leq d}|\hat{U}_x - \alpha_x|^2
         \\& \leq \mc{O}\left(n^d 3^d \epsilon_1^2\right).
     \end{align}
     The inequality follows because there are at most $\mc{O}(n^d3^d)$ Pauli strings $x$ such that $|x| \leq d$.
     Thus, with probability at least $1-\delta$, we have
     \begin{equation}
         d_F(U, V) \leq \opt(U, \mc{L}_{n,d}) + \mc{O}\left(\sqrt{n^d3^d} \epsilon_1\right).
     \end{equation}
     To obtain the 1-agnostic learning guarantee, we choose $\epsilon_1 = \mc{O}(\epsilon/\sqrt{n^d3^d})$. This gives us a query complexity
     \begin{equation}
         N = \mc{O}\left(\frac{d\cdot 9^d n^{2d} \log(n/\delta)}{\epsilon^2}\right)
     \end{equation}
     and time complexity 
     \begin{equation}
         \mc{O}(N)
     \end{equation}
     as stated.
\end{proof}
Now, we will use the non-commutative BH inequality of~\cite{volberg2024noncommutative} to prove Theorem~\ref{thm:aue-low-bh}.
\begin{lemma}[Non-commutative BH inequality; Theorem 1.2 in~\cite{volberg2024noncommutative}]
    \label{lem:non-comm-BH}
    For any $d \geq 1$, there exists $C > 0$, such that for all $n \geq 1$ and all $A \in \mc{B}(\mc{H})$ of degree at most $d$, we have
    \begin{equation}
        \left( \sum_{|x| \leq d} |\hat{A}_x|^{\frac{2d}{d+1}}\right)^{\frac{d+1}{2d}} \leq C^d \|A\|_\infty
    \end{equation}
    for some constant $C$.
\end{lemma}
Now, we can prove Theorem~\ref{thm:aue-low-bh}.
\begin{proof}[Proof of Theorem~\ref{thm:aue-low-bh}]
    Here, we consider the learning algorithm of~\cite{arunachalam2024learning}, and our analysis is similar to theirs.
    The algorithm is sketched in~\Cref{alg:agnostic-unitary-low-2}.
    
    Let $c$ be a constant which we will specify later.
    In Step 1 of~\Cref{alg:agnostic-unitary-low-2}, we prepare $N_1$ copies of the Choi state $|v(U)\rangle$ and measure each of them in the basis $\{|v(\sigma_x)\rangle\}_x$, where again we specify $N_1$ later.
    This allows us to construct empirical estimates $\{|\hat{U}^\prime_x|^2\}_x$ by sampling from the distribution $\{|\hat{U}_x|^2\}_x$. We wish to construct estimates such that, with probability at least $1-\delta/2$, 
    \begin{equation}
        ||\hat{U}^\prime_x|^2 - |\hat{U}_x|^2| \leq c^2, \quad \forall x \in \pauli_n.
    \end{equation}
    Using~\Cref{lem:dist-learning}, we see that this can be done using  
    \begin{equation}
        N_1 = \mc{O}\left(\frac{\log(1/\delta)}{c^4}\right)
    \end{equation}
    copies of the Choi state and $\mathcal{O}(N_1)$ time.
    In what follows, we condition on successful estimation. Then, we have
    \begin{equation}
        \label{eq:4}
        ||\hat{U}^\prime_x| - |\hat{U}_x|| \leq c, \quad \forall x \in \pauli_n.
    \end{equation}
    In Step 2 of~\Cref{alg:agnostic-unitary-low-2}, we denote the set of low-degree strings with high empirical weight by $\chi_c$, i.e.,
    \begin{equation}
        \chi_{c} \triangleq \{x \in \pauli_n :  |\hat{U}^\prime_x| \geq c, |x| \leq d\}
    \end{equation}
    From Parseval's identity (\Cref{lem:parseval-plancherel}), we see that
    \begin{equation}
        |\chi_c| \leq 1/c^2
    \end{equation}
    Note that for any string $x$ with $|x| \leq d$, if $x \notin \chi_c$,
    \begin{equation}
    \label{eq:weight-of-string-not-in-chi}
        |\hat{U}_x| \leq |\hat{U}^\prime_x| + ||\hat{U}_x| - |\hat{U}^\prime_x|| \leq 2c,
    \end{equation}
    where we use \Cref{eq:4} and the definition of $\chi_c$.
    In Step 3 of~\Cref{alg:agnostic-unitary-low-2}, we want to construct estimates $\alpha_x$ for the Pauli coefficients of all strings $x \in \chi_c$, such that with probability at least $1-\delta/2$,
    \begin{equation}
        \label{eq:5}
        |\alpha_x - \hat{U}_x| \leq c\epsilon, \quad \forall x \in \chi_c
    \end{equation}
    This requires $|\chi_c|$ applications of Lemma~\ref{lem:estimate-pauli-coeff-unitary}, using
    \begin{equation}
        N_2 = \mc{O}\left(\frac{|\chi_c|\log(|\chi_c|/\delta)}{c^2\epsilon^2}\right) = \mc{O}\left(\frac{\log(1/c^2\delta)}{c^4\epsilon^2}\right)
    \end{equation}
    querires to the unknown unitary and $\mc{O}(N_2)$ time. Our hypothesis will then be
    \begin{equation}
        V \triangleq \sum_{x \in \chi_c} \alpha_x \sigma_x
    \end{equation}
    It remains to bound the error of this hypothesis, i.e., $d_F(U,V)$. Using the triangle inequality for $d_F$, we have
    \begin{equation}
        d_F(U,V) \leq d_F(U, \Tilde{U}_d) + d_F(\Tilde{U}_d, V)
    \end{equation}
    where
    \begin{equation}
        \Tilde{U}_d = \sum_{|x| \leq d} \hat{U}_x \sigma_x,
    \end{equation}
    as in the proof of Theorem~\ref{thm:aue-low-1}.
    As we showed in the proof of Theorem~\ref{thm:aue-low-1},
    \begin{equation}
        d_F(U, \Tilde{U}_d) \leq \opt (U, \mc{L}_{n,d})
    \end{equation}
    It remains to bound $d_F(\Tilde{U}_d, V)$, for which we follow the arguments of~\cite{arunachalam2024learning}. Denote by $W_x$ the Pauli coefficients of $\Tilde{U}_d$. Then, we have
   \begin{align}
       d_F(\Tilde{U}_d, V)^2 &= \frac{1}{2} \sum_{x \in \pauli_n} |\hat{W}_x - \hat{V}_x|^2
       \\&= \frac{\sum_{x \in \chi_c} |\hat{U}_x - \hat{V}_x|^2}{2} + \frac{\sum_{x \notin \chi_c, |x| \leq d} |\hat{U}_x|^2}{2} + \frac{\sum_{|x| > d} 0}{2}
       \\& \leq \frac{|\chi_c|c^2\epsilon^2}{2} + \frac{\sum_{x \notin \chi_c, |x| \leq d} (2c)^{\frac{2}{d+1}}|\hat{U}_x|^{\frac{2d}{d+1}}}{2} 
       \\&\label{eq:aue-low-deg-error} \leq \frac{\epsilon^2}{2}+\frac{(2c)^{\frac{2}{d+1}} \left(C^d \|\Tilde{U}_d\|_\infty\right)^{\frac{2d}{d+1}}}{2}.
   \end{align}
   The first line follows by \Cref{lem:df-pauli}.
   The third line follows from~\Cref{eq:weight-of-string-not-in-chi} and~\Cref{eq:5}.
   The last line follows from~\Cref{lem:non-comm-BH}.

   In the realizable setting, $\Tilde{U}_d$ is a unitary and thus has operator norm $1$. However, this need not be the case in the agnostic setting, so we need to upper bound $\|\Tilde{U}_d\|_\infty$.
   We state these upper bounds in the following claim, which we will prove later.
    \begin{claim}
    \label{clm:low-deg-truncation-op-norm}
        \begin{equation}
            \|\Tilde{U}_d\|_\infty \leq \min \left[n^{d/2}3^{d/2}, 1 + 2^{(n+1)/2}\cdot \opt(U, \mathcal{L}_{n,d}) \right].
        \end{equation}
    \end{claim}

   Let $u$ be the tightest upper bound from Claim \ref{clm:low-deg-truncation-op-norm}. Hence, by choosing 
   \begin{equation}
   \label{eq:low-deg-unitary-c}
       c = \frac{\epsilon^{d+1}}{2 C^{d^2} u^d},
   \end{equation} we obtain
   \begin{equation}
       d_F(\Tilde{U}_d, V) \leq \epsilon,
   \end{equation}
   giving us
   \begin{equation}
       d_F(U,V) \leq \opt (U, \mc{L}_{n,k}) + \epsilon
   \end{equation}
   as desired. The query complexity is dominated by $N_2$, giving us
   \begin{equation}
       N = \exp \left(\Tilde{\mc{O}} \left( d^2 + d\log(1/\epsilon) + \min\left[d^2 \log n, d \log(1 + 2^{(n+1)/2}\cdot \opt) \right] \right) \right) \cdot \log(1/\delta).
   \end{equation}
   The running time is given by
   \begin{equation}
       \mc{O}(N_1 + N_2) = \mc{O}(N)
   \end{equation}
\end{proof}
It remains to prove Claim \ref{clm:low-deg-truncation-op-norm}.
\begin{proof}[Proof of Claim \ref{clm:low-deg-truncation-op-norm}]
    We start by proving the first bound.
    \begin{align}
        \|\Tilde{U}_d\|_\infty &= \left\| \sum_{|x|\leq d} \hat{U}_x \sigma_x \right\|_\infty
        \\&\leq \sum_{|x|\leq d} \|  \hat{U}_x \sigma_x \|_\infty
        \\&= \sum_{|x|\leq d}|\hat{U}_x|
        \\&\leq \sqrt{n^d 3^d} \sqrt{\sum_{|x|\leq d} |\hat{U}_x|^2} 
        \\& =  n^{d/2}3^{d/2} \sqrt{w^{\leq d}(U)}\\
        &\leq  n^{d/2}3^{d/2} ,
    \end{align}
    where in the second line we use the triangle inequality of $\|\cdot\|_\infty$.
    In the third line we use the fact that $\|\sigma_x\|_\infty = 1$ for all $x \in \pauli_n$.
    In the fourth line we use the Cauchy-Schwarz inequality.
    Finally, in the last line we use the fact that the weight of a unitary up to any degree is at most one.

    We will now prove the second bound which is dependent on $\opt$. Let $W$ be the closest low-degree unitary to $U$, i.e.,
    \begin{equation}
        w^{\leq d}(W) =1 \quad \text{ and } \quad d_F(U,W) = \opt(U, \mc{L}_{n,d}).
    \end{equation}
    We start by lower bounding the normalized Frobenius distance between $W$ and $U$ to obtain a bound on the normalized Frobenius distance between $W$ and $\tilde{U}_d$.
    \begin{align}
        \opt(U, \mc{L}_{n,d})^2 &= d_F(U,W)^2
        \\&= \frac{1}{2}\sum_{|x| \leq d} |\hat{U}_x - \hat{W}_x|^2 + \frac{1}{2}\sum_{|x| > d} |\hat{U}_x|^2
        \\&= d_F(\Tilde{U}_d, W)^2 + w^{>d}(U)
        \\& \geq d_F(\Tilde{U}_d, W)^2
        \\&= \label{eq:low-deg-trunc-opt-dist} \frac{\|\Tilde{U}_d - W\|_F^2}{2^{n+1}}.
    \end{align}
    In the second and third lines, we have used \Cref{lem:df-pauli} and the fact that $W$ has degree at most $d$. In the fourth line, we use the fact that the weight above a degree is non-negative, and in the last line we use the definition of $d_F$. Now, we can directly bound the operator norm of $\Tilde{U}_d$.
    \begin{align}
        \|\Tilde{U}_d\|_\infty &\leq \|\Tilde{U}_d - W\|_\infty + \|W\|_\infty
        \\& \leq \|\Tilde{U}_d - W\|_F + 1
        \\& \leq \sqrt{2^{n+1}} \cdot \opt(U, \mc{L}_{n,d}) + 1.
    \end{align}
    In the first line, we use the triangle inequality of $\|\cdot\|_\infty$. In the second line, we use the fact that $\|\cdot\|_\infty \leq \|\cdot\|_F$ and that $W$ is a unitary. Finally, in the last line, we use \Cref{eq:low-deg-trunc-opt-dist}. This concludes the proof.
\end{proof}

\section{Agnostic Process Tomography (APT)}

In this section, we generalize our agnostic unitary estimation algorithms from \Cref{sec:aue} to obtain algorithms for agnostic process tomography.
Recall that in agnostic process tomography, we have access to an unknown quantum channel rather than a unitary.
This often makes the analysis more complicated.

\label{sec:apt}
\subsection{Bounded-gate circuits}
\label{sec:apt-bounded}
In~\Cref{sec:aue-bounded}, we showed that agnostic unitary estimation can be easily performed efficiently with respect to the number of samples.
However, it can be inefficient with respect to the computational complexity.
This motivates why we should care more about the runtime of algorithms in agnostic unitary estimation.
In this section, we aim to show a similar result for agnostic process tomography.

As in~\Cref{sec:aue-bounded}, we consider APT with respect to any class $\mathcal{C}$ of unitaries with finite size $|\mathcal{C}| < +\infty$.
Here, the error bound we prove is not $1$-agnostic but depends on the fourth root of the optimal error.
Nevertheless, this qualitatively gives the same result that APT is easy with respect to sample complexity but can be more difficult with respect to runtime.

\begin{prop}[Agnostic process tomography of finite-sized unitary classes]
    \label{prop:apt-finite-class}
    Let $\epsilon, \delta > 0$.
    Given query access to an unknown channel $\mc{E}$, for any finite-sized class $\mathcal{C}$ of unitaries, there exists a proper learning algorithm outputting a hypothesis $U \in \mc{C}$, such that 
    \begin{equation}
        \davg(\mc{U}, \mc{E}) \leq \mathsf{opt}(\mc{E}, \mathcal{C})^{1/4}+\epsilon
    \end{equation}
    with probability at least $1-\delta$.
    Here, $\mathsf{opt}$ is measured with respect to $\davg$, and $\mathcal{U}$ denotes the unitary quantum channel defined via $\mathcal{U}(\rho) = U\rho U^\dagger$.
    The algorithm makes
    \begin{equation}
        N = \mathcal{O}\left(\frac{\log (|\mathcal{C}|/\delta)}{\epsilon^4}\right)
    \end{equation}
    queries to the unknown channel $\mc{E}$ and runs in time
    \begin{equation}
    \mathcal{O}\left(T \cdot \frac{|\mathcal{C}|\log(|\mathcal{C}|/\delta)}{\epsilon^4}
        \right),
    \end{equation}
     where $T$ is the time complexity of computing the inner product of two states given their exact description.
\end{prop}

As in \Cref{sec:aue-bounded}, the time complexity in the proposition depends on the complexity of computing the inner products of pairs of states.
This is again due to the use of Clifford classical shadows~\cite{huang2020predicting}, which is not computationally efficient for arbitrary observables (see~\Cref{lem:classical-shadows}).
We do not analyze the time complexity in detail because the scaling with $|\mathcal{C}|$ is already computationally inefficient for most classes of interest.

\begin{proof}
     The algorithm and analysis are similar to the proof of Proposition~\ref{prop:aue-finite-class}.
     Namely, we use Clifford classical shadows~\cite{huang2020predicting} to estimate the distance $\davg(\mc{U}_i, \mc{E})$ between the unknown channel $\mc{E}$ and all unitaries $U_i \in \mathcal{C}$, where again $\mc{U}_i(\rho) = U_i \rho U_i^\dagger$.
    Then, our algorithm outputs the one with the minimal estimated distance.
    
    Again, we prepare random input states from the ensemble
    \begin{equation}
        P = \mathrm{Uniform}(\{|\psi\rangle = U |0\rangle, U \in \clifford(n) \}).
    \end{equation}  
    Note that because our unknown channel may not be a unitary, the quantity estimated is not the root mean squared trace distance but instead the average infidelity.
    This is clear from the proof in Section C.2(a) of~\cite{zhao2023learning}.
    For any measure $\mu$ over input states, we denote the average infidelity by
    \begin{equation}
        \infidel_\mu(\mc{E}_1,\mc{E}_2) \triangleq \mathop{\mbf{E}}\limits_{\rho \sim \mu}\left[1 - F(\mc{E}_1(\rho),\mc{E}_2(\rho))\right].
    \end{equation}
    Now, by Equation (C.8) of~\cite{zhao2023learning}, using
    \begin{equation}
    \label{eq:apt-bounded-samples}
        N = 204 \frac{\log (2|\mathcal{C}|/\delta)}{\epsilon^{\prime2}} = \mathcal{O}\left(\frac{\log(|\mathcal{C}|/\delta)}{\epsilon'^2}\right)
    \end{equation}
    queries, we see that we can construct unbiased estimators $\hat{o}_i$ such that $1-\hat{o}_i$ estimates the average infidelity with respect to $P$ 
    \begin{equation}
        \label{eq:apt-shadow-est}
        |\infidel_P(\mc{U}_i,\mc{E}) - (1-\hat{o}_i)| \leq \epsilon^\prime, \quad \forall i \in [|\mathcal{C}|]
    \end{equation}
    with probability at least $1-\delta$ for some $\epsilon' > 0$.     
    Then, we can select $i^\star = \arg\min_i (1 - \hat{o}_i)$ and output $U_{i^\star}$, i.e., the $U_i \in \mathcal{C}$ with the smallest estimated distance from $\mc{E}$ with respect to $\infidel_P$.
    To relate this back to $\davg$, we recall that $P$ forms a $3$-design~\cite{kueng2015qubit,webb2015clifford,zhu2017multiqubit}, implying that the average infidelity over $P$ is the same as that over the Haar measure.
    This gives us the following relation between the average distances:
    \begin{align}
         d_P(\mc{U}_i, \mc{E})^2 &=   \mathop{\mathbf{E}}\limits_{|\psi\rangle \sim P} [\dtr((\mc{U}_i(\ketbra{\psi}), \mc{E}(\ketbra{\psi})))^2]
         \\& \leq \mathop{\mathbf{E}}\limits_{|\psi\rangle \sim P}[1-F(\mc{U}_i(\ketbra{\psi}), \mc{E}(\ketbra{\psi}))]
         \\&= \mathop{\mathbf{E}}\limits_{|\psi\rangle \sim \mu_S}[1-F(\mc{U}_i(\ketbra{\psi}), \mc{E}(\ketbra{\psi}))]
         \\&\leq \mathop{\mathbf{E}}\limits_{|\psi\rangle \sim \mu_S} [\dtr(\mc{U}_i(\ketbra{\psi}), \mc{E}(\ketbra{\psi}))]
         \\& \leq \sqrt{\mathop{\mathbf{E}}\limits_{|\psi\rangle \sim \mu_S} [\dtr(\mc{U}_i(\ketbra{\psi}), \mc{E}(\ketbra{\psi}))^2]}
         \\&= \label{eq:apt-davg-haar-random-clifford} \davg(\mc{U}_i, \mc{E}),
    \end{align}
    where in the second line, we use~\Cref{def:dtr-infidelity}. In the third line, we use the fact that $P$ is a $2$-design. In the fourth line, we use~\Cref{def:dtr-infidelity} and the fact that $\mc{U}_i(\psi)$ is a pure state.
    In the fifth line we use Jensen's inequality. We can now bound the error of our hypothesis with respect to $\davg$.
    \begin{align}
        \davg(\mc{U}_{i^\star}, \mc{E})^2 &=  \mathop{\mathbf{E}}\limits_{|\psi\rangle \sim \mu_S} [\dtr(\mc{U}_{i^\star}(\ketbra{\psi}), \mc{E}(\ketbra{\psi}))^2]
        \\&\leq \mathop{\mathbf{E}}\limits_{|\psi\rangle \sim \mu_S}[1-F(\mc{U}_{i^\star}(\ketbra{\psi}), \mc{E}(\ketbra{\psi}))]
        \\&=  \mathop{\mathbf{E}}\limits_{|\psi\rangle \sim P}[1-F(\mc{U}_{i^\star}(\ketbra{\psi}), \mc{E}(\ketbra{\psi}))]
        \\&\leq (1-\hat{o}_{i^\star}) + \epsilon^\prime
        \\& = \min_i (1 - \hat{o}_i) + \epsilon^{\prime}
        \\& \leq  \min_i(\infidel_P(\mc{U}_{i}, \mc{E})+\epsilon^\prime) + \epsilon^{\prime}
        \\& \leq \min_i \left(\mathop{\mathbf{E}}\limits_{|\psi\rangle \sim P} [\dtr(\mc{U}_i(\ketbra{\psi}), \mc{E}(\ketbra{\psi}))]\right) + 2\epsilon^{\prime}
        \\&\leq \min_i d_P(\mc{U}_{i}, \mc{E}) + 2\epsilon^{\prime}
        \\&\leq \min_i \sqrt{\davg(\mc{U}_{i}, \mc{E})} + 2\epsilon^{\prime}
        \\&= \label{eq:apt-finite-error-bound}\sqrt{\mathsf{opt}(\mc{E}, \mathcal{C})}+ 2\epsilon^{\prime}
        \\&\leq \left(\mathsf{opt}(\mc{E}, \mathcal{C})^{1/4} + \sqrt{2\epsilon^\prime}\right)^{2}.
    \end{align}
    In the second line, we use~\Cref{def:dtr-infidelity}. In the third line, we use the fact that $P$ is a $2$-design. In the fourth and sixth lines, we use \Cref{eq:apt-shadow-est}. In the seventh line, we use~\Cref{def:dtr-infidelity} and the definition of $\infidel_P$.
    In the eighth line, we use Jensen's inequality. In the ninth line, we use~\Cref{eq:apt-davg-haar-random-clifford}.
    Finally, in the second to last line, we use the definition of $\mathsf{opt}$.
    Thus, by setting $\epsilon^\prime = \epsilon^2/2$, we obtain
    \begin{equation}
         \davg(\mc{U}_{i^\star}, \mc{E}) \leq \mathsf{opt}(\mc{E}, \mathcal{C})^{1/4}+\epsilon,
    \end{equation}
    giving us the desired bound.
    The claimed sample complexity can be obtained from~\Cref{eq:apt-bounded-samples} for our choice of $\epsilon^\prime$, namely,
    \begin{equation}
        N = \mathcal{O}\left(\frac{\log(|\mathcal{C}|/\delta)}{\epsilon^4}\right).
    \end{equation}
    The time complexity is due to the computation of the $\hat{o}_{i}$ for all elements of $\mathcal{C}$, each of which requires computing inner products between $\mathcal{O}(N)$ pairs of states.
\end{proof}
While Proposition~\ref{prop:apt-finite-class} only holds for concept classes of finite size, we now show that this can be easily extended to any class as long as it has bounded metric entropy (see \Cref{def:covering}).
\begin{corollary}
\label{cor:apt-bounded-metric}
    Let $\epsilon, \epsilon^\prime, \delta > 0$. Given query access to an unknown channel $\mc{E}$, for any class of unitaries $\mathcal{C}$ with covering number $\mathcal{N}(\mathcal{C},\davg,\epsilon^\prime)$, there exists a proper learning algorithm outputting a hypothesis $U$ such that 
    \begin{equation}
        \davg(\mc{U}, \mc{E}) \leq \mathsf{opt}(\mc{E}, \mathcal{C})^{1/4}+\epsilon+(\epsilon^\prime)^{1/4}
    \end{equation}
    with probability at least $1-\delta$.
    Here, \textsf{opt} is measured with respect to $\davg$, and $\mc{U}$ denotes the unitary quantum channel defined via $\mathcal{U}(\rho) = U\rho U^\dagger$.
    The algorithm makes
    \begin{equation}
        N = \mathcal{O}\left(\frac{\log (\mathcal{N}(\mathcal{C},\davg,\epsilon^\prime)/\delta)}{\epsilon^4}\right)
    \end{equation}
    queries to the unknown channel $\mc{E}$ and runs in time
    \begin{equation}
    \mathcal{O}\left(T \cdot \mathcal{N}(\mathcal{C},\davg,\epsilon^\prime) \cdot N\right),
    \end{equation}
     where $T$ is the time complexity of computing the inner product of two states given their exact description.
\end{corollary}
\begin{proof}
    Because $\mathcal{C}$ has covering number  $\mathcal{N}(\mathcal{C},\davg,\epsilon^\prime)$, there exists some $\epsilon'$-covering net $M$ with cardinality $\mathcal{N}(\mathcal{C},\davg,\epsilon^\prime)$.
    Consider applying Proposition~\ref{prop:apt-finite-class} to the $\epsilon^\prime$-cover $M$.
    By~\Cref{eq:apt-finite-error-bound} for our choice of $\epsilon' = \epsilon^2/2$, this gives us a hypothesis $U \in M$ satisfying
    \begin{equation}
        \davg(\mc{E}, \mc{U})^2 \leq \min_{V \in M} \sqrt{\davg(\mc{E}, \mc{V})} + \epsilon^2,
    \end{equation}
    where again we use $\mathcal{U}(\rho) = U\rho U^\dagger$ and similarly for $\mathcal{V}$.
    Let $V^*$ be the optimal hypothesis from $\mathcal{C}$, i.e.,
    \begin{equation}
        V^* \triangleq \argmin_{V \in \mathcal{C}} \davg(\mc{E}, \mc{V})\quad \text{ and }\quad \opt(\mc{E}, \mathcal{C}) = \davg(\mc{E}, \mc{V}^*).
    \end{equation}
    Then,
    \begin{align}
        \davg(\mc{E}, \mc{U})^2 &\leq \min_{V \in M} \sqrt{\davg(\mc{E}, \mc{V})} + \epsilon^2
        \\&\leq\min_{V \in M} \sqrt{\davg(\mc{E}, \mc{V}^*) +  \davg(\mc{V}^*, \mc{V})} + \epsilon^2
        \\& \leq \sqrt{\opt(\mc{E}, \mathcal{C})} + \min_{V \in M} \sqrt{\davg(\mc{V}^*, \mc{V})} + \epsilon^2
        \\&\leq \sqrt{\opt(U, \mathcal{C})} + \sqrt{\epsilon^\prime} + \epsilon^2
        \\&\leq \left(\mathsf{opt}(\mc{E}, \mathcal{C})^{1/4}+\epsilon+(\epsilon^\prime)^{1/4}\right)^2
    \end{align}
    where the first inequality follows from~\Cref{eq:apt-finite-error-bound} for our choice of $\epsilon' = \epsilon^2/2$; the second inequality follows from the triangle inequality for $\davg$; the third inequality follows from the fact that $\sqrt{a+b} \leq \sqrt{a} + \sqrt{b}$ for all $a,b \geq 0$. The fourth inequality uses that $M$ $\epsilon^\prime$-covers $\mathcal{C}$. The sample complexity and runtime follow from Proposition~\ref{prop:apt-finite-class} because $|\mathcal{C}| = \mathcal{N}(\mathcal{C},\davg,\epsilon^\prime)$.
\end{proof}
From~\Cref{lem:cover-unitary-gate}, we see that the class of bounded-gate unitaries has bounded metric entropy, which gives us a query-efficient algorithm for agnostic process tomography of this class. However, as the running time is of the order of the covering number, the algorithm is computationally inefficient. From~\Cref{cor:apt-bounded-metric} and~\Cref{lem:cover-unitary-gate}, we have the following corollary.
\begin{corollary}[Agnostic process tomography of bounded-gate unitaries]
\label{cor:apt-gate}
     Let $\epsilon, \delta > 0$. Given query access to an unknown channel $\mc{E}$, for the class $\mc{C}$ of $n$-qubit unitaries of $G$ 2-qubit gates, there exists a (computationally inefficient) proper learning algorithm outputting a hypothesis $U \in \mc{C}$, such that with probability at least $1-\delta$, 
    \begin{equation}
        \davg(\mc{U}, \mc{E}) \leq \mathsf{opt}(\mc{E}, \mathcal{C})^{1/4}+\epsilon.
    \end{equation}
    The algorithm makes
    \begin{equation}
        N = \Tilde{\mathcal{O}}\left(
            \frac{G \log( n/\delta)}{\epsilon^4}
        \right)
    \end{equation}
    queries to the unknown channel $\mc{E}$.
\end{corollary}
\subsection{APT from agnostic state tomography}
\label{sec:apt-clifford}

In~\Cref{sec:aue-clifford}, we showed that we can convert any agnostic state tomography algorithm (\Cref{def:ast}) into one for improper agnostic unitary estimation using ancilla qubits, as long as the concept classes are compatible.
This introduces a quadratic overhead in the sample/time complexity with respect to the error $\epsilon$.

In this section, we prove a similar result for improper agnostic process tomography.
The error bound we prove is not $1$-agnostic but depends on $2^{1/4}$ times the square root of the optimal error, when we measure error with respect to the normalized Frobenius distance for superoperators (\Cref{def:df-superop}).
We state this result formally in the following proposition.

\begin{prop}[Improper APT from agnostic state tomography]
    \label{prop:ast-to-apt}
    Let $1 > \epsilon,\delta > 0$.
    Suppose there exists a learning algorithm for (proper) $(1,\epsilon^2,\delta)$-agnostic state tomography with respect to fidelity $F$ for some class $\mathcal{D}$ of pure states.
    Let $\mc{C}$ be a concept class of unitaries.
    Suppose that $\ket{v(U)} \in \mc{D}$ for all $U \in \mc{C}$.
    Given query access to an unknown channel $\Phi$, there exists an improper learning algorithm outputting a hypothesis $\Phi'$ such that
    \begin{equation}
        d_F(\Phi, \Phi') \leq 2^{1/4} \opt(\Phi, \mc{C})^{1/2} + \epsilon
    \end{equation}
    with probability at least $1-\delta$.
    Here, $\opt$ is measured with respect to $d_F$.
    Moreover, this requires the same sample and time complexity as the $(1,\epsilon^2,\delta)$-agnostic state tomography algorithm.
\end{prop}

As in~\Cref{sec:aue-clifford}, our algorithm simply applies agnostic state tomography to the Choi state of the unknown process, inspired by~\cite{leung2000towards}.
We also note that we can trivially obtain $(1,\epsilon,\delta)$-agnostic process tomography with respect to the entanglement infidelity from $(1,\epsilon,\delta)$-agnostic state tomography (notably, without the quadratic blowup with respect to $\epsilon$) under the same compatibility condition.
Recall that we defined the entanglement infidelity in~\Cref{eq:entangle-infidel}, and this metric has been widely considered in (realizable) unitary estimation~\cite{leung2000towards,haah2023query,acin2001optimal,peres2002covariant,hayashi2006parallel,chiribella2005optimal,kahn2007fast,yang2020optimal}.

\begin{proof}
    The proof is similar to that of Proposition~\ref{prop:ast-to-aue}.
    Let $\Phi$ denote the unknown unitary.
    The algorithm is simple: first, prepare the Choi state $v(\Phi)$ of $\Phi$ and then apply the $(1,\epsilon^2,\delta)$-agnostic state tomography algorithm to $v(\Phi)$.
    Note that the known agnostic state tomography algorithms work for mixed state inputs~\cite{grewal2024agnostic,chen2024stabilizer}.
    The output of our algorithm is the density matrix of the pure state output by the agnostic state tomography algorithm.

    The agnostic state tomography algorithm outputs some state $\ket{\psi}$ such that 
    \begin{equation}
        F(v(\Phi), \ketbra{\psi}) \geq \max_{\ket{\phi} \in \mc{D}} F(v(\Phi), \ketbra{\phi}) - \epsilon^2,
    \end{equation}
    where $v(\Phi)$ denotes the Choi state of $\Phi$.
    We can equivalently write the above guarantee in terms of the infidelity:
    \begin{equation}
        1 - F(v(\Phi), \ketbra{\psi}) \leq \min_{\ket{\phi} \in \mc{D}}(1 - F(v(\Phi), \ketbra{\phi})) + \epsilon^2.
    \end{equation}
    Using the inequalities relating trace distance and fidelity when one of the inputs is pure (see after \Cref{def:dtr-infidelity}), we have
    \begin{equation}
        \dtr(v(\Phi), \ketbra{\psi})^2 \leq \min_{\ket{\phi} \in \mc{D}}\dtr(v(\Phi), \ketbra{\phi}) + \epsilon^2.
    \end{equation}
    We can relate this to the Frobenius norm using a result from~\cite{coles2019strong}.
    Namely, they prove that
    \begin{equation}
        \frac{1}{2}\norm{\rho - \sigma}_F^2 \leq \dtr(\rho, \sigma)^2 \leq \frac{\mathrm{rank}(\rho)\mathrm{rank}(\sigma)}{\mathrm{rank}(\rho) + \mathrm{rank}(\sigma)} \norm{\rho - \sigma}_F^2.
    \end{equation}
    Using this in the previous inequality, we have
    \begin{equation}
        \frac{1}{2}\norm{v(\Phi) - \ketbra{\psi}}_F^2 \leq \min_{\ket{\phi} \in \mc{D}}\sqrt{\frac{\mathrm{rank}(v(\Phi))\mathrm{rank}(\ketbra{\phi})}{\mathrm{rank}(v(\Phi)) + \mathrm{rank}(\ketbra{\phi})}} \norm{v(\Phi) - \ketbra{\phi}}_F + \epsilon^2.
    \end{equation}
    Since $\ketbra{\phi}$ is pure, then $\mathrm{rank}(\ketbra{\phi}) = 1$ so that the term in the square root can be bounded by $1$.
    Moreover, by definition of the normalized Frobenius distance for operators (\Cref{def:dist-frob}),
    \begin{equation}
        d_F(v(\Phi), \ketbra{\psi}) = \frac{1}{2^n\sqrt{2}}\norm{v(\Phi) - \ketbra{\psi}}_F
    \end{equation}
    since $v(\Phi), \ketbra{\psi}$ are on $2n$ qubits.
    Thus, rewriting the above inequality in terms of the normalized Frobenius distance, we have
    \begin{equation}
        4^n d_F(v(\Phi), \ketbra{\psi})^2 \leq \sqrt{2} \cdot 2^n \min_{\ket{\phi} \in \mc{D}} d_F(v(\Phi), \ketbra{\phi}) + \epsilon^2.
    \end{equation}
    Taking the square root of both sides, we have
    \begin{equation}
        2^n d_F(v(\Phi), \ketbra{\psi}) \leq 2^{1/4} \sqrt{2^n \min_{\ket{\phi} \in \mc{D}} d_F(v(\Phi), \ketbra{\phi})} + \epsilon.
    \end{equation}
    Now, by the same calculation as in~\Cref{sec:aue-clifford}, agnostically learning the Choi state $v(\Phi)$ with respect to $2^n d_F$ is equivalent to agnostically learning the channel $\Phi$ with respect to $d_F$.
    Note that the first $d_F$ is the normalized Frobenius distance for operators (\Cref{def:dist-frob}) while the second $d_F$ is the normalized Frobenius norm for superoperators (\Cref{def:df-superop}).
    We repeat the calculation here since it is short:
    \begin{align}
        d_F(\Phi, \Psi) &= \frac{1}{2^n\sqrt{2}}\norm{\Phi - \Psi}_F\\
        &= \frac{1}{2^n\sqrt{2}}\norm{J(\Phi) - J(\Psi)}_F\\
        &= \frac{1}{2}\norm{v(\Phi) - v(\Psi)}_F\\
        &= 2^n d_F(v(\Phi), v(\Psi)).
    \end{align}
    In the first and second lines, we use the definition of $d_F$ for superoperators (\Cref{def:df-superop}).
    In the third line, we use that the Choi representation is simply the unnormalized Choi state.
    Finally, in the last line, we use the definition of the normalized Frobenius distance for operators (\Cref{def:dist-frob}).
    This completes the proof.
\end{proof}

This gives a fairly general result for converting an agnostic state tomography algorithm with respect to $\mathcal{D}$ into an improper agnostic process tomography algorithm with respect to $\mathcal{C}$ when $\ket{v(U)} \in \mathcal{D}$ for all $U \in \mathcal{C}$.
However, our result is weaker than that of Proposition~\ref{prop:ast-to-aue}, 
as our algorithm is no longer $1$-agnostic but has an error bound of $2^{1/4}\sqrt{\mathsf{opt}}$ instead.

As remarked earlier, the compatibility condition holds for many classes of unitaries for which agnostic state tomography algorithms already exist with respect to a suitable class.
In particular, there exist agnostic state tomography algorithms for stabilizer states, states with high stabilizer dimension, and discrete product states~\cite{grewal2024agnostic,chen2024stabilizer}.
Note that the corresponding classes of unitaries, i.e., Clifford circuits~\cite{gottesman1997stabilizer}, Clifford + T circuits, and unitaries consisting of a tensor product of single-qubit gates, respectively, all satisfy the required property of Proposition~\ref{prop:ast-to-apt}.
Thus, similarly to \Cref{sec:aue-clifford}, we can directly apply Proposition~\ref{prop:ast-to-aue} with the state-of-the-art sample/time complexity bounds for agnostic state tomography to obtain improper APT algorithms for the respective unitary classes.
These are each direct corollaries of Corollary 6.3, Theorem 7.1, Corollary 8.3, and in~\cite{chen2024stabilizer}.

\begin{corollary}[Improper APT of Clifford Circuits; via Corollary 6.3 in~\cite{chen2024stabilizer}]
    Let $1 > \epsilon, \delta > 0$.
    Let $\mc{C}$ be the class of $n$-qubit Clifford unitaries.
    Given query access to an unknown channel $\Phi$, there exists an improper learning algorithm outputting a hypothesis $\Phi'$ such that
    \begin{equation}
        d_F(\Phi, \Phi') \leq 2^{1/4} \opt(\Phi, \mc{C})^{1/2} + \epsilon,
    \end{equation}
    with probability at least $1-\delta$, using
    \begin{equation}
        N = 2n\log(1/\delta)\left(\frac{1}{\epsilon}\right)^{\mathcal{O}(\log 1/\epsilon)} + \mathcal{O}\left(\frac{\log(1/\delta) + \log^2(1/\epsilon)}{\epsilon^4}\right)
    \end{equation}
    queries to the unknown unitary and running in time
    \begin{equation}
        \mathcal{O}\left(n^2\log(1/\delta)\left(n + \frac{\log(1/\delta)}{\epsilon^4}\right)\right) \cdot \left(\frac{1}{\epsilon}\right)^{\mathcal{O}(\log 1/\epsilon)}.
    \end{equation}
\end{corollary}

\begin{corollary}[Improper APT of Clifford + T Circuits; via Theorem 7.1 in~\cite{chen2024stabilizer}]
    Let $1 > \epsilon, \delta > 0$ and $t \geq 0$.
    Let $\mc{C}$ be the class of $n$-qubit unitaries consisting of Clifford gates and $t$ T gates.
    Given query access to an unknown channel $\Phi$, there exists an improper learning algorithm outputting a hypothesis $\Phi'$ such that
    \begin{equation}
        d_F(\Phi, \Phi') \leq 2^{1/4} \opt(\Phi, \mc{C})^{1/2} + \epsilon,
    \end{equation}
    with probability at least $1-\delta$, using
    \begin{equation}
        N = 2n\log(1/\delta)\left(\frac{2^t}{\epsilon^2}\right)^{\mathcal{O}(\log 1/\epsilon)}
    \end{equation}
    queries to the unknown unitary and running in time
    \begin{equation}
        \mathcal{O}(n^2 \log(1/\delta) N).
    \end{equation}
\end{corollary}

\begin{corollary}[Improper APT of Product Circuits; via Corollary 8.3 in~\cite{chen2024stabilizer}]
    Let $\mathcal{K}$ be a set of single-qubit gates for which $|\braket{v(U_1)}{v(U_2)}|^2 \leq 1 - \mu$ for any distinct $U_1, U_2 \in \mathcal{K}$.
    Let $\mc{C} = \mc{K}^{\otimes n}$.
    Given query access to an unknown channel $\Phi$, there exists an improper learning algorithm outputting a hypothesis $\Phi'$ such that
    \begin{equation}
        d_F(\Phi, \Phi') \leq 2^{1/4} \opt(\Phi, \mc{C})^{1/2} + \epsilon,
    \end{equation}
    with probability at least $1-\delta$, using
    \begin{equation}
        N = \frac{\log^2(1/\delta)(2n|\mathcal{K}|)^{\mathcal{O}\left(\log(1/\epsilon)/\mu\right)}}{\epsilon^4}
    \end{equation}
    queries to the unknown unitary and running in time $\mathcal{O}(N)$.
\end{corollary}

\begin{corollary}[Improper APT of Clifford Product Circuits; via Corollary 9.3 in~\cite{chen2024stabilizer}]
    Let $1 > \epsilon, \delta > 0$.
    Let $\mc{C}$ be the class of tensor products of single-qubit Clifford gates.
    Given query access to an unknown channel $\Phi$, there exists an improper learning algorithm outputting a hypothesis $\Phi'$ such that
    \begin{equation}
        d_F(\Phi, \Phi') \leq 2^{1/4} \opt(\Phi, \mc{C})^{1/2} + \epsilon,
    \end{equation}
    with probability at least $1-\delta$, using
    \begin{equation}
        N = \log (2n) \log(1/\delta) \left(\frac{1}{\epsilon}\right)^{\mathcal{O}(\log 1/\epsilon)} + \mathcal{O}\left(\frac{\log^2(1/\epsilon) + \log(1/\delta)}{\epsilon^4}\right)
    \end{equation}
    and running in time
    \begin{equation}
        \frac{4n^2 \log^2(1/\delta) (1/\epsilon)^{\mathcal{O}(\log 1/\epsilon)}}{\epsilon^4}.
    \end{equation}
\end{corollary}

Note that all of these algorithms are improper, while for applications, one would prefer proper learning algorithms.
We leave it to future work to make these algorithms proper.
\subsection{Pauli strings and Pauli channels}
\label{sec:apt-pauli}
In this section, we consider agnostic process tomography of the class of $n$-qubit Pauli strings $\mathcal{C}_{\pauli_n} = \{I, X, Y, Z\}^{\otimes n}$ as well as the class of $n$-qubit Pauli channels $\mc{C}_{\mc{PC}}$. A Pauli channel is defined as follows.
\begin{definition}[Pauli Channel] A quantum channel $\Phi$ on $\mc{B}(\mc{H})$ is a \emph{Pauli channel} if
\begin{equation}
    \Phi(\rho) = \sum_{x \in \pauli_n} p_x \sigma_x \rho \sigma_x,
\end{equation}
where $\{p(x)\}_{x\in \pauli_n}$ forms a probability distribution.
\end{definition}
Pauli channels form a naturally occurring class of noise channels, where the probabilities $p_x$ denote the chance of a particular Pauli error occurring. 
Clearly, Pauli channels have Fourier coefficients of the following form
\begin{equation}
    \hat{\Phi}(x,x) = p_x, \quad \forall x \in \pauli_n
\end{equation}
and
\begin{equation}
    \hat{\Phi}(x,y) = 0,\quad x \neq y
\end{equation}
For both classes, our results are with respect to $d_F$, the normalized Frobenius distance for superoperators (\Cref{def:df-superop}).
Our proof techniques are similar to those used in~\Cref{sec:aue-pauli}, but we use the more general Fourier spectrum of superoperators, as introduced in~\Cref{sec:fourier-superop}. We start with the class of Pauli strings.
\begin{theorem}
    [Agnostic Process Tomography of Pauli strings in $d_F$]\label{thm:apt-pauli-df} 
There exists a learning algorithm for proper $(1,\epsilon,\delta)$-agnostic process tomography of the class of $n$-qubit Pauli strings $\mathcal{C}_{\pauli_n}$ with respect to error in $d_F$ using 
\begin{equation}
    N = 
    \mathcal{O}\left(
        \frac{ \log(1/\delta)}{\epsilon^4}
    \right)
\end{equation}
samples and running in time
\begin{equation}
    \mathcal{O}\left(
        N
    \right).
\end{equation}
\end{theorem}
\begin{proof}
    The algorithm is the same as~\Cref{alg:agnostic-unitary-paulis}, but we instead prepare the Choi state $v(\Phi)$ of the unknown quantum channel $\Phi$. Recall that we can write the Choi state as
    \begin{equation}
        v(\Phi) = \sum_{x,y \in \pauli_n} \hat{\Phi}(x,y) |v(\sigma_x)\rangle \langle v(\sigma_y)|.
    \end{equation}
    Then, by measuring $v(\Phi)$ in the basis $\{|v(\sigma_x)\rangle\}_x$, we measure $x$ with probability
    \begin{equation}
        p(x) \triangleq \langle v(\sigma_x) | v(\Phi) | v(\sigma_x) \rangle = \hat{\Phi}(x,x).
    \end{equation}
    Now, our analysis proceeds similarly to the proof of~\Cref{thm:aue-pauli-davg}. Denote the true most likely Pauli string by
    \begin{equation}
        z^\star \triangleq \arg \max_{z \in \pauli_n} p(z).
    \end{equation}
    We compute the optimal error for the class of Pauli strings. Recall from~\Cref{lem:df-superop-fourier} that for quantum channels $\Phi, \Psi$,
    \begin{equation}
        d_F(\Phi, \Psi)^2 = \frac{1}{2}\sum_{x,y \in \pauli_n} |\hat{\Phi}(x,y) - \hat{\Psi}(x,y)|^2.
    \end{equation}
    Then, we have
    \begin{align}
            \opt(\Phi, \mathcal{C}_{\pauli_n})^2
            &= \min_{z \in \pauli_n} d_F(\Phi, \Phi_{z,z})^2 
            \\&= \frac{1}{2}\left(\min_{z \in \pauli_n} \sum_{\substack{x,y \in \pauli_n\\x \neq z \text{ or } y \neq z}} |\hat{\Phi}(x,y)|^2  + |\hat{\Phi}(z,z) - 1|^2\right)
            \\&=  \frac{1+ \|\hat{\Phi}\|^2_F}{2} - \max_{z \in \pauli_n} \hat{\Phi}(z,z)
            \\&= \frac{1+ \|\hat{\Phi}\|^2_F}{2} - \hat{\Phi}(z^*,z^*)\label{eq:opt-pauli-string-superop},
    \end{align}
    where we use $\Phi_{z,z}$ to denote the channel $\Phi_{z,z}(\rho) = \sigma_z \rho \sigma_z$ so that $\hat{\Phi}_{z,z}(z,z) = 1$ and zero otherwise.
    The second equality follows from~\Cref{lem:df-superop-fourier}.
    In the third line, note that $\|\hat{\Phi}\|_F^2$ denotes $\|\hat{\Phi}\|_F^2 \triangleq \sum_{x,y \in \pauli_n} |\hat{\Phi}(x,y)|^2$, and we use the fact that the diagonal elements of $\hat{\Phi}$ are all real and non-negative.
    In the last line, we use the definition of $z^\star$.
    
    Hence, we see that the hypothesis $\sigma_{z^\star}$ achieves the optimal error for the class $\mathcal{C}_{\mathcal{P}_n}$.
    Our algorithm aims to estimate the true most likely Pauli string $z^\star$ by the one most frequently occurring in our measurements.
    Let $\{\hat{p}(x)\}_x$ be the empirical distribution obtained from our measurement outcomes, which estimates $p$.
    Then, denote the most frequent outcome out of our measurements by
    \begin{equation}
        x^*  \triangleq \argmax_{x \in \{x_i\}_{i \in [N]}} \hat{p}(x).
    \end{equation}
     Using~\Cref{lem:dist-learning}, for some $\epsilon_1 > 0$, the empirical estimates $\hat{p}(x)$ of $p(x) = \hat{\Phi}(x,x)$ satisfy, with probability at least $1-\delta$,
    \begin{equation}
        |\hat{p}(x) -\hat{\Phi}(x,x)| \leq \epsilon_1, \quad \forall x \in \pauli_n,
    \end{equation}
    using $\mathcal{O}\left(\frac{ \log(1/\delta)}{\epsilon_1^2}\right)$ copies of the Choi state $|v(U)\rangle$. Now, since $x^*$ is the most frequent string observed,
    \begin{equation}
        \hat{p}(x^*) \geq \hat{p}(z^\star).
    \end{equation}
    Conditioned on obtaining estimates $\hat{p}(x)$ that are $\epsilon_1$-close to $|\hat{U}_x|^2$ for all $x \in \pauli_n$, we have
    \begin{equation}
        \hat{\Phi}(x^*,x^*) + \epsilon_1 \geq \hat{p}(x^*)\quad \text{ and }\quad \hat{p}(z^\star) \geq \hat{\Phi}(z^*,z^*) - \epsilon_1.
    \end{equation}
    Thus, we have
    \begin{equation}
        \label{eq:pauli-string-apt-1}
       \hat{\Phi}(z^*,z^*) - \hat{\Phi}(x^*,x^*) \leq 2\epsilon_1.
    \end{equation}
    Now, we can bound the error of the hypothesis as follows
    \begin{align}
        d_F(\Phi, \sigma_{x^*})^2 &= \frac{1+ \|\hat{\Phi}\|^2_F}{2} - \hat{\Phi}(x^*,x^*)
        \\&= \frac{1+ \|\hat{\Phi}\|^2_F}{2} - \hat{\Phi}(z^*,z^*) + \hat{\Phi}(z^*,z^*) - \hat{\Phi}(x^*,x^*)
        \\& \leq \opt(\Phi, \mathcal{C}_{\pauli_n})^2 + 2\epsilon_1,
    \end{align}
    where the first line is true by~\Cref{lem:df-superop-fourier} (a similar calculation as before for $\mathsf{opt}$), and the last line follows by~\Cref{eq:opt-pauli-string-superop} and \Cref{eq:pauli-string-apt-1}.
    Thus, by setting $\epsilon_1 = \epsilon^2/2$, we obtain
    \begin{equation}
        d_F(\Phi, \sigma_{x^*})^2 \leq \opt(\Phi, \mathcal{C}_{\pauli_n})^2 + \epsilon^2 \leq (\opt(\Phi, \mathcal{C}_{\pauli_n}) + \epsilon)^2,
    \end{equation}
    which is the desired bound.
    Now, by our choice of $\epsilon_1 = \epsilon^2/2$, we use
    \begin{equation}
        N = \mathcal{O}\left(\frac{\log (1/\delta)}{\epsilon_1^2}\right) = \mathcal{O}\left(\frac{ \log(1/\delta)}{\epsilon^4}\right)
    \end{equation}
    samples of the Choi state or queries to the channel $\Phi$. The time complexity is that of identifying the most frequent string out of $N$ strings, which can be done in time $\mathcal{O}(N)$.
\end{proof}
Now, we present a proper 1-agnostic learning algorithm for the class of Pauli channels.

\begin{algorithm}
   \caption{1-Agnostic Proper Learning Pauli Channels} 
   \label{alg:apt-pauli-channel}
   \begin{algorithmic}[1]
   \State Prepare $N = \mathcal{O}(\log(1/\delta)/\epsilon^2)$ copies of $v(\Phi)$ and measure them in the $\{\ket{v(\sigma_x)}\}_x$ basis to construct estimates $\hat{p}(x)$ of $\hat{\Phi}(x,x)$.
    \State \Return $\Phi = \sum_{x \in \pauli_n} \hat{p}(x) \sigma_x (\cdot) \sigma_x$.
   \end{algorithmic}
\end{algorithm}

\begin{theorem}
     [Agnostic Process Tomography of Pauli channels in $d_F$]\label{thm:apt-pauli-channel-df} 
\Cref{alg:apt-pauli-channel} performs proper $(1,\epsilon,\delta)$-agnostic process tomography of the class of $n$-qubit Pauli channels $\mathcal{C}_{\text{PC}}$ with respect to error in $d_F$ using 
\begin{equation}
    N = 
    \mathcal{O}\left(
        \frac{ \log(1/\delta)}{\epsilon^2}
    \right)
\end{equation}
samples and running in time
\begin{equation}
    \mathcal{O}\left(
        N \log(N)
    \right).
\end{equation}
\end{theorem}
\begin{proof}
    The algorithm is detailed in \Cref{alg:apt-pauli-channel}.
    We start by computing the optimal error for the class $\mc{C}_{\mc{PC}}$.
    \begin{align}
        \opt(\Phi, \mc{C}_{\mc{PC}})^2 &= \min_{\Psi \in \mc{C}_{\mc{PC}}} d_F(\Phi, \Psi)^2
        \\&= \frac{1}{2}\left(\min_{\Psi \in \mc{C}_{\mc{PC}}} \sum_{x \in \pauli_n} |\hat{\Phi}(x,x) - \hat{\Psi}(x,x)|^2 + \sum_{\substack{x,y\in \pauli_n\\x \neq y}}|\hat{\Phi}(x,y) - 0|^2\right)
        \\& \geq \frac{1}{2} \sum_{\substack{x,y\in \pauli_n\\x \neq y}} |\hat{\Phi}(x,y)|^2 ,
    \end{align}
    where the second line follows from~\Cref{lem:df-superop-fourier} and the fact that $\Psi$ is a Pauli channel. As $\{\hat{\Phi}(x,x)\}_{x \in \pauli_n}$ form a probability distribution, the superoperator
    \begin{equation}
        \Phi^\prime = \sum_{x \in \pauli_n} \hat{\Phi}(x,x) \sigma_x (\cdot) \sigma_x
    \end{equation}
    is a Pauli channel. Clearly, 
    \begin{equation}
        d_F(\Phi, \Phi^\prime)^2 = \frac{1}{2} \sum_{\substack{x,y\in \pauli_n\\x \neq y}} |\hat{\Phi}(x,y)|^2.
    \end{equation}
    Thus, from the lower bound on \textsf{opt} from the above discussion, we have
    \begin{equation}
        \label{eq:apt-pauli-channel-opt} \opt(\Phi, \mc{C}_{\mc{PC}})^2 = \frac{1}{2} \sum_{\substack{x,y\in \pauli_n\\x \neq y}} |\hat{\Phi}(x,y)|^2 .
    \end{equation}
    Since measuring $v(\Phi)$ in the basis $\{|v(\sigma_x)\rangle\}_{x\in\pauli_n}$ returns samples from the distribution $\{\hat{\Phi}(x,x)\}_x$, agnostic process tomography of Pauli channels reduces to the task of learning this distribution. From~\Cref{lem:dist-learning}, we see that $\mc{O}\left(\log(1/\delta)/\epsilon^2\right)$ samples allow us to empirically construct a distribution $\{\hat{p}(x)\}_x$, that satisfies, with probability at least $1-\delta$,
    \begin{equation}
    \label{eq:apt-pauli-channels-error}
        \sum_{x \in \pauli_n} |\hat{\Phi}(x,x) - \hat{p}(x)|^2 \leq 2\epsilon^2.
    \end{equation}
    Now, let our hypothesis channel be defined by
    \begin{equation}
        \Phi_{\hat{p}} \triangleq \sum_{x \in \pauli_n} \hat{p}(x) \sigma_x (\cdot) \sigma_x.
    \end{equation}
    The error of this hypothesis is bounded by
    \begin{align}
        d_F(\Phi, \Phi_{\hat{p}})^2  &= \frac{1}{2}\sum_{\substack{x,y\in \pauli_n\\x \neq y}} |\hat{\Phi}(x,y)|^2
      + \frac{1}{2}
            \sum_{x \in \pauli_n} |\hat{\Phi}(x,x) - \hat{p}(x)|^2
        \\& \leq  \opt(\Phi, \mc{C}_{\mc{PC}})^2 + \epsilon^2
        \\& \leq ( \opt(\Phi, \mc{C}_{\mc{PC}}) + \epsilon)^2,
    \end{align}
    where the second line follows from~\Cref{eq:apt-pauli-channel-opt,eq:apt-pauli-channels-error},
    giving us the desired 1-agnostic learning guarantee. The running time is that of counting the number of occurrences of each measurement outcome, which can be done in time $\mc{O}(N \log(N))$.
\end{proof}

\subsection{Junta channels}
\label{sec:apt-junta}
In this section, we present an algorithm for agnostic process tomography of junta channels. Similar to the definition of unitary juntas, $k$-junta channels on $n$-qubits are defined as quantum channels that act non-trivially only on $k$ qubits.

\begin{definition}[$k$-junta channel]
    \label{def:channel-junta}
    A quantum channel $\Phi$ on $n$ qubits is a \emph{$k$-junta channel} if there exists a set $S\subseteq [n]$ with $|S| = k$ such that $\Phi = \Psi_S \otimes \mc{I}_{S^c}$, for some $k$-qubit quantum channel $\Psi_S$.
\end{definition}

In \Cref{sec:aue-junta}, for the problem of agnostic unitary estimation of unitary $k$-juntas, we identified the subset of qubits with the highest Pauli weight and then learned all Pauli coefficients restricted to this subset. We were able to identify this subset efficiently because, for an $n$-qubit unitary $U$, we could efficiently sample from the distribution $\{|\hat{U}_x^2|\}_{x \in \pauli_n}$.

However, in the case of a quantum channel $\Phi$, a similar method allows us to sample from the distribution $\{\hat{\Phi}(x,x)\}_{x \in \pauli_n}$. This does not allow us to estimate the weight of a subset $S$, which is given by $w_S(\Phi) = \sum_{x,y \in \pauli_S} |\hat{\Phi}(x,y)|^2$. To overcome this issue, we learn the Fourier coefficients associated with each subset and use them to estimate the weight. Unsurprisingly, this results in a worse sample complexity than the unitary junta case (\Cref{thm:aue-juntas-df}). 
Once we have estimated the weights, we can construct an improper hypothesis by restricting to the Fourier coefficients corresponding to the heaviest estimated subset.
The algorithm is sketched in \Cref{alg:apt-junta-channel}.

\begin{algorithm}
   \caption{1-Agnostic Improper Learning $k$-Junta Channels} 
   \label{alg:apt-junta-channel}
   \begin{algorithmic}[1]
    \State Use $N$ queries to $\Phi$ to construct estimates $\alpha_{x,y}$ of the Fourier coefficients with $x,y \in \pauli_S$, for all subsets $S \subseteq [n], |S| = k$.
    \State Construct estimates of weights of all subsets of $[n]$ of cardinality $k$ using $\alpha_{x,y}$.
    \State Select $S'$ to be the subset with the greatest estimated weight.
    \State \Return $\Psi = \sum_{x,y\in \pauli_{S'}} \alpha_{x,y} \sigma_x (\cdot) \sigma_y$.
   \end{algorithmic}
\end{algorithm}

\begin{theorem}[Agnostic Process Tomography of $k$-Junta Channels in $d_F$.]
    \label{thm:apt-juntas}
    \Cref{alg:apt-junta-channel} performs improper $(1,\epsilon,\delta)$-agnostic process tomography of the class of quantum $k$-junta channels on $\mc{B}(\mc{H})$, $\mathcal{JC}_{n,k}$, with respect to error in $d_F$ using 
\begin{equation}
    N = \mc{O}\left(\frac{k 2^{12k} n^k \log(n/\delta)}{\epsilon^4}\right)
\end{equation}
queries to the unknown channel $\Phi$ and running in time
\begin{equation}
   \mc{O}(N).
\end{equation}
\end{theorem}
\begin{proof}
    The algorithm is detailed in \Cref{alg:apt-junta-channel}.
    We start by lower bounding the optimal error for the class of junta channels. For a junta channel $\Psi_S \in \mc{JC}_{n,k}$ acting only on a set $S$ with $|S| = k$, we have
    \begin{align}
        d_F(\Phi,\Psi_S)^2 &= \frac{1}{2} \sum_{x,y \in \pauli_n} |\hat{\Phi}(x,y) - \hat{\Psi}_S(x,y)|^2
        \\&= \frac{1}{2}\sum_{x,y \in \pauli_S} |\hat{\Phi}(x,y) - \hat{\Psi}_S(x,y)|^2  + \frac{1}{2}\sum_{\substack{x, y\in \pauli_n\\x \notin \pauli_S \text{ or } y \notin \pauli_S}}|\hat{\Phi}(x,y)|^2
        \\& \geq 0 + \frac{1}{2}\sum_{\substack{x, y\in \pauli_n\\x \notin \pauli_S \text{ or } y \notin \pauli_S}}|\hat{\Phi}(x,y)|^2
        \\& = \frac{\|\hat{\Phi}\|_F^2 - w_S(\Phi)}{2},
    \end{align}
    where the first line follows from Lemma~\ref{lem:df-superop-fourier}.
    The second line uses that $\Psi_S$ is a $k$-junta on the set $S$.
    The final equality uses the definition of weight and $\|\hat{\Phi}\|_F^2 = \sum_{x,y \in \pauli_n} |\hat{\Phi}(x,y)|^2$.
    Then, we can obtain a lower bound on $\opt$ by minimizing over the choice of set $S$ and junta channel $\Psi_S$.
    \begin{align}
        \opt(\Phi, \mc{JC}_{n,k})^2 &= \min_{S \subseteq [n], |S| = k} \min_{\Psi_S}  d_F(\Phi,\Psi_S)^2
        \\& \geq \min_{S \subseteq [n], |S| = k} \min_{\Psi_S} \frac{\|\hat{\Phi}\|_F^2 - w_S(\Phi)}{2}
        \\& =  \min_{S \subseteq [n], |S| = k} \frac{\|\hat{\Phi}\|_F^2 - w_S(\Phi)}{2}.
    \end{align}
    Let $S^* \triangleq \argmax_{S \subseteq [n], |S| = k} w_{S}(\Phi)$ be the set of cardinality $k$ with the highest Fourier weight. Then, we have
    \begin{equation}
        \label{eq:apt-juntas-opt-weight}
        \opt(\Phi, \mc{JC}_{n,k})^2 \geq \frac{\|\hat{\Phi}\|_F^2 - w_{S^*}(\Phi)}{2}.
    \end{equation}
    Now that we have lower bounded opt, we analyze our algorithm (\Cref{alg:apt-junta-channel}), which starts by estimating all Fourier coefficients associated with subsets of size $k$.
    Note that there are at most $n^k 16^k$ such Fourier coefficients.
    Thus, we use~\Cref{lem:estimate-fourier-coeff-channel} $\mc{O}(n^k 16^k)$ times, to obtain estimates $\alpha_{x,y}$, such that with probability at least $1-\delta$,
     \begin{equation}
     \label{eq:apt-juntas-fourier-estimates}
         |\alpha_{x,y}-\hat{\Phi}(x,y)| \leq \epsilon_1, \quad \forall x,y \in \pauli_S,\; S \subseteq[n],\; |S| = k.
     \end{equation}
     Using a union bound over failure probabilities of repeated applications of Lemma~\ref{lem:estimate-fourier-coeff-channel} for each Fourier coefficient, we see that this requires
     \begin{equation}
         N = \mc{O}\left(\frac{k \cdot 16^kn^k \log(n/\delta)}{\epsilon_1^2}\right).
     \end{equation}
     queries to $\Phi$ and runs in time $\mc{O}(N)$. In the following, we condition on successful estimation. Now, we use the estimated coefficients to compute estimates $\hat{w}_S$ for all subsets 
     of cardinality $k$. 
     \begin{equation}
         \hat{w}_S \triangleq \sum_{x, y \in \pauli_S} |\alpha_{x,y}|^2, \quad \forall S \subseteq [n], |S| =k.
     \end{equation}
     As we have already estimated all Fourier coefficients associated with subsets of cardinality $k$ by $\alpha_{x,y}$, we can now compute $\hat{w}_S$. As there are $\mc{O}(n^k)$ subsets of cardinality $k$, and computing each weight requires adding weights of $16^k$ coefficients, finding all the weights takes time $\mc{O}(n^k \cdot 16^k)$. 
     We now bound the error of these estimates.
     \begin{align}
          |\hat{w}_S - w_S(\Phi)| &= \left|\sum_{x,y \in \pauli_S} |\alpha_{x,y}|^2 - |\hat{\Phi}(x,y)|^2\right|
          \\&\leq \sum_{x,y \in \pauli_S} ||\alpha_{x,y}|^2 - |\hat{\Phi}(x,y)|^2|
          \\&= \sum_{x,y \in \pauli_S} ||\alpha_{x,y}| - |\hat{\Phi}(x,y)|| \cdot ||\alpha_{x,y}| + |\hat{\Phi}(x,y)||
          \\& \leq \sum_{x,y \in \pauli_S} |\alpha_{x,y} - \hat{\Phi}(x,y)| \cdot ||\alpha_{x,y}| + |\hat{\Phi}(x,y)||
          \\& \leq \sum_{x,y \in \pauli_S} \epsilon_1 \cdot 2
          \\& = \label{eq:apt-juntas-weight-estimates} 2 \cdot 16^k \epsilon_1, 
     \end{align}
     The first line follows from the definitions of the weights.
     The second line follows by the triangle inequality.
     The fourth line follows by the reverse triangle inequality.
     In the fifth line, we use \Cref{eq:apt-juntas-fourier-estimates} and the fact that the absolute value of a Fourier coefficient is at most 1.
     In the last line, we sum over all $16^k$ Fourier coefficients with support in $S$.
     
     Let the subset with the highest estimated weight be $S^\prime$.
     \begin{equation}
         S^\prime \triangleq \argmax_{S \subseteq [n], |S| = k} \hat{w}_S.
     \end{equation}
     Then, our hypothesis will be
     \begin{equation}
         \Psi \triangleq \sum_{x,y \in \pauli_{S^\prime}} \alpha_{x,y} \Phi_{x,y},
     \end{equation}
     where recall we denote $\Phi_{x,y} = \sigma_x (\cdot)\sigma_y$.
     This simply aims to approximate $\Phi$ by estimating all Fourier coefficients of the highest weight subset. It remains to bound the error of this hypothesis, i.e., $d_F(\Phi, \Psi)$. Denote by
     \begin{equation}
         \Phi_S \triangleq \sum_{x,y \in \pauli_S} \hat{\Phi}(x,y) \Phi_{x,y},
     \end{equation}
     the Fourier truncation of $\Phi$ restricted to the subset $S$, for some subset $S \subseteq [n]$.
     Now, using the triangle inequality for $d_F$, we have
     \begin{equation}
         d_F(\Phi, \Psi) \leq d_F(\Phi,  \Phi_{S^\prime}) + d_F(\Phi_{S^\prime}, \Psi).
     \end{equation}
     We can upper bound $d_F(\Phi,\Phi_{S^\prime})$ as 
     \begin{align}
         d_F(\Phi,\Phi_{S^\prime})^2 &= \frac{1}{2}\sum_{x,y \in \pauli_n} |\hat{\Phi}(x,y) - \hat{\Phi}_{S^\prime}(x,y)|^2
         \\&= \frac{1}{2}\sum_{x,y \in \pauli_{S^\prime}} |\hat{\Phi}(x,y) - \hat{\Phi}(x,y)|^2 + \frac{1}{2}\sum_{\substack{x,y \in \pauli_n\\x \notin \pauli_{S^\prime} \text{ or } y \notin \pauli_{S^\prime}}} |\hat{\Phi}(x,y)|^2
         \\&= \frac{\|\hat{\Phi}\|_F^2 - w_{S^\prime}(\Phi)}{2}
         \\&= \frac{\|\hat{\Phi}\|_F^2 - w_{S^*}(\Phi)}{2} +  \frac{w_{S^*}(\Phi) -  w_{S^\prime}(\Phi)}{2}
         \\&\leq \opt(\Phi,\mc{JC}_{n,k})^2 + \frac{w_{S^*}(\Phi) -  w_{S^\prime}(\Phi)}{2},
     \end{align}
     where the last inequality follows from \Cref{eq:apt-juntas-opt-weight}. To bound the differences in weights, observe that from \Cref{eq:apt-juntas-weight-estimates},
    \begin{equation}
        w_{S^\prime}(\Phi)+2\cdot16^k\epsilon_1 \geq \hat{w}_{S^\prime}\quad \text{ and } \quad \hat{w}_{S^\star} \geq w_{S^\star}(\Phi)-2\cdot16^k\epsilon_1.
    \end{equation}
     As $S^\prime$ is the set with the heaviest estimated weight, 
     \begin{equation}
         \hat{w}_{S^\prime} \geq \hat{w}_{S^\star}.
     \end{equation}
     Thus, we finally obtain
     \begin{equation}
     \label{eq:apt-juntas-claim-1}
         d_F(\Phi,\Phi_{S^\prime})^2 \leq \opt(\Phi,\mc{JC}_{n,k})^2 + 2\cdot16^k\epsilon_1 \leq \left(\opt(\Phi,\mc{JC}_{n,k}) + 4^k \sqrt{2\epsilon_1} \right)^2.
     \end{equation}
     Now, it remains to bound $d_F(\Phi_{S^\prime}, \Psi)$.
     \begin{align}
         d_F(\Phi_{S'}, \Psi)^2 = \frac{1}{2}\sum_{x,y \in \pauli_{S^\prime}} |\hat{\Phi}(x,y) - \alpha_{x,y}|^2 \leq \frac{16^k \epsilon_1^2}{2},
     \end{align}
     where the inequality follows by \Cref{eq:apt-juntas-fourier-estimates} and the fact that there are at most $16^k$ Fourier coefficients for $x,y \in\pauli_{S'}$ since $|S'| = k$. 
     Thus, with probability at least $1-\delta$, we have
     \begin{equation}
         d_F(\Phi, \Psi) \leq \opt(\Phi,\mc{JC}_{n,k}) + 4^k \sqrt{2\epsilon_1} + \frac{4^k \epsilon_1}{\sqrt{2}}.
     \end{equation}
     To obtain the 1-agnostic learning guarantee, we choose $\epsilon_1 = \mc{O}(\epsilon^2/16^k)$. This gives us the stated query complexity. To show the stated time complexity, we observe that it is dominated by that of learning the Fourier coefficients.
\end{proof}
\begin{remark}
\label{rmk:improper-junta-APT}
    When considering improper learning, we remark that APT with respect to the class of junta channels $\mc{JC}_{n,k}$ immediately implies APT with respect to the class of unitary juntas $\mc{J}_{n,k}$. To see this, we note that the class $\mc{J}_{n,k}$ is contained within $\mc{JC}_{n,k}$, and thus must have a higher optimal error.
\end{remark}
We discuss a potential method for taking this improper hypothesis and obtaining a proper one in \Cref{sec:proper-apt}.
\subsection{Low-degree channels}
\label{sec:apt-low-deg}
In this section, we present an algorithm for agnostic process tomography of low-degree channels.
We define a low-degree channel similarly to the definition of a low-degree unitary.
\begin{definition}[Low-degree channels]
    \label{def:channel-low-degree}
     A quantum channel $\Phi$ is said to have \emph{degree} $d$ if 
    \begin{equation}
       w^{> d}(\Phi) = 0,
    \end{equation}
    where $w^{>d}$ is defined in \Cref{def:weight-superop}.
    In other words, the degree of a channel $\Phi$ is the minimum integer $d$ such that if $|x|, |y| > d$, then $\hat{\Phi}(x,y) = 0$.
\end{definition}
We note a difference between our definition of degree-$d$ channels and that of~\cite{arunachalam2024learning}. Our definition requires that for all $\hat{\Phi}(x,y) \neq 0$, both $|x|$ and $|y|$ are at most $d$. On the other hand, the definition of~\cite{arunachalam2024learning} requires that for all non-zero $\hat{\Phi}(x,y)$, $|x| + |y| \leq d$. With our chosen definition, the channels corresponding to degree-$d$ unitaries (\Cref{def:unitary-low-degree}) are also channels of degree-$d$.
Thus, we find this to be a more natural definition, and we also use this fact in \Cref{rmk:improper-low-APT}. However, the definitions do not differ significantly, and one can show that an approach similar to \Cref{alg:apt-low-degree-channel} allows for agnostic tomography of low-degree channels adhering to the definition in~\cite{arunachalam2024learning}.  

Now, we will show that by learning all low-degree Fourier coefficients of the channel, we obtain an improper $1$-agnostic hypothesis.
This is similar to the first step of the junta channel learning algorithm (\Cref{thm:apt-juntas}).
The algorithm is detailed in \Cref{alg:apt-low-degree-channel}.

\begin{algorithm}
   \caption{1-Agnostic Improper Learning Degree-$d$ Channels} 
   \label{alg:apt-low-degree-channel}
   \begin{algorithmic}[1]
    \State Use $N$ queries to $\Phi$ to construct estimates $\alpha_{x,y}$ of the Fourier coefficients for all $x,y \in \pauli_n$ such that $|x|,|y| \leq d$.
    \State \Return $\Psi = \sum_{\substack{x,y\in \pauli_n\\|x|,|y|\leq d}} \alpha_{x,y} \sigma_x (\cdot) \sigma_y$.
   \end{algorithmic}
\end{algorithm}

\begin{theorem}[Agnostic Process Tomography of Low-Degree Channels in $d_F$.]
    \label{thm:apt-low-1}
    \Cref{alg:apt-low-degree-channel} performs improper $(1,\epsilon,\delta)$-agnostic process tomography of the class of quantum channels on $\mc{B}(\mc{H})$ of degree-$d$, $\mathcal{LC}_{n,d}$, with respect to error in $d_F$ using 
\begin{equation}
    N = \mc{O}\left(\frac{d \cdot 81^d n^{4d} \log(n/\delta)}{\epsilon^2}\right)
\end{equation}
queries to the unknown channel $\Phi$ and running in time
\begin{equation}
    \mathcal{O}(N).
\end{equation}
\end{theorem}
\begin{proof}
    The algorithm is detailed in \Cref{alg:apt-low-degree-channel}, and the analysis is similar to that of \Cref{thm:apt-juntas}.
    We start by lower bounding the optimal error for the class of low-degree channels. We have
    \begin{align}
        \opt(\Phi, \mc{LC}_{n,d})^2 &= \min_{\Psi \in \mc{LC}_{n,d}} d_F^2(\Phi,\Psi)  
        \\&= \frac{1}{2} \min_{\Psi \in \mc{LC}_{n,d}} \sum_{x,y \in \pauli_n} |\hat{\Phi}(x,y) - \hat{\Psi}(x,y)|^2
        \\&= \frac{1}{2}\min_{\Psi \in \mc{LC}_{n,d}} \sum_{\substack{x,y\in\pauli_n\\|x|,|y| \leq d}} |\hat{\Phi}(x,y) - \hat{\Psi}(x,y)|^2  + \frac{1}{2}\sum_{\substack{x,y\in\pauli_n\\|x| > d \text{ or } |y| > d}}|\hat{\Phi}(x,y)|^2
        \\& \geq 0 + \frac{1}{2}\sum_{\substack{x,y\in\pauli_n\\|x| > d \text{ or } |y| > d}}|\hat{\Phi}(x,y)|^2
        \\& = \frac{\|\hat{\Phi}\|_F^2 - w^{\leq d}(\Phi)}{2}\label{eq:apt-low-opt-lower}
    \end{align}
    where the second line follows from Lemma~\ref{lem:df-superop-fourier}.
    In the third line, we use that $\Psi$ has degree $d$.
    The final equality uses the definition of weight (\Cref{def:weight-superop}) and $\|\hat{\Phi}\|_F^2 = \sum_{x,y \in \pauli_n} |\hat{\Phi}(x,y)|^2$.
    
    Our algorithm (\Cref{alg:apt-low-degree-channel}) estimates all Fourier coefficients of degree at most $d$. Note that there are at most $\mc{O}(n^{2d} 9^d)$ of them.
    We use~\Cref{lem:estimate-fourier-coeff-channel} $\mc{O}(n^{2d} 9^d)$ times, to obtain estimates $\alpha_{x,y}$ such that
     \begin{equation}
        \label{eq:apt-low-estimate-coeff}
         |\alpha_{x,y}-\hat{\Phi}(x,y)| \leq \epsilon_1, \quad \forall x,y \in \pauli_n \text{ s.t. } |x|,|y| \leq d,
     \end{equation}
     with probability at least $1-\delta$.
     Using a union bound over failure probabilities of repeated applications of Lemma~\ref{lem:estimate-fourier-coeff-channel} for each Fourier coefficient, we see that this requires
     \begin{equation}
         N = \mc{O}\left(\frac{d \cdot 9^d n^{2d} \log(n/\delta)}{\epsilon_1^2}\right).
     \end{equation}
     queries to $\Phi$ and runs in time $\mc{O}(N)$. Now, we construct our hypothesis
     \begin{equation}
         \Psi \triangleq \sum_{\substack{x,y\in\pauli_n\\|x|,|y| \leq d}} \alpha_{x,y} \Phi_{x,y}.
     \end{equation}
     This simply aims to approximate $\Phi$ by estimating all low-degree coefficients. It remains to bound the error of this hypothesis, i.e., $d_F(\Phi, \Psi)$. Denote the low-degree Fourier truncation of $\Phi$ by
     \begin{equation}
         \Phi_d \triangleq \sum_{\substack{x,y\in\pauli_n\\|x|,|y| \leq d}} \hat{\Phi}(x,y) \Phi_{x,y}.
     \end{equation}
     Now, using the triangle inequality for $d_F$, we have
     \begin{equation}
         d_F(\Phi, \Psi) \leq d_F(\Phi,  \Phi_d) + d_F(\Phi_d, \Psi).
     \end{equation}
     We can upper bound $d_F(\Phi,\Phi_d)$ as 
     \begin{align}
         d_F(\Phi,\Phi_d)^2 &= \frac{1}{2}\sum_{x,y \in \pauli_n} |\hat{\Phi}(x,y) - \hat{\Phi}_d(x,y)|^2
         \\&= \frac{1}{2}\sum_{\substack{x,y\in\pauli_n\\|x|,|y| \leq d}} |\hat{\Phi}(x,y) - \hat{\Phi}(x,y)|^2 + \frac{1}{2}\sum_{\substack{x,y\in\pauli_n\\|x| > d \text{ or } |y| > d}} |\hat{\Phi}(x,y)|^2
         \\&= \frac{\|\hat{\Phi}\|_F^2 - w^{\leq d}(\Phi)}{2}
         \\&\leq \opt(\Phi,\mc{LC}_{n,d})^2.
     \end{align}
     The first line follows by \Cref{lem:df-superop-fourier}.
     The second line uses the definition of $\Phi_d$.
     The third line uses the definition of weight (\Cref{def:weight-superop}).
     Finally, the last line uses \Cref{eq:apt-low-opt-lower}.
     Now, conditioned on the successful estimation of $\alpha_{x,y}$, we can bound $d_F(\Phi_d, \Psi)$.
     \begin{align}
         d_F(\Phi_d, \Psi)^2 &= \frac{1}{2}\sum_{x,y \in \pauli_n} |\hat{\Phi}_d(x,y) - \alpha_{x,y}|^2
         \\&= \frac{1}{2}\sum_{\substack{x,y\in\pauli_n\\|x|, |y| \leq d}} |\hat{\Phi}(x,y) - \alpha_{x,y}|^2
         \\& \leq \mc{O}\left(n^{2d} 9^d\epsilon_1^2\right).
     \end{align}
     In the second line, we use the definition of $\Phi_d$.
     In the third line, we use \Cref{eq:apt-low-estimate-coeff} and that there are at most $\mc{O}(n^{2d} 9^d)$ Fourier coefficients of degree at most $d$.
     Thus, with probability at least $1-\delta$, we have
     \begin{equation}
         d_F(\Phi, \Psi) \leq \opt(\Phi, \mc{LC}_{n,d}) + \mc{O}\left(n^d3^d \epsilon_1\right).
     \end{equation}
     To obtain the 1-agnostic learning guarantee, we choose $\epsilon_1 = \mc{O}(\epsilon/n^d3^d)$. This gives us the stated query and time complexities.
\end{proof}
\begin{remark}
\label{rmk:improper-low-APT}
    Using a similar argument to \Cref{rmk:improper-junta-APT}, we note that improper APT with respect to the class of low-degree channels $\mc{LC}_{n,d}$ implies improper APT with respect to the class of low-degree unitaries $\mc{L}_{n,d}$.
\end{remark}
Similar to the case of junta channels, we will discuss a potential method for taking this improper hypothesis and obtaining a proper one in \Cref{sec:proper-apt}.
\subsection{$n$-to-1 QAC0 channels}
\label{sec:apt-qac0}
In this section, we consider the class of channels $\qacz$ formed by constant-depth circuits with arbitrary single-qubit gates and controlled phase-flip gates of arbitrary width~\cite{moore1999quantum}.
Following standard terminology, we refer to the \emph{size} of a quantum circuit as the number of multi-qubit gates, and the \emph {depth} as the number of layers of multi-qubit gates in the circuit.
We consider the algorithm of~\cite{nadimpalli2024pauli} and show that it is an improper 1-agnostic learning algorithm.
We follow the definition of~\cite{nadimpalli2024pauli} for the circuit class $\qacz$. 
\begin{definition}[$\qacz$ circuits; Definition 16 in~\cite{nadimpalli2024pauli}]
    \label{def:qac0-circuits}
    The class $\qacz$ contains quantum circuit families consisting of multiple-qubit $\mathsf{CZ}$ gates and arbitrary single-qubit gates, with depth bounded below a constant. That is, the circuit family $\{C_n\}_{n \in \mathbb{N}}$ is in $\qacz$ if there exists a $d \in \mathbb{N}$ such that for each $n \in \mathbb{N}$, $C_n$ is a quantum circuit on $n$ qubits consisting only of multiple-qubit $\mathsf{CZ}$ gates and arbitrary single-qubit gates, with depth at most $d$.
\end{definition}
The algorithm of~\cite{nadimpalli2024pauli} focuses on learning a specific class of channels formed by $\qacz$ circuits.
In particular, they consider the class of channels formed by tracing out all except a single qubit.
\begin{definition}[$n$-to-1 $\qacz$ channels; Definition 12 in~\cite{nadimpalli2024pauli}]
    \label{def:qac0-channels}
    Let $n, a \in \mathbb{N}$. Let $U$ be an $(n+a)$-qubit unitary implemented by a $\qacz$ circuit, and let $\psi$ be an $a$-qubit quantum state. We define the \emph{$n$-to-$1$ $\qacz$ channel $\mathcal{E}_{U, \psi}$} to be the following channel mapping $n$ qubits to $1$ qubit
    \begin{equation}
        \mathcal{E}_{U, \psi}(\rho) \triangleq \tr_{[n+a-1]}\left(U (\rho \otimes \psi) U^\dagger \right),
    \end{equation}
    where $\tr_{[n+a-1]}$ denotes the partial trace over all except the last qubit.
\end{definition}
We consider the Pauli spectrum expansion (see \Cref{def:pauli-spectrum}) of the Choi representation of such channels.
Namely, for an $n$-to-$1$ $\qacz$ channel $\mathcal{E}_{U, \psi}$, the Choi representation $J(\mathcal{E}_{U, \psi}) \in \mathcal{B}((\mathbb{C}^2)^{\otimes n+1})$ can be expressed as
\begin{equation}
    \label{def:choi-qacz-pauli}
    J(\mathcal{E}_{U, \psi}) = \sum_{x \in \pauli_{n+1}} \alpha_x \sigma_x
\end{equation}
One of the key technical results of~\cite{nadimpalli2024pauli} is that when expressing the Choi representation of $n$-to-1 $\qacz$ channels in the Pauli spectrum, the weight is concentrated on the low-degree terms.
Formally, we have the following lemma.
\begin{lemma}[Low-degree concentration of $\qacz$ channels; Theorem 18 in~\cite{nadimpalli2024pauli}]
\label{lem:qac0-pauli-conc}
    Let $U$ be a unitary implemented by an $(n+a)$-qubit $\qacz$ circuit of depth $d$ and size $s$. Let $\psi \in \mathcal{B}((\mathbb{C}^2)^{\otimes a})$ be an $a$-qubit quantum state. For each $k \in [n+1]$,
    \begin{equation}
        w^{> k}(J(\mathcal{E}_{U, \psi})) \triangleq \sum_{\substack{x \in \pauli_{n+1}\\|x| > k}} |\alpha_x|^2 \leq \mathcal{O}\left( s^2 2^{-k^{1/d}}\right) \cdot 2^a \| \psi\|_F^2.
    \end{equation}
\end{lemma}
Our distance metric for learning $\qacz$ channels is the normalized Frobenius distance, as defined in~\Cref{def:df-superop}, which is equal to the normalized Frobenius distance between Choi representations.
From~\Cref{lem:qac0-pauli-conc}, we see that for improper learning of the Choi representation within normalized Frobenius distance, one simply needs to learn all of its low-degree Pauli coefficients. 
We describe the algorithm in \Cref{alg:apt-qac0}.

\begin{algorithm}
   \caption{1-Agnostic Improper Learning $n$-to-$1$ $\qacz$ Channels} 
   \label{alg:apt-qac0}
   \begin{algorithmic}[1]
    \State Prepare $N$ copies of $v(\Phi)$.
    \State Use classical shadows for $v(\Phi)$ to obtain estimates $\hat{\alpha}_x$ of the Pauli coefficients for $x \in \pauli_{n+1}$ s.t. $|x| \leq \widetilde{k} = \log^d(s^4/\epsilon^4)$.
    \State \Return $A = \sum_{\substack{x \in \pauli_{n+1}\\ |x| \leq \widetilde{k}}} \hat{\alpha_x}\sigma_x$.
   \end{algorithmic}
\end{algorithm}

\begin{theorem}[Improper 1-APT of n-to-1 $\qacz$ channels in $d_F$.]
    \label{thm:qac0-apt}
    There exists a learning algorithm for improper $(1,\epsilon,\delta)$-agnostic process tomography of $\qacz(n, a, s, d)$, the class of n-to-1 $\qacz$ channels $\{\mathcal{E}_{U, |0^a\rangle}\}$, where the unitaries are implemented by $(n+a)$-qubit $\qacz$ circuits with size $s$ and depth $d$, and $a \leq 2 \log (s/\epsilon)$, with respect to error in $d_F$, using
    \begin{equation}
        N = n^{\mathcal{O}(\log^d(s^4/\epsilon^4))} \cdot \log(1/\delta)
    \end{equation}
    queries to the unknown channel $\Phi$ and running in time
    \begin{equation}
        n^{\mathcal{O}(\log^d(s^4/\epsilon^4))} \cdot \log(1/\delta).
    \end{equation}
\end{theorem}
We define the following classes of low-degree operators, which we agnostically learn as part of our proof of~\Cref{thm:qac0-apt}.
\begin{definition}[Low-degree operators]
    \label{def:op-low-degree}
    Define the class of \emph{$k$-degree, $n+1$-qubit bounded linear operators} $\mathcal{O}_k \subseteq \mc{B}((\mathbb{C}^2)^{\otimes n+1})$ as
    \begin{equation}
        \mathcal{O}_k \triangleq \{A \in \mc{B}((\mathbb{C}^2)^{\otimes n+1}) : w^{>k}(A) = 0\}.
    \end{equation}
\end{definition}
In order to prove~\Cref{thm:qac0-apt}, we will make use of the following lemma on learning the closest low-degree operator to the Choi representation of an unknown channel.
\begin{lemma}[Agnostic-learning a low-degree operator]
    \label{lem:agnostic-operator}
    Let $\epsilon, \delta > 0$. Given copies of the Choi-state $v(\Phi)$ of an unknown quantum channel $\Phi$ from $\mathcal{B}((\mathbb{C}^2)^{\otimes n})$ to $\mathcal{B}(\mathbb{C}^2)$, there exists an algorithm that outputs a hypothesis $A \in \mathcal{O}_k$, such that
    \begin{equation}
        d_F(J(\Phi), A) \leq \min_{B \in \mathcal{O}_k} d_F(J(\Phi), B) + \epsilon,
    \end{equation}
    with probability at least $1-\delta$. Moreover, this algorithm uses
    \begin{equation}
        N = \frac{n^{\mc{O}(k)} \log(1/\delta)}{\epsilon^2}
    \end{equation}
    copies of $v(\Phi)$ and runs in time
    \begin{equation}
         k n^{\mc{O}(k)}N.
    \end{equation}
\end{lemma}
\begin{proof}
    Recall from~\Cref{def:choi-qacz-pauli} that we can write $J({\Phi})$ in the Pauli basis as
    \begin{equation}
        J(\Phi) = \sum_{x \in \pauli_{n+1}} \alpha_x \sigma_x.
    \end{equation}
    Now, we compute the optimal error over the class of low-degree operators $\mathcal{O}_k$.
    \begin{align}
        \min_{B \in \mathcal{O}_k} d_F(J(\Phi), B)^2 &= \min_{B \in \mathcal{O}_k} \sum_{x \in \pauli_{n+1}} \frac{|\alpha_x - \hat{B}_x|^2}{2}
        \\&= \min_{B \in \mathcal{O}_k} \sum_{\substack{x \in \pauli_{n+1}\\|x| \leq k}} \frac{|\alpha_x - \hat{B}_x|^2}{2} + \sum_{\substack{x \in \pauli_{n+1}\\|x| >  k}} \frac{|\alpha_x|^2}{2}
        \\& \geq \min_{B \in \mathcal{O}_k} \sum_{\substack{x \in \pauli_{n+1}\\|x| >  k}} \frac{|\alpha_x|^2}{2}
        \\&= \frac{w^{> k}(J(\Phi))}{2}.
    \end{align}
    In the first line, we use \Cref{lem:df-pauli}.
    In the second line, we use that $B$ is low-degree.
    One can check that this lower bound is achieved by the low-degree truncation of $J(\Phi)$, i.e., the operator
    \begin{equation}
        B^* \triangleq \sum_{\substack{x\in\pauli_{n+1}\\|x| \leq k}} \alpha_x \sigma_x
    \end{equation}
    has error
    \begin{equation}
    \label{eq:low-deg-operator-opt}
        \min_{B \in \mathcal{O}_k} d_F(J(\Phi), B) = d_F(J(\Phi), B^*) = \sqrt{\frac{w^{> k}(J(\Phi))}{2}}.
    \end{equation}
    Thus, it suffices to learn the low-degree truncation of $J(\Phi)$.
    First, we observe that because $\Phi$ is trace-preserving, $\tr(J(\Phi)) = 2^n$, the dimension of the input system. Thus, we have
    \begin{equation}
        v(\Phi) = \frac{J(\Phi)}{2^n} = \frac{1}{2^n}\sum_{x \in \pauli_{n+1}} \alpha_x \sigma_x.
    \end{equation}
    To estimate the Pauli coefficients $\alpha_x$, we use classical shadows (\Cref{lem:classical-shadows}) to estimate the expectation values of the corresponding Pauli $\sigma_x$ with respect to the Choi state $v(\Phi)$.
    These expectation values indeed tell us information about the Pauli coefficients: for any $y \in \pauli_{n+1}$,
    \begin{align}
        \tr(v(\Phi) \sigma_y) &= \sum_{x \in \pauli_{n+1}} \frac{\alpha_x}{2^n} \tr(\sigma_x \sigma_y) \\&= \sum_{x \in \pauli_{n+1}}  \frac{\alpha_x}{2^n} \delta_{x, y} \tr(I^{\otimes n+1})
        \\&= 2 \alpha_y.\label{eq:exp-val}
    \end{align}
    Thus, by estimating the expectation values of low-degree Paulis with respect to the Choi state $v(\Phi)$, we can obtain estimates of the Pauli coefficients of $J(\Phi)$.
    We need to estimate $n^{\mathcal{O}(k)}$ Paulis of degree at most $k$. Let the desired error in each estimate be at most $\epsilon_1 > 0$ and the overall failure probability be at most $\delta > 0$.
    Then, from~\Cref{lem:classical-shadows}, we see that this can be done using
    \begin{equation}
        N = \mc{O}\left(\frac{k \cdot 3^k \log(n/\delta)}{\epsilon_1^2}\right)
    \end{equation}
    copies of $v(\Phi)$ and in time
    \begin{equation}
        k n^{\mc{O}(k)}N.
    \end{equation}
    Suppose this procedure gives us estimates $\hat{\beta}_x$, for all strings $x \in \pauli_{n+1}$ of degree at most $k$. By \Cref{lem:classical-shadows}, we have
    \begin{equation}
        |\hat{\beta}_x - \tr(v(\Phi)\sigma_x)| \leq \epsilon_1, \quad \forall x \in \pauli_{n+1} \text{ s.t. } |x| \leq k
    \end{equation}
    with probability at least $1-\delta$.
    By \Cref{eq:exp-val}, we then have
    \begin{equation}
        \label{eq:apt-qac0-est-pauli}
        |\hat{\beta}_x - 2\alpha_x| \leq \epsilon_1, \quad \forall x \in \pauli_{n+1} \text{ s.t. } |x| \leq k
    \end{equation}
    Then, define our hypothesis by
    \begin{equation}
        A \triangleq \sum_{\substack{x \in \pauli_{n+1}\\|x| \leq k}} \frac{\hat{\beta}_x}{2} \sigma_x.
    \end{equation}
    It remains to bound the error of this hypothesis.
    \begin{align}
        d_F(J(\Phi), A) &\leq d_F(J(\Phi), B^*) + d_F(B^*, A)
        \\& \leq \min_{B \in \mathcal{O}_k} d_F(J(\Phi), B) + \frac{1}{\sqrt{2}} \sqrt{\sum_{\substack{x \in \pauli_{n+1}\\|x| \leq k}}\left|\alpha_x - \frac{\hat{\beta}_x}{2}\right|^2}
        \\& \leq \min_{B \in \mathcal{O}_k} d_F(J(\Phi), B) + n^{\mathcal{O}(k)}\frac{\epsilon_1}{\sqrt{2}}.
    \end{align}
    In the first line, we use the triangle inequality for $d_F$.
    In the second line, we use the definition of $B^*$ and $A$ and \Cref{lem:df-pauli}.
    In the third line, we use \Cref{eq:apt-qac0-est-pauli} and the fact that there are at most $n^{\mathcal{O}(k)}$ Pauli coefficients of degree at most $k$.
    Then, setting $\epsilon_1 = 
    \epsilon/n^{\mathcal{O}(k)}$, we obtain the desired bound
    \begin{equation}
        d_F(J(\Phi), A) \leq \min_{B \in \mathcal{O}_k} d_F(J(\Phi), B) + \epsilon.
    \end{equation}
    Our choice of $\epsilon_1$ can be used to derive the sample and time complexities.
\end{proof}

With this, we can prove Theorem~\ref{thm:qac0-apt}.

\begin{proof}[Proof of Theorem~\ref{thm:qac0-apt}]
    The algorithm is described in \Cref{alg:apt-qac0}.
    For the unknown quantum channel $\Phi$, let $\Phi^*$ be the closest channel from the class of channels $\qacz(n,a,s,d)$, as defined in the theorem statement. 
    In other words, $\Phi^*$ achieves $\mathsf{opt}(\Phi, \qacz(n,a,s,d))$.
    Then, from~\Cref{lem:qac0-pauli-conc}, we know that the Choi representation of this channel satisfies
    \begin{equation}
        w^{>k}(J (\Phi^*)) \leq \mc{O}\left(s^2 2^{-k^{1/d}+a} \right).
    \end{equation}
    Then, setting $\widetilde{k} \triangleq \log^d(s^4/\epsilon^4)$, and using $a \leq 2\log (s / \epsilon) = \frac{1}{2} \widetilde{k}^{1/d}$, we have
    \begin{equation}
    \label{eq:qac0-choi-weight}
        w^{> \widetilde{k}}(J (\Phi^*)) \leq \mc{O}(\epsilon^2).
    \end{equation}
    Our algorithm (\Cref{alg:apt-qac0}) is simply to apply~\Cref{lem:agnostic-operator} to the Choi state $v(\Phi)$ of the channel $\Phi$.
    This outputs a low-degree operator $A \in \mc{O}_{\widetilde{k}}$, which is also our final hypothesis.
    We can bound the error $d_F(J(\mathcal{E}), A)$ as
    \begin{align}
        d_F(J(\Phi), A) &\leq \min_{B \in \mc{O}_{\widetilde{k}}} d_F(J(\Phi), B) + \epsilon
        \\&\leq \min_{B \in \mc{O}_{\widetilde{k}}} d_F(J(\Phi), J(\Phi^*)) + d_F(J(\Phi^*), B) + \epsilon
        \\& \leq d_F(J(\Phi), J(\Phi^*)) + \min_{B \in \mc{O}_{\widetilde{k}}} d_F(J(\Phi^*), B) + \epsilon
        \\& = d_F(J(\Phi), J(\Phi^*)) + \sqrt{\frac{w^{> \widetilde{k}}(J (\Phi^*))}{2}} + \epsilon
        \\& \leq d_F(J(\Phi), J(\Phi^*)) + \mc{O}(\epsilon)\\
        &= \min_{\Psi \in \qacz(n,a,s,d)} d_F(J(\Phi), J(\Psi)) + \mathcal{O}(\epsilon).
    \end{align}
    where the second line is by triangle inequality.
    The fourth line follows from~\Cref{eq:low-deg-operator-opt}, and the last line is due to~\Cref{eq:qac0-choi-weight}.
    Due to the equivalence of $d_F$ between channels and their Choi representations, agnostic learning the Choi representation is the same as agnostic learning a description of the channel.
    Thus, by rescaling $\epsilon$ by an appropriate constant, we obtain the required improper $1$-agnostic learning bound.
    To obtain the stated sample and time complexities, we simply plug in the values of $\widetilde{k}$ and $\epsilon$ in the bounds of~\Cref{lem:agnostic-operator}. 
\end{proof}
Unlike the case for APT of junta channels and low-degree channels, the method discussed in \Cref{sec:proper-apt} does not help us obtain a proper hypothesis for $\qacz$ channels. As discussed in \cite{nadimpalli2024pauli}, one can map the problem of obtaining the nearest quantum channel from the improper hypothesis to a convex optimization problem. However, it is not clear how to obtain the nearest quantum channel that can be implemented using a $\qacz$ circuit. As a result, we only achieve improper agnostic process tomography of $\qacz$ channels.
\subsection{Proper APT}
\label{sec:proper-apt}
In this section, we discuss an approach towards proper learning for APT based on convex optimization.
In particular, we aim to devise a projection algorithm which takes as input an improper hypothesis output by one of our algorithms (e.g., \Cref{sec:apt-junta,sec:apt-low-deg}) and returns a valid quantum channel.
We are unable to fully solve this problem due to a difficult-to-verify condition from~\cite{wang2024convergence}, which we rephrase in our setting in as Conjecture~\ref{conj:unsolved-condition}.
Nonetheless, we consider this as a significant first step towards resolving this open question, which is also recurrent in much of the realizable process tomography literature that uses Pauli spectrum analysis~\cite{nadimpalli2024pauli,chen2023testing,bao2023testing}.
Note that ideas from the quasiprobability decomposition and probabilistic error cancellation which approximate linear superoperators by linear combinations of quantum channels~\cite{jiang2021physical,regula2021operational,temme2017error,van2023probabilistic,zhao2023power,horodecki2002method,horodecki2003limits,fiuravsek2002structural,korbicz2008structural} do not help because we need to find exactly a quantum channel.

Suppose that we are given an improper hypothesis $\Psi_1$, from an agnostic process tomography algorithm of a concept class $\mc{C}$ and an unknown channel $\Phi$, with respect to the normalized Frobenius distance $d_F$.
Denote the Fourier coefficient matrix of $\Phi$ by $\hat{\Phi}$. 
Moreover, suppose that our hypothesis is given to us as a Hermitian matrix of Fourier coefficients, $\hat{\Psi}_1$.
This aligns with our learning algorithms from \Cref{sec:apt-junta,sec:apt-low-deg}, where we learn the Fourier coefficients corresponding to a principle submatrix of $\hat{\Phi}$ and simply set the rest to $0$.
To ensure that $\hat{\Psi}_1$ is Hermitian, we can instead only learn the upper triangular coefficients of this principle submatrix, and set the corresponding coefficients in the lower triangle as their complex conjugates.
From the improper APT guarantee, we have
\begin{equation}
\label{eq:proper-initial-bound}
    d_F(\Phi, \Psi_1) \leq \opt(\Phi, \mc{C}) + \epsilon_1.
\end{equation}
for some $\epsilon_1 > 0$.
Our goal in the rest of this section will be to construct a \emph{proper} hypothesis satisfying a similar error guarantee.
We break this up into several steps in the following discussion, which we outline below.
\begin{enumerate}
    \item First, we derive constraints that the Fourier coefficient matrix must satisfy in order to be a valid quantum channel.
    Thus, we can formulate our projection task as a (complex) convex optimization problem.
    \item Second, we map this problem to another convex optimization problem over real, symmetric matrices to allow us to leverage powerful tools in the optimization literature to solve it.
    \item\label{item:key-cond} Next, we phrase our problem as a well-studied convex optimization problem with known algorithms and discuss the key condition needed for fast convergence.
    \item Although we may have fast convergence, in a finite number of steps, this does not guarantee \emph{exact} convergence to a quantum channel. By mixing the output of the optimization algorithm with the maximally mixed state, we can guarantee that the resulting matrix corresponds to a quantum channel.
    \item Finally, we complete the rigorous error analysis for this proper hypothesis.
\end{enumerate}

Note that Step~\ref{item:key-cond} is the only step where our proof is incomplete.
Namely, we identify this key mathematical condition (see Conjecture~\ref{conj:unsolved-condition}) but are unable to prove (or disprove) that it holds for our problem.

\paragraph{Optimization constraints for quantum channels.}
First, we derive the constraints that the Fourier coefficients of any quantum channel must satisfy.
Recall that the Choi representation of a quantum channel can be written as
\begin{equation}
    J(\Phi) = \sum_{x,y} \hat{\Phi}(x,y) J(\Phi_{x,y}), 
\end{equation}
where $\Phi_{x,y} \triangleq \sigma_x (\cdot) \sigma_y$.
From~\Cref{lem:channel-choi-equiv}, we have that 
\begin{equation}
    \tr_1(J(\Phi)) = I.
\end{equation}
We first show the effect of tracing out the first register on the Choi representations $J(\Phi_{x,y})$:
\begin{align}
    \tr_1(J(\Phi_{x,y})) &= \tr_1 \left(\sum_{a,b} \sigma_x | a \rangle \langle b |  \sigma_y \otimes | a \rangle \langle b |\right)
    \\&= \sum_{a,b} \tr(\sigma_x | a \rangle \langle b | \sigma_y) | a \rangle \langle b |
    \\&=  \sum_{a,b} (\langle b | \sigma_y \sigma_x | a \rangle) | a \rangle \langle b |
    \\&= (\sigma_y \sigma_x)^\intercal 
    \\&= \sigma_x^\intercal \sigma_y^\intercal 
    \\&= \sigma_x^* \sigma_y^*,
\end{align}
where in the last line we use the fact that all Pauli strings are Hermitian.
It is easy to verify that complex conjugation leaves the Paulis $I, X,$ and $Z$ invariant, while $Y^* = -Y$.
Thus, we see that $\sigma_x^* \sigma_y^*$ is equal to $\sigma_x \sigma_y$ up to a phase.
We define the phase $\theta(x,y)$ by the following equation
\begin{equation}
    \sigma_x^* \sigma_y^* = e^{i\theta(x,y)} \sigma_x \sigma_y.
\end{equation}
Further, we define the phase $\phi(x,y)$ by the following equation.
\begin{equation}
    \sigma_x \sigma_y = e^{i \phi(x,y)} \sigma_z.
\end{equation} which is associated with a unique Pauli string $z$ for all pairs $x,y$. Now, we can rewrite the partial trace of the Choi representation $J(\Phi)$ as follows.
\begin{align}
    I &= \tr_1(J(\Phi)) \\&= \sum_{x,y} \hat{\Phi}(x,y) \tr_1(J(\Phi_{x,y}))
    \\&= \sum_{x,y} \hat{\Phi}(x,y) e^{i \theta(x,y)} \sigma_x \sigma_y. 
\end{align}
By summing over the terms with $x = y$, we obtain the following constraint
\begin{equation}
    \label{eq:cptp-proj-constr-1}
    \sum_{x} \hat{\Phi}(x,x) = 1, 
\end{equation}
which can also be seen from~\Cref{lem:superop-fourier-properties}. Now, for each non-identity Pauli string $z$, we obtain the following constraint
\begin{equation}
\label{eq:cptp-proj-constr-2}
    \sum_{\substack{x,y\\ \sigma_x \sigma_y = e^{i \phi(x,y)} \sigma_z}} e^{i\theta(x,y)}e^{i \phi(x,y)}\hat{\Phi}(x,y) = 0.
\end{equation}
Finally, from~\Cref{lem:channel-choi-equiv}, we have
\begin{equation}
\label{eq:cptp-proj-constr-3}
    \hat{\Phi} \succcurlyeq 0.
\end{equation}
Now, the original task of projecting $\hat{\Psi}$ reduces to the following optimization problem with respect to the matrix $\hat{\Phi}$.
\begin{equation}
    \min_{\hat{\Phi}} \|\hat{\Psi} - \hat{\Phi}\|_F^2
\end{equation}
subject to the constraints given by~\Cref{eq:cptp-proj-constr-1,eq:cptp-proj-constr-2,eq:cptp-proj-constr-3}. For ease of notation, we will denote the dimensions of $\hat{\Psi}$ as $D \times D$ for the rest of this discussion and specify them for the concept classes we consider later.

\paragraph{Mapping to Real, Symmetric Matrices.}
This problem can be seen as the complex, Hermitian generalization of a corresponding problem in convex optimization over real, symmetric matrices.
We wish to map the problem to one with respect to real, symmetric matrices to leverage tools in convex optimization literature, which typically only deal with this case.
To this end, note that any complex matrix $A$ can be decomposed as a sum of its real and complex parts, $A \triangleq X + iY$, where $X$ and $Y$ are both real.
Then, we can map $A$ to the following real operator
\begin{equation}
    M = \begin{pmatrix}
        X & -Y \\
        Y & X
    \end{pmatrix}
\end{equation}
This transformation is well-known, first noted in~\cite{goemans2001approximation}, and can be used to transform any complex semidefinite programming problem into its real analogue.
It can be shown that this transformation preserves the properties of the original operator.
For example, a Hermitian operator is mapped to a symmetric one and a Hermitian positive semi-definite operator is mapped to a symmetric positive semi-definite one. Using this transformation, we see that it suffices to solve the following convex optimization problem over reals:
\begin{equation}
    \min_{Q} \|P - Q\|_F^2 \quad \text{s.t.}\quad Q \succcurlyeq 0,\; \tr(A_i^\intercal Q) = b_i,\;\forall i\in[2D],
\end{equation}
where $P,Q, A_i \in \mathbb{R}^{2D \times 2D}$ and $b \in \mathbb{R}^{2D}$, and $P$ is a symmetric operator.
Here, the $A_i$ matrices and the $b$ vector need to be chosen suitably to encompass our constraints in~\Cref{eq:cptp-proj-constr-1,eq:cptp-proj-constr-2,eq:cptp-proj-constr-3} after the real mapping.
It is clear that our problem can be written this way for a suitable choice of $A_i, b$.
This is the problem of finding the nearest projection (with respect to the Frobenius norm) of a symmetric operator onto the intersection of the positive semidefinite cone and an affine subspace.

\paragraph{Convex optimization algorithms and convergence.}
In particular, we wish to project onto the intersection of two convex sets, which we denote as $C_1$ and $C_2$.
Explicitly,
\begin{equation}
    C_1 \triangleq S_{2D}^+,\quad C_2 \triangleq \{Q \in \mathbb{R}^{2D \times 2D} : \tr(A_i^\intercal Q) = b_i,\; \forall i \in [2D]\},
\end{equation}
where $S_{2D}^+$ denotes the positive semidefinite (PSD) cone of $2D \times 2D$ matrices.
We can equivalently consider the vectorizations of these sets, where the vectorization of a matrix is defined by combining all columns of the matrix into a single vector.
Vectorization can be inverted when the dimensions of the original matrix are known.
We denote the vectorization operation as $\mathsf{vec}$ and its inverse as $\mathsf{vec}^{-1}$.
Then, the sets $C_1$ and $C_2$ can be written as
\begin{equation}
    C_1 = \{x \in \mathbb{R}^{4D^2} : \mathsf{vec}^{-1}(x) \in S_{2D}^+\},\quad C_2 = \{x \in \mathbb{R}^{4D^2} : \tilde{A}x = \tilde{b}\},
\end{equation}
where we use $\tilde{A}$ and $\tilde{b}$ to denote the constraints for the vectorized version of $Q$ (i.e., $x$ should be though of as $\mathsf{vec}(Q)$ for $C_2$).

Note that projecting onto either $C_1$ or $C_2$ individually is easy (running in time $\mathrm{poly}(D)$) since they are simply the PSD cone and a polyhedral set (see, e.g., \cite{henrion2012projection}).
Note that the runtime being in terms of $D$ is not a problem for our applications, as we discuss at the end of this section.
Thus, the problem is as follows: how can we use projections onto $C_1$ and $C_2$ individually to project onto $C_1 \cap C_2$?
This problem is well-studied in convex optimization, where researchers have developed methods such as alternating projections~\cite{von1949rings,von1950functional} and Dykstra's algorithm~\cite{dykstra1983algorithm}.
As the name suggests, at a high level, these algorithms work by projecting the input onto $C_1$ and $C_2$ in turn.
Classical results in optimization theory show that Dykstra's algorithm converges to the desired projection~\cite{dykstra1983algorithm,boyle1986method,bauschke1994dykstra,han1988successive,bauschke2000dykstras}, and its variants are also widely studied~\cite{bauschke2020dykstra,deutsch1994rate,gaffke1989cyclic,hundal1997two,pang2015set}; however, researchers have only recently rigorously analyzed its convergence \emph{rate} for non-polyhedral sets~\cite{wang2024convergence}.
Convergence rates have previously been analyzed for the basic alternating projections algorithm~\cite{deutsch2001best,bauschke1993convergence,lewis2009local} and Dykstra's algorithm for projection onto the intersection of polyhedral sets~\cite{deutsch1994rate,tseng2009coordinate,luo1993error}.
However, the former only guarantees convergence to a point in $C_1 \cap C_2$, rather than the closest one, which would not help for our purposes since we require an error bound.
Also, the latter does not apply to our setting, as $C_1 = S_{2D}^+$ is clearly not polyhedral.
In our setting, using Dykstra's algorithm, one would hope for linear convergence, as this leads to a computationally efficient (in the size of the input matrix) projection algorithm.
In fact,~\cite{wang2024convergence} shows that Dykstra's algorithm achieves a linear convergence rate under certain conditions.
Namely, they prove the following theorem.

\begin{theorem}[Theorem 5.3(i) in~\cite{wang2024convergence}; Simplified]
    \label{thm:converge}
    Consider the optimization problem
    \begin{equation}
        \min_{x \in \mathbb{R}^n} \frac{1}{2}\norm{x - x_0}_2^2 \quad \text{s.t.}\quad x \in \bigcap_{i=1}^\ell C_i,
    \end{equation}
    where $\ell$ is the number of convex sets $C_i$ and $x_0 \in \mathbb{R}^n$ is the initial point to be projected.
    Suppose the following three conditions hold.
    \begin{enumerate}
        \item \label{item:1} Each $C_i$ is a $C^{1,1}$-cone reducible closed convex set.
        \item \label{item:2} $\bigcap_{i=1}^\ell \mathrm{ri}(C_i) \neq \emptyset$, where $\mathrm{ri}$ denotes the relative interior.
        \item \label{item:3} $0 \in x^* - x_0 + \mathrm{ri}\left(\partial\left(\sum_{i=1}^\ell \delta_{C_i}\right)(x^*)\right)$, where $x^*$ is the optimal projected point, $\partial$ denotes the subdifferential, and $\delta_{C_i}$ is the indicator function of the set $C_i$ ($0$ for inputs inside $C_i$ and $+\infty$ otherwise).
    \end{enumerate}
    Then, Dykstra's algorithm produces iterates that linearly converge to the optimal point $x^*$.
\end{theorem}

Note that~\cite{wang2024convergence} considers a more general problem, in which they wish to project onto a point such that a linear transformation of this point lies in the intersection of the convex sets.
For our setting, we can take this transformation to be the identity, in which case the algorithm from~\cite{wang2024convergence} reduces to Dykstra's algorithm.
Thus, this theorem would give the convergence result we need if its conditions hold.

Now, let us discuss each of the conditions.
First, for Condition~\ref{item:1}, the exact definition of $C^{1,\alpha}$-cone reducibility, for $\alpha \in (0,1]$, is outside of the scope of this work, and we refer to Definition 4.1 in~\cite{wang2024convergence} for the definition.
Importantly, this is an extension of the more common notion of $C^2$-cone reducibility~\cite{shapiro2003sensitivity,bonnans2013perturbation}.
Specifically, ~\cite{wang2024convergence} states that any $C^2$-cone reducible set is also $C^{1,\alpha}$-cone reducible for all $\alpha \in (0, 1]$.
Moreover,~\cite{shapiro2003sensitivity} shows that all polyhedrons and the cone of positive semidefinite matrices are both $C^2$-cone reducible.
Hence, our convex sets $C_1$ and $C_2$ are both $C^2$-cone reducible and hence $C^{1,\alpha}$-cone reducible for all $\alpha \in (0,1]$.
In particular, $C_1$ and $C_2$ are both $C^{1,1}$-cone reducible, satisfying Condition~\ref{item:1}.

For Condition~\ref{item:2}, this is a standard assumption in optimization theory, which simply says that there is a strictly feasible point.
This is satisfied for our setting.

Finally, we turn to Condition~\ref{item:3}.
Let us define some of the notation more properly. Here, $\delta_{C_i}$ is the indicator function of the set $C_i$:
\begin{equation}
    \delta_{C_i}(x) \triangleq \begin{cases}
    0 & x \in C_i\\
    +\infty & x \notin C_i
    \end{cases}.
\end{equation}
Also, $\partial$ denotes the subdifferential.
For a function $f: \mathbb{R}^{4D^2} \to \mathbb{R}$, the subdifferential is defined as
\begin{equation}
    \partial f(x) \triangleq \{z \in \mathbb{R}^{4D^2} : f(y) \geq f(x) + (y - x)^\intercal z \quad \forall y \in \mathbb{R}^{4D^2}\}.
\end{equation}
Finally, the relative interior for a set $S \subseteq \mathbb{R}^{4D^2}$ is defined as
\begin{equation}
    \mathrm{ri}(S) \triangleq \{x \in S : \forall y \in S, \; \exists \lambda > 1 \text{ s.t. } \lambda x + (1-\lambda)y \in S\}.
\end{equation}

Because of Condition~\ref{item:2}, standard results in subdifferential calculus (see, e.g.,~\cite{nesterov2018lectures}) state that
\begin{equation}
    \partial (\delta_{C_1} + \delta_{C_2}) = \partial \delta_{C_1} + \partial \delta_{C_2}.
\end{equation}
Thus, we need to compute the subdifferentials of the indicator functions of our convex sets $C_1$ and $C_2$.
It is well-known in convex optimization that the subdifferential of an indicator function of a set $S$ is the normal cone of $S \subseteq \mathbb{R}^{4D^2}$, where the normal cone is defined as
\begin{equation}
    N_S(y) \triangleq \{a \in \mathbb{R}^{4D^2} : a^\intercal y \geq a^\intercal x, \quad \forall x \in S\}.
\end{equation}
The normal cone of the cone of positive semidefinite matrices is known to be
\begin{equation}
    \delta_{C_1}(X) = N_{C_{1}}(X) = \{C \in \mathbb{R}^{2D \times 2D} : C \preccurlyeq 0, \tr(CX) = 0\},
\end{equation}
where here $X \in \mathbb{R}^{2D \times 2D}$ are matrices.
Writing in terms of vectors using vectorization, we can write
\begin{equation}
    \delta_{C_1}(x) = \{c \in \mathbb{R}^{4D^2} : \mathsf{vec}^{-1}(c) \preccurlyeq 0, \tr(\mathsf{vec}^{-1}(c)\mathsf{vec}^{-1}(x)) = 0\},
\end{equation}
where $x \in \mathbb{R}^{4D^2}$.
Moreover, we can compute the normal cone of the polyhedral set $C_2 = \{x \in \mathbb{R}^{4D^2} : \tilde{A}x = \tilde{b}\}$:
\begin{align}
    \delta_{C_2}(x) &= N_{C_2}(x)\\
    &= \{g : g^\intercal (x - y) \geq 0, \quad \forall y\;\text{ s.t. } \tilde{A}y = \tilde{b}\}\\
    &= \{g : g^\intercal z \geq 0, \quad \forall z \in \mathrm{ker}(\tilde{A})\}\\
    &= (\mathrm{ker}(\tilde{A}))^\perp\\
    &= \mathrm{im}(\tilde{A}^\intercal).
\end{align}
Here, the second line is the definition of the normal cone.
The third line follows because $x - y$ is in the kernel of the matrix $\tilde{A}$ since both $x,y \in C_2$.
The fifth line follows by the standard result in linear algebra that the orthogonal complement of the kernel space is the image of the transpose.

Now, putting everything together, Condition~\ref{item:3} is as follows.
\begin{conjecture}
\label{conj:unsolved-condition}
Let $x_0 \in \mathbb{R}^{4D^2}$ be the vectorization of the real-mapped matrix of Fourier coefficients that we wish to project onto that of a valid quantum channel. Let $x^* \in \mathbb{R}^{4D^2}$ denote the vectorization of the real-mapped matrix of Fourier coefficients of the closest valid quantum channel to $x_0$ in $2$-norm.
Then, it remains to show the following condition holds.
    \begin{equation}
        x_0 - x^* \in \mathrm{ri}\left(\left\{c + \tilde{A}^\intercal y : c,y \in \mathbb{R}^{4D^2}, \mathsf{vec}^{-1}(c) \preccurlyeq 0, \tr(\mathsf{vec}^{-1}(c)\mathsf{vec}^{-1}(x^*)) = 0\right\}\right),
    \end{equation}
where $\mathrm{ri}$ denotes the relative interior.
\end{conjecture}

We are unsure if this condition holds (or does not hold) in our setting, but suppose for the rest of this discussion that it does.
Now, by~\Cref{thm:converge}, Dykstra's algorithm converges linearly to the point $x^*$, which is the vectorization of the matrix of Fourier coefficients (after being mapped to a real, symmetric matrix) of the closest valid quantum channel to $x_0$.
Recall that Dykstra's algorithm works by repeatedly projecting alternately onto $C_1$ and then $C_2$ (with a correction term).
By linear convergence, this converges to some point $\epsilon_2$-close in $2$-norm to $x^*$ in $\mc{O}(\log(1/\epsilon_2))$ iterations, for some $\epsilon_2 > 0$.
Moreover, suppose without loss of generality that the final projection applied is the one projecting onto $C_2$. We will make use of this property in the next step.

Mapping the output of the algorithm back to our original problem for Fourier coefficients of channels gives us a Fourier coefficient matrix $\hat{\Psi}_2$. Then, we obtain the following guarantee for $\hat{\Psi}_2$.
\begin{equation}
\label{eq:proper-dykstra-guarantee-1}
    \frac{\|\hat{\Psi}_2 - \hat{\Psi}^*\|_F}{2} \leq \epsilon_2,
\end{equation}
where
\begin{equation}
    \hat{\Psi}^* = \argmin_{\Psi \in \text{CPTP}} \|\hat{\Psi}_1 - \hat{\Psi}\|_F.
\end{equation}
An immediate application of the triangle inequality for the Frobenius norm then gives us
\begin{equation}
\label{eq:proper-dykstra-guarantee-2}
    \frac{\|\hat{\Psi}_1 - \hat{\Psi}_2\|_F}{2} \leq \frac{\|\hat{\Psi}_1 - \hat{\Psi}\|_F}{2} + \epsilon_2, \quad \forall \text{ CPTP maps } \Psi.
\end{equation}

\paragraph{Mixing to guarantee that the output is a quantum channel.}
Even though $\hat{\Psi}_2$ has desirable distance properties, as it does not achieve \emph{exact} convergence to the projection, it may not correspond to a CPTP map.
However, as our final projection was onto the set $C_2$, $\hat{\Psi}_2$ is Hermitian and satisfies the affine constraints.
Thus, the only remaining condition necessary to obtain a proper hypothesis is to obtain a positive semidefinite hypothesis.
To ensure that our final hypothesis is PSD while still retaining the other properties, we take inspiration from \cite{surawy2022projected} and mix it with the maximally mixed state.
In particular, for a suitably chosen $p > 0$, our output is
\begin{equation}
\label{eq:proper-mixer}
    \hat{\Psi}_3 = (1-p) \hat{\Psi}_2 + p \frac{I}{D},
\end{equation}
where $I$ is the $D \times D$ identity operator.
To understand the intuition behind mixing, we first note that the maximally mixed state can be viewed as the Fourier coefficient matrix of a CPTP map that applies a Pauli operator uniformly at random.
Thus, it must satisfy all required constraints.
Then, we note that if $p$ is taken large enough, $\hat{\Psi}_3$ also satisfies the constraints.
This can be seen by noting that the convex combination of two points satisfying the affine constraints will also satisfy them. For the positive semidefiniteness, we note that $\hat{\Psi}_2$ is Hermitian, and we only need to take $p$ large enough to make the most negative eigenvalue of $\hat{\Psi}_2$ map to $0$ in the eigenspectrum of $\hat{\Psi_3}$.
Using the fact that $\hat{\Psi}_2$ is not too far from a PSD matrix (\Cref{eq:proper-dykstra-guarantee-1}), we will then show that $p$ does not need to be too large. 

First, suppose that $\hat{\Psi}_2$ does have some negative eigenvalue.
Otherwise, our constraints are already satisfied, and we do not need to proceed further.
Let the negative eigenvalue of $\hat{\Psi}_2$ with the largest absolute value be $\lambda^\star$.
Clearly, by \Cref{eq:proper-mixer}, we need to take $p$ large enough such that
\begin{equation}
    (1-p)\lambda^\star + p/D \geq 0.
\end{equation}
This can be rewritten as
\begin{equation}
\label{eq:proper-p-lower}
    p \geq \frac{D|\lambda^\star|}{1 + D|\lambda^\star|}.
\end{equation}
We can now upper bound $|\lambda^\star|$ as follows.
First, recall that projection of a Hermitian operator onto the positive semidefinite cone is well-known (see, e.g.,~\cite{henrion2012projection}).
In particular, this projection works by taking the positive component of the Hermitian operator's eigenspectrum and setting the negative eigenvalues to $0$.
For a Hermitian $A$, we denote this by $(A)^+$. Then, we have
\begin{align}
    \epsilon_2 & \geq \frac{\|\hat{\Psi}_2 - \hat{\Psi}^*\|_F}{2}
    \\& \geq \frac{\|\hat{\Psi}_2 - (\hat{\Psi}_2)^+\|_F}{2}
    \\&= \frac{\sqrt{\sum_{\lambda_i < 0} \lambda_{i}^2}}{2}
    \\&\geq \frac{|\lambda^\star|}{2}
\end{align}
where the first line follows from \Cref{eq:proper-dykstra-guarantee-1}.
The second line follows from the fact that $\hat{\Psi}^*$ is PSD and $(\hat{\Psi}_2)^+$ is the nearest PSD operator to $\hat{\Psi}_2$.
In the third line, we use $\lambda_i$ to denote the eigenvalues of $\hat{\Psi}_2$ and only sum over the negative ones. In the final line, we ignore all except the negative eigenvalue with the largest absolute value. Thus, we set
\begin{equation}
\label{eq:proper-p-actual}
    p = \frac{2D\epsilon_2}{1 + 2D\epsilon_2},
\end{equation}
which satisfies \Cref{eq:proper-p-lower}.

\paragraph{Error analysis.}
Finally, it remains to bound the error between $\hat{\Psi}_2$ and $\hat{\Psi}_3$.

\begin{align}
    \frac{\|\hat{\Psi}_2 - \hat{\Psi}_3\|_F}{2} &= \frac{\|p(\hat{\Psi}_2 - I/D)\|_F}{2}
    \\&= \frac{p}{2} \norm{\hat{\Psi}_2 - \frac{I}{D}}_F
    \\&\leq \frac{p}{2} \left(\|\hat{\Psi}_2\|_F + \norm{\frac{I}{D}}_F\right)
    \\&\leq \frac{p}{2} \left(\|\hat{\Psi}_2 - \hat{\Psi}^*\|_F + \|\hat{\Psi}^*\|_F + \norm{\frac{I}{D}}_F\right)
    \\& \leq \frac{p}{2} \left(2\epsilon_2 + 1 + \frac{1}{\sqrt{D}}\right)
    \\& \leq 2p
    \\&= \frac{4D\epsilon_2}{1 + 2D\epsilon_2},
\end{align}
where in the first line we use \Cref{eq:proper-mixer}. In the third and fourth lines, we use the triangle inequality for the Frobenius norm. In the fifth line, we use \Cref{eq:proper-dykstra-guarantee-1}, the fact that $\|\hat{\Psi}^*\|_F \leq 1$ for a CPTP map 1, and we compute the Frobenius norm of the identity operator. In the sixth line, we take a loose upper bound on these quantities, and finally, in the last line we use \Cref{eq:proper-p-actual}. Finally, we need this error to be at most $\mc{O}(\epsilon_1)$. Then, by taking $\epsilon_2 \leq \epsilon_1/4D$, we obtain
\begin{equation}
\label{eq:proper-final-guarantee}
    \frac{\|\hat{\Psi}_2 - \hat{\Psi}_3\|_F}{2} \leq \epsilon_1.
\end{equation}
Then, using the triangle inequality for $d_F$ and combining \Cref{eq:proper-initial-bound,eq:proper-dykstra-guarantee-2,eq:proper-final-guarantee}, we obtain
\begin{equation}
     d_F(\Phi, \Psi_3) \leq 2 \cdot \opt(\Phi, \mc{C}) + 3\epsilon_1,
\end{equation}
which is the desired bound, up to constant factors for $\epsilon_1$. Here, we implicitly make use of the fact that $d_F(\Phi, \Psi) = \frac{\|\hat{\Phi} - \hat{\Psi}\|_F}{2}$ for any quantum channels $\Phi, \Psi$, which can be seen from \Cref{lem:superop-parseval-plancherel}.

We can now bound the time complexity of this method. In order to mix the Fourier coefficient matrix, we need to compute the smallest eigenvalue of $\hat{\Psi}_2$. We also require making alternating projections. Each of these steps can be done in time $\mathsf{poly}(D)$.
Moreover, due to linear convergence, each projection needs to be carried out $\mc{O}(\log(1/\epsilon_2)) = \mc{O}(\log(D/\epsilon_1))$ times.
This gives us a final time complexity of $\mathsf{poly}(D) \log(1/\epsilon_1)$.
Again, this analysis only holds assuming that Conjecture~\ref{conj:unsolved-condition} holds.

For $k$-junta channels, the principle submatrix that we optimize over would be of dimension $4^k \times 4^k$, giving us a time complexity $\mathsf{poly}(4^k)\log(1/\epsilon_1)$. Similarly, the dimension for low-degree channels would be at most $n^k4^k \times n^k4^k$, giving us a time complexity $\mathsf{poly}(n^k4^k)\log(1/\epsilon_1)$.

Thus, assuming Conjecture~\ref{conj:unsolved-condition} holds, we can efficiently obtain proper $2$-agnostic process tomography of junta channels and low-degree channels by applying this method to the outputs of~\Cref{thm:apt-juntas,thm:apt-low-1} respectively.

\newpage
\bibliographystyle{unsrt}
\bibliography{biblio}

@article{chen2024stabilizer,
  title={Stabilizer bootstrapping: A recipe for efficient agnostic tomography and magic estimation},
  author={Chen, Sitan and Gong, Weiyuan and Ye, Qi and Zhang, Zhihan},
  journal={arXiv preprint arXiv:2408.06967},
  year={2024}
}

@book{watrous2018theory,
  title={The theory of quantum information},
  author={Watrous, John},
  year={2018},
  publisher={Cambridge university press}
}

@article{canonne2020short,
  title={A short note on learning discrete distributions},
  author={Canonne, Cl{\'e}ment L},
  journal={arXiv preprint arXiv:2002.11457},
  year={2020}
}

@article{volberg2024noncommutative,
  title={Noncommutative Bohnenblust--Hille inequalities},
  author={Volberg, Alexander and Zhang, Haonan},
  journal={Mathematische Annalen},
  volume={389},
  number={2},
  pages={1657--1676},
  year={2024},
  publisher={Springer}
}

@inproceedings{nadimpalli2024pauli,
  title={On the Pauli Spectrum of QAC0},
  author={Nadimpalli, Shivam and Parham, Natalie and Vasconcelos, Francisca and Yuen, Henry},
  booktitle={Proceedings of the 56th Annual ACM Symposium on Theory of Computing},
  pages={1498--1506},
  year={2024}
}

@inproceedings{bao2023testing,
  title={On Testing and Learning Quantum Junta Channels},
  author={Bao, Zongbo and Yao, Penghui},
  booktitle={The Thirty Sixth Annual Conference on Learning Theory},
  pages={1064--1094},
  year={2023},
  organization={PMLR}
}

@article{grewal2024agnostic,
  title={Agnostic Tomography of Stabilizer Product States},
  author={Grewal, Sabee and Iyer, Vishnu and Kretschmer, William and Liang, Daniel},
  journal={arXiv preprint arXiv:2404.03813},
  year={2024}
}

@article{wadhwa2024noise,
  title={Noise-tolerant learnability of shallow quantum circuits from statistics and the cost of quantum pseudorandomness},
  author={Wadhwa, Chirag and Doosti, Mina},
  journal={arXiv preprint arXiv:2405.12085},
  year={2024}
}

@article{wadhwa2023learning,
  title={Learning Quantum Processes with Quantum Statistical Queries},
  author={Wadhwa, Chirag and Doosti, Mina},
  journal={arXiv preprint arXiv:2310.02075},
  year={2023}
}

@inproceedings{caro2023classical,
  title={Classical Verification of Quantum Learning},
  author={Caro, Matthias C and Hinsche, Marcel and Ioannou, Marios and Nietner, Alexander and Sweke, Ryan},
  booktitle={15th Innovations in Theoretical Computer Science Conference (ITCS 2024)},
  year={2024},
  organization={Schloss-Dagstuhl-Leibniz Zentrum f{\"u}r Informatik}
}

@article{zhao2023learning,
  title={Learning quantum states and unitaries of bounded gate complexity},
  author={Zhao, Haimeng and Lewis, Laura and Kannan, Ishaan and Quek, Yihui and Huang, Hsin-Yuan and Caro, Matthias C},
  journal={arXiv preprint arXiv:2310.19882},
  year={2023}
}

@article{arunachalam2024learning,
  title={Learning low-degree quantum objects},
  author={Arunachalam, Srinivasan and Dutt, Arkopal and Guti{\'e}rrez, Francisco Escudero and Palazuelos, Carlos},
  journal={arXiv preprint arXiv:2405.10933},
  year={2024}
}

@inproceedings{chen2023testing,
  title={Testing and learning quantum juntas nearly optimally},
  author={Chen, Thomas and Nadimpalli, Shivam and Yuen, Henry},
  booktitle={Proceedings of the 2023 Annual ACM-SIAM Symposium on Discrete Algorithms (SODA)},
  pages={1163--1185},
  year={2023},
  organization={SIAM}
}

@article{montanaro2008quantum,
  title={Quantum boolean functions},
  author={Montanaro, Ashley and Osborne, Tobias J},
  journal={Chicago Journal OF Theoretical Computer Science},
  volume={1},
  pages={1--45},
  year={2010}
}

@article{rouze2024quantum,
  title={Quantum Talagrand, KKL and Friedgut’s theorems and the learnability of quantum Boolean functions},
  author={Rouz{\'e}, Cambyse and Wirth, Melchior and Zhang, Haonan},
  journal={Communications in Mathematical Physics},
  volume={405},
  number={4},
  pages={95},
  year={2024},
  publisher={Springer}
}

@inproceedings{gur2024power,
  title={On the Power of Interactive Proofs for Learning},
  author={Gur, Tom and Jahanara, Mohammad Mahdi and Khodabandeh, Mohammad Mahdi and Rajgopal, Ninad and Salamatian, Bahar and Shinkar, Igor},
  booktitle={Proceedings of the 56th Annual ACM Symposium on Theory of Computing},
  pages={1063--1070},
  year={2024}
}

@inproceedings{goldwasser2021interactive,
  title={Interactive proofs for verifying machine learning},
  author={Goldwasser, Shafi and Rothblum, Guy N and Shafer, Jonathan and Yehudayoff, Amir},
  booktitle={12th Innovations in Theoretical Computer Science Conference (ITCS 2021)},
  year={2021},
  organization={Schloss-Dagstuhl-Leibniz Zentrum f{\"u}r Informatik}
}

@article{arunachalam2024role,
  title={On the role of entanglement and statistics in learning},
  author={Arunachalam, Srinivasan and Havlicek, Vojtech and Schatzki, Louis},
  journal={Advances in Neural Information Processing Systems},
  volume={36},
  year={2024}
}

@article{nietner2023unifying,
  title={Unifying (Quantum) Statistical and Parametrized (Quantum) Algorithms},
  author={Nietner, Alexander},
  journal={arXiv preprint arXiv:2310.17716},
  year={2023}
}

@article{huang2024learning,
  title={Learning shallow quantum circuits},
  author={Huang, Hsin-Yuan and Liu, Yunchao and Broughton, Michael and Kim, Isaac and Anshu, Anurag and Landau, Zeph and McClean, Jarrod R},
  journal={arXiv preprint arXiv:2401.10095},
  year={2024}
}

@article{nietner2023average,
  title={On the average-case complexity of learning output distributions of quantum circuits},
  author={Nietner, Alexander and Ioannou, Marios and Sweke, Ryan and Kueng, Richard and Eisert, Jens and Hinsche, Marcel and Haferkamp, Jonas},
  journal={arXiv preprint arXiv:2305.05765},
  year={2023}
}

@book{o2014analysis,
  title={Analysis of boolean functions},
  author={O'Donnell, Ryan},
  year={2014},
  publisher={Cambridge University Press}
}

@article{choi1975completely,
  title={Completely positive linear maps on complex matrices},
  author={Choi, Man-Duen},
  journal={Linear algebra and its applications},
  volume={10},
  number={3},
  pages={285--290},
  year={1975},
  publisher={Elsevier}
}

@article{jamiolkowski1972linear,
  title={Linear transformations which preserve trace and positive semidefiniteness of operators},
  author={Jamio{\l}kowski, Andrzej},
  journal={Reports on mathematical physics},
  volume={3},
  number={4},
  pages={275--278},
  year={1972},
  publisher={Elsevier}
}

@article{kuo2020markovian,
  title={Markovian entanglement dynamics under locally scrambled quantum evolution},
  author={Kuo, Wei-Ting and Akhtar, AA and Arovas, Daniel P and You, Yi-Zhuang},
  journal={Physical Review B},
  volume={101},
  number={22},
  pages={224202},
  year={2020},
  publisher={APS}
}

@article{hu2023classical,
  title={Classical shadow tomography with locally scrambled quantum dynamics},
  author={Hu, Hong-Ye and Choi, Soonwon and You, Yi-Zhuang},
  journal={Physical Review Research},
  volume={5},
  number={2},
  pages={023027},
  year={2023},
  publisher={APS}
}

@article{mele2024introduction,
  title={Introduction to Haar Measure Tools in Quantum Information: A Beginner's Tutorial},
  author={Mele, Antonio Anna},
  journal={Quantum},
  volume={8},
  pages={1340},
  year={2024},
  publisher={Verein zur F{\"o}rderung des Open Access Publizierens in den Quantenwissenschaften}
}

@inproceedings{kearns1992toward,
  title={Toward efficient agnostic learning},
  author={Kearns, Michael J and Schapire, Robert E and Sellie, Linda M},
  booktitle={Proceedings of the fifth annual workshop on Computational learning theory},
  pages={341--352},
  year={1992}
}

@article{mohseni2008quantum,
  title={Quantum-process tomography: Resource analysis of different strategies},
  author={Mohseni, Masoud and Rezakhani, Ali T and Lidar, Daniel A},
  journal={Physical Review A—Atomic, Molecular, and Optical Physics},
  volume={77},
  number={3},
  pages={032322},
  year={2008},
  publisher={APS}
}

@article{o2004quantum,
  title={Quantum process tomography of a controlled-NOT gate},
  author={O'Brien, Jeremy L and Pryde, Geoff J and Gilchrist, Alexei and James, Daniel FV and Langford, Nathan K and Ralph, Timothy C and White, Andrew G},
  journal={Physical review letters},
  volume={93},
  number={8},
  pages={080502},
  year={2004},
  publisher={APS}
}

@article{scott2008optimizing,
  title={Optimizing quantum process tomography with unitary 2-designs},
  author={Scott, Andrew James},
  journal={Journal of Physics A: Mathematical and Theoretical},
  volume={41},
  number={5},
  pages={055308},
  year={2008},
  publisher={IOP Publishing}
}

@article{chuang1997prescription,
  title={Prescription for experimental determination of the dynamics of a quantum black box},
  author={Chuang, Isaac L and Nielsen, Michael A},
  journal={Journal of Modern Optics},
  volume={44},
  number={11-12},
  pages={2455--2467},
  year={1997},
  publisher={Taylor \& Francis}
}

@article{d2001quantum,
  title={Quantum tomography for measuring experimentally the matrix elements of an arbitrary quantum operation},
  author={D'Ariano, GM and Presti, P Lo},
  journal={Physical review letters},
  volume={86},
  number={19},
  pages={4195},
  year={2001},
  publisher={APS}
}

@inproceedings{haah2023query,
  title={Query-optimal estimation of unitary channels in diamond distance},
  author={Haah, Jeongwan and Kothari, Robin and O’Donnell, Ryan and Tang, Ewin},
  booktitle={2023 IEEE 64th Annual Symposium on Foundations of Computer Science (FOCS)},
  pages={363--390},
  year={2023},
  organization={IEEE}
}

@inproceedings{aaronson2018shadow,
  title={Shadow tomography of quantum states},
  author={Aaronson, Scott},
  booktitle={Proceedings of the 50th annual ACM SIGACT symposium on theory of computing},
  pages={325--338},
  year={2018}
}

@article{huang2020predicting,
  title={Predicting many properties of a quantum system from very few measurements},
  author={Huang, Hsin-Yuan and Kueng, Richard and Preskill, John},
  journal={Nature Physics},
  volume={16},
  number={10},
  pages={1050--1057},
  year={2020},
  publisher={Nature Publishing Group UK London}
}

@article{atici2007quantum,
  title={Quantum algorithms for learning and testing juntas},
  author={At{\i}c{\i}, Alp and Servedio, Rocco A},
  journal={Quantum Information Processing},
  volume={6},
  number={5},
  pages={323--348},
  year={2007},
  publisher={Springer}
}

@article{haussler1992decision,
  title={Decision theoretic generalizations of the PAC model for neural net and other learning applications},
  author={Haussler, David},
  journal={Information and computation},
  volume={100},
  number={1},
  pages={78--150},
  year={1992},
  publisher={Elsevier}
}

@article{anshu2024survey,
  title={A survey on the complexity of learning quantum states},
  author={Anshu, Anurag and Arunachalam, Srinivasan},
  journal={Nature Reviews Physics},
  volume={6},
  number={1},
  pages={59--69},
  year={2024},
  publisher={Nature Publishing Group UK London}
}

@inproceedings{badescu2021improved,
  title={Improved quantum data analysis},
  author={B{\u{a}}descu, Costin and O'Donnell, Ryan},
  booktitle={Proceedings of the 53rd Annual ACM SIGACT Symposium on Theory of Computing},
  pages={1398--1411},
  year={2021}
}

@inproceedings{grewal2024improved,
  title={Improved stabilizer estimation via bell difference sampling},
  author={Grewal, Sabee and Iyer, Vishnu and Kretschmer, William and Liang, Daniel},
  booktitle={Proceedings of the 56th Annual ACM Symposium on Theory of Computing},
  pages={1352--1363},
  year={2024}
}

@inproceedings{kobayashi2003quantum,
  title={Quantum Merlin-Arthur proof systems: Are multiple Merlins more helpful to Arthur?},
  author={Kobayashi, Hirotada and Matsumoto, Keiji and Yamakami, Tomoyuki},
  booktitle={Algorithms and Computation: 14th International Symposium, ISAAC 2003, Kyoto, Japan, December 15-17, 2003. Proceedings 14},
  pages={189--198},
  year={2003},
  organization={Springer}
}

@article{barenco1997stabilization,
  title={Stabilization of quantum computations by symmetrization},
  author={Barenco, Adriano and Berthiaume, Andre and Deutsch, David and Ekert, Artur and Jozsa, Richard and Macchiavello, Chiara},
  journal={SIAM Journal on Computing},
  volume={26},
  number={5},
  pages={1541--1557},
  year={1997},
  publisher={SIAM}
}

@article{buhrman2001quantum,
  title={Quantum fingerprinting},
  author={Buhrman, Harry and Cleve, Richard and Watrous, John and De Wolf, Ronald},
  journal={Physical review letters},
  volume={87},
  number={16},
  pages={167902},
  year={2001},
  publisher={APS}
}

@book{nielsen2010quantum,
  title={Quantum computation and quantum information},
  author={Nielsen, Michael A and Chuang, Isaac L},
  year={2010},
  publisher={Cambridge university press}
}

@article{kueng2015qubit,
  title={Qubit stabilizer states are complex projective 3-designs},
  author={Kueng, Richard and Gross, David},
  journal={arXiv preprint arXiv:1510.02767},
  year={2015}
}

@article{webb2015clifford,
  title={The Clifford group forms a unitary 3-design},
  author={Webb, Zak},
  journal={arXiv preprint arXiv:1510.02769},
  year={2015}
}

@article{zhu2017multiqubit,
  title={Multiqubit Clifford groups are unitary 3-designs},
  author={Zhu, Huangjun},
  journal={Physical Review A},
  volume={96},
  number={6},
  pages={062336},
  year={2017},
  publisher={APS}
}

@inproceedings{lyubashevsky2010ideal,
  title={On ideal lattices and learning with errors over rings},
  author={Lyubashevsky, Vadim and Peikert, Chris and Regev, Oded},
  booktitle={Advances in Cryptology--EUROCRYPT 2010: 29th Annual International Conference on the Theory and Applications of Cryptographic Techniques, French Riviera, May 30--June 3, 2010. Proceedings 29},
  pages={1--23},
  year={2010},
  organization={Springer}
}

@book{gottesman1997stabilizer,
  title={Stabilizer codes and quantum error correction},
  author={Gottesman, Daniel},
  year={1997},
  publisher={California Institute of Technology}
}

@article{aaronson2004improved,
  title={Improved simulation of stabilizer circuits},
  author={Aaronson, Scott and Gottesman, Daniel},
  journal={Physical Review A—Atomic, Molecular, and Optical Physics},
  volume={70},
  number={5},
  pages={052328},
  year={2004},
  publisher={APS}
}

@article{gottesman1998theory,
  title={Theory of fault-tolerant quantum computation},
  author={Gottesman, Daniel},
  journal={Physical Review A},
  volume={57},
  number={1},
  pages={127},
  year={1998},
  publisher={APS}
}

@article{giovannetti2011advances,
  title={Advances in quantum metrology},
  author={Giovannetti, Vittorio and Lloyd, Seth and Maccone, Lorenzo},
  journal={Nature photonics},
  volume={5},
  number={4},
  pages={222--229},
  year={2011},
  publisher={Nature Publishing Group UK London}
}

@article{gorecki2022quantum,
  title={Quantum metrology of noisy spreading channels},
  author={G{\'o}recki, Wojciech and Riccardi, Alberto and Maccone, Lorenzo},
  journal={Physical Review Letters},
  volume={129},
  number={24},
  pages={240503},
  year={2022},
  publisher={APS}
}

@article{demkowicz2012elusive,
  title={The elusive Heisenberg limit in quantum-enhanced metrology},
  author={Demkowicz-Dobrza{\'n}ski, Rafa{\l} and Ko{\l}ody{\'n}ski, Jan and Gu{\c{t}}{\u{a}}, M{\u{a}}d{\u{a}}lin},
  journal={Nature communications},
  volume={3},
  number={1},
  pages={1063},
  year={2012},
  publisher={Nature Publishing Group UK London}
}

@article{escher2011general,
  title={General framework for estimating the ultimate precision limit in noisy quantum-enhanced metrology},
  author={Escher, BM and de Matos Filho, Ruynet Lima and Davidovich, Luiz},
  journal={Nature Physics},
  volume={7},
  number={5},
  pages={406--411},
  year={2011},
  publisher={Nature Publishing Group UK London}
}

@article{shettell2022cryptographic,
  title={Cryptographic approach to quantum metrology},
  author={Shettell, Nathan and Kashefi, Elham and Markham, Damian},
  journal={Physical Review A},
  volume={105},
  number={1},
  pages={L010401},
  year={2022},
  publisher={APS}
}

@article{huang2019cryptographic,
  title={Cryptographic quantum metrology},
  author={Huang, Zixin and Macchiavello, Chiara and Maccone, Lorenzo},
  journal={Physical Review A},
  volume={99},
  number={2},
  pages={022314},
  year={2019},
  publisher={APS}
}

@article{van2023probabilistic,
  title={Probabilistic error cancellation with sparse Pauli--Lindblad models on noisy quantum processors},
  author={Van Den Berg, Ewout and Minev, Zlatko K and Kandala, Abhinav and Temme, Kristan},
  journal={Nature physics},
  volume={19},
  number={8},
  pages={1116--1121},
  year={2023},
  publisher={Nature Publishing Group UK London}
}

@article{ferracin2024efficiently,
  title={Efficiently improving the performance of noisy quantum computers},
  author={Ferracin, Samuele and Hashim, Akel and Ville, Jean-Loup and Naik, Ravi and Carignan-Dugas, Arnaud and Qassim, Hammam and Morvan, Alexis and Santiago, David I and Siddiqi, Irfan and Wallman, Joel J},
  journal={Quantum},
  volume={8},
  pages={1410},
  year={2024},
  publisher={Verein zur F{\"o}rderung des Open Access Publizierens in den Quantenwissenschaften}
}

@article{kim2023evidence,
  title={Evidence for the utility of quantum computing before fault tolerance},
  author={Kim, Youngseok and Eddins, Andrew and Anand, Sajant and Wei, Ken Xuan and Van Den Berg, Ewout and Rosenblatt, Sami and Nayfeh, Hasan and Wu, Yantao and Zaletel, Michael and Temme, Kristan and others},
  journal={Nature},
  volume={618},
  number={7965},
  pages={500--505},
  year={2023},
  publisher={Nature Publishing Group UK London}
}

@article{tuckett2018ultrahigh,
  title={Ultrahigh error threshold for surface codes with biased noise},
  author={Tuckett, David K and Bartlett, Stephen D and Flammia, Steven T},
  journal={Physical review letters},
  volume={120},
  number={5},
  pages={050505},
  year={2018},
  publisher={APS}
}

@article{erhard2019characterizing,
  title={Characterizing large-scale quantum computers via cycle benchmarking},
  author={Erhard, Alexander and Wallman, Joel J and Postler, Lukas and Meth, Michael and Stricker, Roman and Martinez, Esteban A and Schindler, Philipp and Monz, Thomas and Emerson, Joseph and Blatt, Rainer},
  journal={Nature communications},
  volume={10},
  number={1},
  pages={5347},
  year={2019},
  publisher={Nature Publishing Group UK London}
}

@article{carignan2023error,
  title={The error reconstruction and compiled calibration of quantum computing cycles},
  author={Carignan-Dugas, Arnaud and Dahlen, Dar and Hincks, Ian and Ospadov, Egor and Beale, Stefanie J and Ferracin, Samuele and Skanes-Norman, Joshua and Emerson, Joseph and Wallman, Joel J},
  journal={arXiv preprint arXiv:2303.17714},
  year={2023}
}

@article{harper2020efficient,
  title={Efficient learning of quantum noise},
  author={Harper, Robin and Flammia, Steven T and Wallman, Joel J},
  journal={Nature Physics},
  volume={16},
  number={12},
  pages={1184--1188},
  year={2020},
  publisher={Nature Publishing Group UK London}
}

@article{flammia2020efficient,
  title={Efficient estimation of Pauli channels},
  author={Flammia, Steven T and Wallman, Joel J},
  journal={ACM Transactions on Quantum Computing},
  volume={1},
  number={1},
  pages={1--32},
  year={2020},
  publisher={ACM New York, NY, USA}
}

@article{chen2022quantum,
  title={Quantum advantages for Pauli channel estimation},
  author={Chen, Senrui and Zhou, Sisi and Seif, Alireza and Jiang, Liang},
  journal={Physical Review A},
  volume={105},
  number={3},
  pages={032435},
  year={2022},
  publisher={APS}
}

@article{flammia2021pauli,
  title={Pauli error estimation via population recovery},
  author={Flammia, Steven T and O'Donnell, Ryan},
  journal={Quantum},
  volume={5},
  pages={549},
  year={2021},
  publisher={Verein zur F{\"o}rderung des Open Access Publizierens in den Quantenwissenschaften}
}

@article{fawzi2023lower,
  title={Lower bounds on learning pauli channels},
  author={Fawzi, Omar and Oufkir, Aadil and Fran{\c{c}}a, Daniel Stilck},
  journal={arXiv preprint arXiv:2301.09192},
  year={2023}
}

@article{chen2024tight,
  title={Tight bounds on Pauli channel learning without entanglement},
  author={Chen, Senrui and Oh, Changhun and Zhou, Sisi and Huang, Hsin-Yuan and Jiang, Liang},
  journal={Physical Review Letters},
  volume={132},
  number={18},
  pages={180805},
  year={2024},
  publisher={APS}
}

@article{chen2024predicting,
  title={Predicting quantum channels over general product distributions},
  author={Chen, Sitan and Pont, Jaume de Dios and Hsieh, Jun-Ting and Huang, Hsin-Yuan and Lange, Jane and Li, Jerry},
  journal={arXiv preprint arXiv:2409.03684},
  year={2024}
}

@article{moore1999quantum,
  title={Quantum circuits: Fanout, parity, and counting},
  author={Moore, Cristopher},
  journal={arXiv preprint quant-ph/9903046},
  year={1999}
}

@article{lai2022learning,
  title={Learning quantum circuits of some T gates},
  author={Lai, Ching-Yi and Cheng, Hao-Chung},
  journal={IEEE Transactions on Information Theory},
  volume={68},
  number={6},
  pages={3951--3964},
  year={2022},
  publisher={IEEE}
}

@article{cai2023quantum,
  title={Quantum error mitigation},
  author={Cai, Zhenyu and Babbush, Ryan and Benjamin, Simon C and Endo, Suguru and Huggins, William J and Li, Ying and McClean, Jarrod R and O’Brien, Thomas E},
  journal={Reviews of Modern Physics},
  volume={95},
  number={4},
  pages={045005},
  year={2023},
  publisher={APS}
}

@article{endo2021hybrid,
  title={Hybrid quantum-classical algorithms and quantum error mitigation},
  author={Endo, Suguru and Cai, Zhenyu and Benjamin, Simon C and Yuan, Xiao},
  journal={Journal of the Physical Society of Japan},
  volume={90},
  number={3},
  pages={032001},
  year={2021},
  publisher={The Physical Society of Japan}
}

@article{seif2023shadow,
  title={Shadow distillation: Quantum error mitigation with classical shadows for near-term quantum processors},
  author={Seif, Alireza and Cian, Ze-Pei and Zhou, Sisi and Chen, Senrui and Jiang, Liang},
  journal={PRX Quantum},
  volume={4},
  number={1},
  pages={010303},
  year={2023},
  publisher={APS}
}

@article{levy2024classical,
  title={Classical shadows for quantum process tomography on near-term quantum computers},
  author={Levy, Ryan and Luo, Di and Clark, Bryan K},
  journal={Physical Review Research},
  volume={6},
  number={1},
  pages={013029},
  year={2024},
  publisher={APS}
}

@article{kunjummen2023shadow,
  title={Shadow process tomography of quantum channels},
  author={Kunjummen, Jonathan and Tran, Minh C and Carney, Daniel and Taylor, Jacob M},
  journal={Physical Review A},
  volume={107},
  number={4},
  pages={042403},
  year={2023},
  publisher={APS}
}

@article{huang2023learning,
  title={Learning to predict arbitrary quantum processes},
  author={Huang, Hsin-Yuan and Chen, Sitan and Preskill, John},
  journal={PRX Quantum},
  volume={4},
  number={4},
  pages={040337},
  year={2023},
  publisher={APS}
}

@article{cerezo2021variational,
  title={Variational quantum algorithms},
  author={Cerezo, Marco and Arrasmith, Andrew and Babbush, Ryan and Benjamin, Simon C and Endo, Suguru and Fujii, Keisuke and McClean, Jarrod R and Mitarai, Kosuke and Yuan, Xiao and Cincio, Lukasz and others},
  journal={Nature Reviews Physics},
  volume={3},
  number={9},
  pages={625--644},
  year={2021},
  publisher={Nature Publishing Group UK London}
}

@inproceedings{goldreich1989hard,
  title={A hard-core predicate for all one-way functions},
  author={Goldreich, Oded and Levin, Leonid A},
  booktitle={Proceedings of the twenty-first annual ACM symposium on Theory of computing},
  pages={25--32},
  year={1989}
}

@inproceedings{kushilevitz1991learning,
  title={Learning decision trees using the Fourier spectrum},
  author={Kushilevitz, Eyal and Mansour, Yishay},
  booktitle={Proceedings of the twenty-third annual ACM symposium on Theory of computing},
  pages={455--464},
  year={1991}
}

@article{linial1993constant,
  title={Constant depth circuits, Fourier transform, and learnability},
  author={Linial, Nathan and Mansour, Yishay and Nisan, Noam},
  journal={Journal of the ACM (JACM)},
  volume={40},
  number={3},
  pages={607--620},
  year={1993},
  publisher={ACM New York, NY, USA}
}

@inproceedings{heidari2023agnostic,
  title={Agnostic PAC Learning of $ k $-juntas Using $ L\_2 $-Polynomial Regression},
  author={Heidari, Mohsen and Szpankowski, Wojciech},
  booktitle={International Conference on Artificial Intelligence and Statistics},
  pages={2922--2938},
  year={2023},
  organization={PMLR}
}

@article{czarnik2021error,
  title={Error mitigation with Clifford quantum-circuit data},
  author={Czarnik, Piotr and Arrasmith, Andrew and Coles, Patrick J and Cincio, Lukasz},
  journal={Quantum},
  volume={5},
  pages={592},
  year={2021},
  publisher={Verein zur F{\"o}rderung des Open Access Publizierens in den Quantenwissenschaften}
}

@article{huang2022foundations,
  title={Foundations for learning from noisy quantum experiments},
  author={Huang, Hsin-Yuan and Flammia, Steven T and Preskill, John},
  journal={arXiv preprint arXiv:2204.13691},
  year={2022}
}

@article{merkel2013self,
  title={Self-consistent quantum process tomography},
  author={Merkel, Seth T and Gambetta, Jay M and Smolin, John A and Poletto, Stefano and C{\'o}rcoles, Antonio D and Johnson, Blake R and Ryan, Colm A and Steffen, Matthias},
  journal={Physical Review A—Atomic, Molecular, and Optical Physics},
  volume={87},
  number={6},
  pages={062119},
  year={2013},
  publisher={APS}
}

@article{blume2017demonstration,
  title={Demonstration of qubit operations below a rigorous fault tolerance threshold with gate set tomography},
  author={Blume-Kohout, Robin and Gamble, John King and Nielsen, Erik and Rudinger, Kenneth and Mizrahi, Jonathan and Fortier, Kevin and Maunz, Peter},
  journal={Nature communications},
  volume={8},
  number={1},
  pages={14485},
  year={2017},
  publisher={Nature Publishing Group UK London}
}

@article{giovannetti2006quantum,
  title={Quantum metrology},
  author={Giovannetti, Vittorio and Lloyd, Seth and Maccone, Lorenzo},
  journal={Physical review letters},
  volume={96},
  number={1},
  pages={010401},
  year={2006},
  publisher={APS}
}

@article{low2009learning,
  title={Learning and testing algorithms for the Clifford group},
  author={Low, Richard A},
  journal={Physical Review A—Atomic, Molecular, and Optical Physics},
  volume={80},
  number={5},
  pages={052314},
  year={2009},
  publisher={APS}
}

@article{torlai2023quantum,
  title={Quantum process tomography with unsupervised learning and tensor networks},
  author={Torlai, Giacomo and Wood, Christopher J and Acharya, Atithi and Carleo, Giuseppe and Carrasquilla, Juan and Aolita, Leandro},
  journal={Nature Communications},
  volume={14},
  number={1},
  pages={2858},
  year={2023},
  publisher={Nature Publishing Group UK London}
}

@article{harper2021fast,
  title={Fast estimation of sparse quantum noise},
  author={Harper, Robin and Yu, Wenjun and Flammia, Steven T},
  journal={PRX Quantum},
  volume={2},
  number={1},
  pages={010322},
  year={2021},
  publisher={APS}
}

@article{chung2018sample,
  title={Sample efficient algorithms for learning quantum channels in PAC model and the approximate state discrimination problem},
  author={Chung, Kai-Min and Lin, Han-Hsuan},
  journal={arXiv preprint arXiv:1810.10938},
  year={2018}
}

@article{ono2010effects,
  title={Effects of photon losses on phase estimation near the Heisenberg limit using coherent light and squeezed vacuum},
  author={Ono, Takafumi and Hofmann, Holger F},
  journal={Physical Review A—Atomic, Molecular, and Optical Physics},
  volume={81},
  number={3},
  pages={033819},
  year={2010},
  publisher={APS}
}

@article{gilbert2008use,
  title={Use of maximally entangled N-photon states for practical quantum interferometry},
  author={Gilbert, Gerald and Hamrick, Michael and Weinstein, Yaakov S},
  journal={JOSA B},
  volume={25},
  number={8},
  pages={1336--1340},
  year={2008},
  publisher={Optica Publishing Group}
}

@article{rubin2007loss,
  title={Loss-induced limits to phase measurement precision with maximally entangled states},
  author={Rubin, Mark A and Kaushik, Sumanth},
  journal={Physical Review A—Atomic, Molecular, and Optical Physics},
  volume={75},
  number={5},
  pages={053805},
  year={2007},
  publisher={APS}
}

@article{ji2008parameter,
  title={Parameter estimation of quantum channels},
  author={Ji, Zhengfeng and Wang, Guoming and Duan, Runyao and Feng, Yuan and Ying, Mingsheng},
  journal={IEEE Transactions on Information Theory},
  volume={54},
  number={11},
  pages={5172--5185},
  year={2008},
  publisher={IEEE}
}

@article{hinsche2023one,
  title={One T Gate Makes Distribution Learning Hard.},
  author={Hinsche, M and Ioannou, M and Nietner, A and Haferkamp, J and Quek, Y and Hangleiter, D and Seifert, JP and Eisert, J and Sweke, R},
  journal={Physical Review Letters},
  volume={130},
  number={24},
  pages={240602--240602},
  year={2023}
}

@article{arapinis2021quantum,
  title={Quantum physical unclonable functions: Possibilities and impossibilities},
  author={Arapinis, Myrto and Delavar, Mahshid and Doosti, Mina and Kashefi, Elham},
  journal={Quantum},
  volume={5},
  pages={475},
  year={2021},
  publisher={Verein zur F{\"o}rderung des Open Access Publizierens in den Quantenwissenschaften}
}

@article{schuster2024random,
  title={Random unitaries in extremely low depth},
  author={Schuster, Thomas and Haferkamp, Jonas and Huang, Hsin-Yuan},
  journal={arXiv preprint arXiv:2407.07754},
  year={2024}
}

@book{leung2000towards,
  title={Towards robust quantum computation},
  author={Leung, Debbie Wun Chi},
  year={2000},
  publisher={stanford university}
}

@article{acin2001optimal,
  title={Optimal estimation of quantum dynamics},
  author={Ac{\'\i}n, Antonio and Jan{\'e}, E and Vidal, Guifr{\'e}},
  journal={Physical Review A},
  volume={64},
  number={5},
  pages={050302},
  year={2001},
  publisher={APS}
}

@article{peres2002covariant,
  title={Covariant quantum measurements may not be optimal},
  author={Peres, Asher and Scudo, Petra},
  journal={Journal of Modern Optics},
  volume={49},
  number={8},
  pages={1235--1243},
  year={2002},
  publisher={Taylor \& Francis}
}

@article{hayashi2006parallel,
  title={Parallel treatment of estimation of SU (2) and phase estimation},
  author={Hayashi, Masahito},
  journal={Physics Letters A},
  volume={354},
  number={3},
  pages={183--189},
  year={2006},
  publisher={Elsevier}
}

@article{chiribella2005optimal,
  title={Optimal estimation of group transformations using entanglement},
  author={Chiribella, Giulio and D’ariano, GM and Sacchi, Massimiliano Federico},
  journal={Physical Review A—Atomic, Molecular, and Optical Physics},
  volume={72},
  number={4},
  pages={042338},
  year={2005},
  publisher={APS}
}

@article{kahn2007fast,
  title={Fast rate estimation of a unitary operation in SU (d)},
  author={Kahn, Jonas},
  journal={Physical Review A—Atomic, Molecular, and Optical Physics},
  volume={75},
  number={2},
  pages={022326},
  year={2007},
  publisher={APS}
}

@article{yang2020optimal,
  title={Optimal universal programming of unitary gates},
  author={Yang, Yuxiang and Renner, Renato and Chiribella, Giulio},
  journal={Physical review letters},
  volume={125},
  number={21},
  pages={210501},
  year={2020},
  publisher={APS}
}

@article{wang2024convergence,
  title={Convergence rate analysis of a Dykstra-type projection algorithm},
  author={Wang, Xiaozhou and Pong, Ting Kei},
  journal={SIAM Journal on Optimization},
  volume={34},
  number={1},
  pages={563--589},
  year={2024},
  publisher={SIAM}
}

@article{coles2019strong,
  title={Strong bound between trace distance and Hilbert-Schmidt distance for low-rank states},
  author={Coles, Patrick J and Cerezo, M and Cincio, Lukasz},
  journal={Physical Review A},
  volume={100},
  number={2},
  pages={022103},
  year={2019},
  publisher={APS}
}

@article{poyatos1997complete,
  title={Complete characterization of a quantum process: the two-bit quantum gate},
  author={Poyatos, JF and Cirac, J Ignacio and Zoller, Peter},
  journal={Physical Review Letters},
  volume={78},
  number={2},
  pages={390},
  year={1997},
  publisher={APS}
}

@article{strikis2021learning,
  title={Learning-based quantum error mitigation},
  author={Strikis, Armands and Qin, Dayue and Chen, Yanzhu and Benjamin, Simon C and Li, Ying},
  journal={PRX Quantum},
  volume={2},
  number={4},
  pages={040330},
  year={2021},
  publisher={APS}
}

@article{jiang2021physical,
  title={Physical implementability of linear maps and its application in error mitigation},
  author={Jiang, Jiaqing and Wang, Kun and Wang, Xin},
  journal={Quantum},
  volume={5},
  pages={600},
  year={2021},
  publisher={Verein zur F{\"o}rderung des Open Access Publizierens in den Quantenwissenschaften}
}

@article{temme2017error,
  title={Error mitigation for short-depth quantum circuits},
  author={Temme, Kristan and Bravyi, Sergey and Gambetta, Jay M},
  journal={Physical review letters},
  volume={119},
  number={18},
  pages={180509},
  year={2017},
  publisher={APS}
}

@article{regula2021operational,
  title={Operational applications of the diamond norm and related measures in quantifying the non-physicality of quantum maps},
  author={Regula, Bartosz and Takagi, Ryuji and Gu, Mile},
  journal={Quantum},
  volume={5},
  pages={522},
  year={2021},
  publisher={Verein zur F{\"o}rderung des Open Access Publizierens in den Quantenwissenschaften}
}

@article{zhao2023power,
  title={Power of quantum measurement in simulating unphysical operations},
  author={Zhao, Xuanqiang and Zhang, Lei and Zhao, Benchi and Wang, Xin},
  journal={arXiv preprint arXiv:2309.09963},
  year={2023}
}

@article{horodecki2003limits,
  title={From limits of quantum operations to multicopy entanglement witnesses and state-spectrum estimation},
  author={Horodecki, Pawe{\l}},
  journal={Physical Review A},
  volume={68},
  number={5},
  pages={052101},
  year={2003},
  publisher={APS}
}

@article{horodecki2002method,
  title={Method for direct detection of quantum entanglement},
  author={Horodecki, Pawe{\l} and Ekert, Artur},
  journal={Physical review letters},
  volume={89},
  number={12},
  pages={127902},
  year={2002},
  publisher={APS}
}

@article{fiuravsek2002structural,
  title={Structural physical approximations of unphysical maps and generalized quantum measurements},
  author={Fiur{\'a}{\v{s}}ek, Jarom{\'\i}r},
  journal={Physical Review A},
  volume={66},
  number={5},
  pages={052315},
  year={2002},
  publisher={APS}
}

@article{korbicz2008structural,
  title={Structural approximations to positive maps and entanglement-breaking channels},
  author={Korbicz, JK and Almeida, ML and Bae, Joonwoo and Lewenstein, M and Acin, A},
  journal={Physical Review A—Atomic, Molecular, and Optical Physics},
  volume={78},
  number={6},
  pages={062105},
  year={2008},
  publisher={APS}
}

@article{henrion2012projection,
  title={Projection methods in conic optimization},
  author={Henrion, Didier and Malick, J{\'e}r{\^o}me},
  journal={Handbook on Semidefinite, Conic and Polynomial Optimization},
  pages={565--600},
  year={2012},
  publisher={Springer}
}

@article{von1949rings,
  title={On rings of operators. Reduction theory},
  author={Von Neumann, John},
  journal={Annals of Mathematics},
  volume={50},
  number={2},
  pages={401--485},
  year={1949},
  publisher={JSTOR}
}

@misc{von1950functional,
  title={Functional Operators, Vol. II. Number 22 in Annals of Mathematics Studies},
  author={von Neumann, John},
  year={1950},
  publisher={Princeton University Press}
}

@article{dykstra1983algorithm,
  title={An algorithm for restricted least squares regression},
  author={Dykstra, Richard L},
  journal={Journal of the American Statistical Association},
  volume={78},
  number={384},
  pages={837--842},
  year={1983},
  publisher={Taylor \& Francis}
}

@article{han1988successive,
  title={A successive projection method},
  author={Han, Shih-Ping},
  journal={Mathematical Programming},
  volume={40},
  number={1},
  pages={1--14},
  year={1988},
  publisher={Springer}
}

@article{bauschke1994dykstra,
  title={Dykstra's alternating projection algorithm for two sets},
  author={Bauschke, Heinz H and Borwein, Jonathan M},
  journal={Journal of Approximation Theory},
  volume={79},
  number={3},
  pages={418--443},
  year={1994},
  publisher={Elsevier}
}

@inproceedings{boyle1986method,
  title={A method for finding projections onto the intersection of convex sets in Hilbert spaces},
  author={Boyle, James P and Dykstra, Richard L},
  booktitle={Advances in Order Restricted Statistical Inference: Proceedings of the Symposium on Order Restricted Statistical Inference held in Iowa City, Iowa, September 11--13, 1985},
  pages={28--47},
  year={1986},
  organization={Springer}
}

@book{deutsch2001best,
  title={Best approximation in inner product spaces},
  author={Deutsch, Frank and Deutsch, F},
  volume={7},
  year={2001},
  publisher={Springer}
}

@article{lewis2009local,
  title={Local linear convergence for alternating and averaged nonconvex projections},
  author={Lewis, Adrian S and Luke, D Russell and Malick, J{\'e}r{\^o}me},
  journal={Foundations of Computational Mathematics},
  volume={9},
  number={4},
  pages={485--513},
  year={2009},
  publisher={Springer}
}

@article{bauschke1993convergence,
  title={On the convergence of von Neumann's alternating projection algorithm for two sets},
  author={Bauschke, Heinz H and Borwein, Jonathan M},
  journal={Set-Valued Analysis},
  volume={1},
  pages={185--212},
  year={1993},
  publisher={Springer}
}

@article{bauschke2020dykstra,
  title={On Dykstra’s algorithm: finite convergence, stalling, and the method of alternating projections},
  author={Bauschke, Heinz H and Burachik, Regina S and Herman, Daniel B and Kaya, C Yal{\c{c}}{\i}n},
  journal={Optimization Letters},
  volume={14},
  pages={1975--1987},
  year={2020},
  publisher={Springer}
}

@article{bauschke2000dykstras,
  title={Dykstras algorithm with bregman projections: A convergence proof},
  author={Bauschke, Heinz H and Lewis, Adrian S},
  journal={Optimization},
  volume={48},
  number={4},
  pages={409--427},
  year={2000},
  publisher={Taylor \& Francis}
}

@article{deutsch1994rate,
  title={The rate of convergence of Dykstra's cyclic projections algorithm: The polyhedral case},
  author={Deutsch, Frank and Hundal, Hein},
  journal={Numerical Functional Analysis and Optimization},
  volume={15},
  number={5-6},
  pages={537--565},
  year={1994},
  publisher={Taylor \& Francis}
}

@article{gaffke1989cyclic,
  title={A cyclic projection algorithm via duality},
  author={Gaffke, Norbert and Mathar, Rudolf},
  journal={Metrika},
  volume={36},
  number={1},
  pages={29--54},
  year={1989},
  publisher={Springer}
}

@article{hundal1997two,
  title={Two generalizations of Dykstra’s cyclic projections algorithm},
  author={Hundal, Hein and Deutsch, Frank},
  journal={Mathematical programming},
  volume={77},
  pages={335--355},
  year={1997},
  publisher={Springer}
}

@article{pang2015set,
  title={Set intersection problems: supporting hyperplanes and quadratic programming},
  author={Pang, CH Jeffrey},
  journal={Mathematical Programming},
  volume={149},
  number={1},
  pages={329--359},
  year={2015},
  publisher={Springer}
}

@article{luo1993error,
  title={Error bounds and convergence analysis of feasible descent methods: a general approach},
  author={Luo, Zhi-Quan and Tseng, Paul},
  journal={Annals of Operations Research},
  volume={46},
  number={1},
  pages={157--178},
  year={1993},
  publisher={Springer}
}

@article{tseng2009coordinate,
  title={A coordinate gradient descent method for nonsmooth separable minimization},
  author={Tseng, Paul and Yun, Sangwoon},
  journal={Mathematical Programming},
  volume={117},
  pages={387--423},
  year={2009},
  publisher={Springer}
}

@article{shapiro2003sensitivity,
  title={Sensitivity Analysis of Generalized Equations.},
  author={Shapiro, Alexander},
  journal={Journal of Mathematical Sciences},
  volume={115},
  number={4},
  year={2003}
}

@book{bonnans2013perturbation,
  title={Perturbation analysis of optimization problems},
  author={Bonnans, J Fr{\'e}d{\'e}ric and Shapiro, Alexander},
  year={2013},
  publisher={Springer Science \& Business Media}
}

@book{nesterov2018lectures,
  title={Lectures on convex optimization},
  author={Nesterov, Yurii and others},
  volume={137},
  year={2018},
  publisher={Springer}
}

@article{surawy2022projected,
  title={Projected least-squares quantum process tomography},
  author={Surawy-Stepney, Trystan and Kahn, Jonas and Kueng, Richard and Guta, Madalin},
  journal={Quantum},
  volume={6},
  pages={844},
  year={2022},
  publisher={Verein zur F{\"o}rderung des Open Access Publizierens in den Quantenwissenschaften}
}

@inproceedings{goemans2001approximation,
  title={Approximation algorithms for MAX-3-CUT and other problems via complex semidefinite programming},
  author={Goemans, Michel X and Williamson, David},
  booktitle={Proceedings of the thirty-third annual ACM symposium on Theory of computing},
  pages={443--452},
  year={2001}
}

@article{kim2014quantum,
  title={Quantum control and process tomography of a semiconductor quantum dot hybrid qubit},
  author={Kim, Dohun and Shi, Zhan and Simmons, CB and Ward, DR and Prance, JR and Koh, Teck Seng and Gamble, John King and Savage, DE and Lagally, MG and Friesen, Mark and others},
  journal={Nature},
  volume={511},
  number={7507},
  pages={70--74},
  year={2014},
  publisher={Nature Publishing Group UK London}
}

@article{childs2012hamiltonian,
  title={Hamiltonian simulation using linear combinations of unitary operations},
  author={Childs, Andrew M and Wiebe, Nathan},
  journal={arXiv preprint arXiv:1202.5822},
  year={2012}
}

@article{heredge2024non,
  title={Non-Unitary Quantum Machine Learning},
  author={Heredge, Jamie and West, Maxwell and Hollenberg, Lloyd and Sevior, Martin},
  journal={arXiv preprint arXiv:2405.17388},
  year={2024}
}

@article{coyle2024training,
  title={Training-efficient density quantum machine learning},
  author={Coyle, Brian and Cherrat, El Amine and Jain, Nishant and Mathur, Natansh and Raj, Snehal and Kazdaghli, Skander and Kerenidis, Iordanis},
  journal={arXiv preprint arXiv:2405.20237},
  year={2024}
}

@article{clinton2021hamiltonian,
  title={Hamiltonian simulation algorithms for near-term quantum hardware},
  author={Clinton, Laura and Bausch, Johannes and Cubitt, Toby},
  journal={Nature communications},
  volume={12},
  number={1},
  pages={4989},
  year={2021},
  publisher={Nature Publishing Group UK London}
}

@article{chatterjee2024comprehensive,
  title={A Comprehensive Cross-Model Framework for Benchmarking the Performance of Quantum Hamiltonian Simulations},
  author={Chatterjee, Avimita and Rappaport, Sonny and Giri, Anish and Johri, Sonika and Proctor, Timothy and Neira, David E Bernal and Sathe, Pratik and Lubinski, Thomas},
  journal={arXiv preprint arXiv:2409.06919},
  year={2024}
}

@article{schmitt2022quantum,
  title={Quantum phase transition dynamics in the two-dimensional transverse-field Ising model},
  author={Schmitt, Markus and Rams, Marek M and Dziarmaga, Jacek and Heyl, Markus and Zurek, Wojciech H},
  journal={Science Advances},
  volume={8},
  number={37},
  pages={eabl6850},
  year={2022},
  publisher={American Association for the Advancement of Science}
}

@article{yanay2020two,
  title={Two-dimensional hard-core Bose--Hubbard model with superconducting qubits},
  author={Yanay, Yariv and Braum{\"u}ller, Jochen and Gustavsson, Simon and Oliver, William D and Tahan, Charles},
  journal={npj Quantum Information},
  volume={6},
  number={1},
  pages={58},
  year={2020},
  publisher={Nature Publishing Group UK London}
}

@article{havlivcek2019supervised,
  title={Supervised learning with quantum-enhanced feature spaces},
  author={Havl{\'\i}{\v{c}}ek, Vojt{\v{e}}ch and C{\'o}rcoles, Antonio D and Temme, Kristan and Harrow, Aram W and Kandala, Abhinav and Chow, Jerry M and Gambetta, Jay M},
  journal={Nature},
  volume={567},
  number={7747},
  pages={209--212},
  year={2019},
  publisher={Nature Publishing Group}
}

@article{abbas2021power,
  title={The power of quantum neural networks},
  author={Abbas, Amira and Sutter, David and Zoufal, Christa and Lucchi, Aur{\'e}lien and Figalli, Alessio and Woerner, Stefan},
  journal={Nature Computational Science},
  volume={1},
  number={6},
  pages={403--409},
  year={2021},
  publisher={Nature Publishing Group US New York}
}

@article{romero2017quantum,
  title={Quantum autoencoders for efficient compression of quantum data},
  author={Romero, Jonathan and Olson, Jonathan P and Aspuru-Guzik, Alan},
  journal={Quantum Science and Technology},
  volume={2},
  number={4},
  pages={045001},
  year={2017},
  publisher={IOP Publishing}
}

@article{golub1965calculating,
  title={Calculating the singular values and pseudo-inverse of a matrix},
  author={Golub, Gene and Kahan, William},
  journal={Journal of the Society for Industrial and Applied Mathematics, Series B: Numerical Analysis},
  volume={2},
  number={2},
  pages={205--224},
  year={1965},
  publisher={SIAM}
}

@book{trefethen2022numerical,
  title={Numerical linear algebra},
  author={Trefethen, Lloyd N and Bau, David},
  year={2022},
  publisher={SIAM}
}

@article{gupta2024probabilistic,
  title={Probabilistic error cancellation for dynamic quantum circuits},
  author={Gupta, Riddhi S and Van Den Berg, Ewout and Takita, Maika and Rist{\`e}, Diego and Temme, Kristan and Kandala, Abhinav},
  journal={Physical Review A},
  volume={109},
  number={6},
  pages={062617},
  year={2024},
  publisher={APS}
}

@article{geller2013efficient,
  title={Efficient error models for fault-tolerant architectures and the Pauli twirling approximation},
  author={Geller, Michael R and Zhou, Zhongyuan},
  journal={Physical Review A—Atomic, Molecular, and Optical Physics},
  volume={88},
  number={1},
  pages={012314},
  year={2013},
  publisher={APS}
}

@article{wallman2016noise,
  title={Noise tailoring for scalable quantum computation via randomized compiling},
  author={Wallman, Joel J and Emerson, Joseph},
  journal={Physical Review A},
  volume={94},
  number={5},
  pages={052325},
  year={2016},
  publisher={APS}
}

@article{arunachalam2020quantum,
  title={Quantum statistical query learning},
  author={Arunachalam, Srinivasan and Grilo, Alex B and Yuen, Henry},
  journal={arXiv preprint arXiv:2002.08240},
  year={2020}
}

@article{angrisani2023learning,
  title={Learning unitaries with quantum statistical queries},
  author={Angrisani, Armando},
  journal={arXiv preprint arXiv:2310.02254},
  year={2023}
}

@article{montanaro2013survey,
  title={A survey of quantum property testing},
  author={Montanaro, Ashley and de Wolf, Ronald},
  journal={arXiv preprint arXiv:1310.2035},
  year={2013}
}

@article{wang2011property,
  title={Property testing of unitary operators},
  author={Wang, Guoming},
  journal={Physical Review A—Atomic, Molecular, and Optical Physics},
  volume={84},
  number={5},
  pages={052328},
  year={2011},
  publisher={APS}
}

@article{demkowicz2017adaptive,
  title={Adaptive quantum metrology under general markovian noise},
  author={Demkowicz-Dobrza{\'n}ski, Rafa{\l} and Czajkowski, Jan and Sekatski, Pavel},
  journal={Physical Review X},
  volume={7},
  number={4},
  pages={041009},
  year={2017},
  publisher={APS}
}

@article{zhou2018achieving,
  title={Achieving the Heisenberg limit in quantum metrology using quantum error correction},
  author={Zhou, Sisi and Zhang, Mengzhen and Preskill, John and Jiang, Liang},
  journal={Nature communications},
  volume={9},
  number={1},
  pages={78},
  year={2018},
  publisher={Nature Publishing Group UK London}
}

@article{zhou2024achieving,
  title={Achieving metrological limits using ancilla-free quantum error-correcting codes},
  author={Zhou, Sisi and Manes, Argyris Giannisis and Jiang, Liang},
  journal={Physical Review A},
  volume={109},
  number={4},
  pages={042406},
  year={2024},
  publisher={APS}
}

@article{zhou2024limits,
  title={Limits of noisy quantum metrology with restricted quantum controls},
  author={Zhou, Sisi},
  journal={arXiv preprint arXiv:2402.18765},
  year={2024}
}
\end{document}